\documentclass[preprintnumbers,amsmath,11pt,amssymb,floatfix,superscriptaddress,nofootinbib]{article}
\usepackage{array}
\usepackage{relsize}
\usepackage{cite}
\usepackage{amsmath}
\usepackage{amssymb}
\usepackage{graphicx}
\usepackage{amsmath}
\usepackage{breqn}
\usepackage{booktabs}
\usepackage{soul}
\def\mytitle#1{\setcounter{equation}{0}
\setcounter{footnote}{0}
\begin{flushleft}\Large\textbf{#1}\end{flushleft}
\vspace{0.25cm}}
\def\myname#1{\leftline{{\large #1}}\vspace{-0.13cm}}
\def\myplace#1#2{\small\begin{flushleft}\textit{#1}\\
\texttt{#2}\end{flushleft}}

\def\myclassification#1{\small\noindent
PACS:
       #1\vspace{0.5cm}}
\usepackage{graphicx}
\begin{document}
\mytitle{Pseudo Newtonian Potential For a Rotating Kerr Black Hole Embedded in Quintessence}

\myname{Siddhartha Sankar Sarkar \footnote{sarkar.siddharth6@gmail.com} and Ritabrata Biswas\footnote{biswas.ritabrata@gmail.com}}
\myplace{Department of Mathematics, The University of Burdwan, Burdwan -713104, West Bengal, India.}{} 
 
\begin{abstract}
Pseudo-Newtonian Potential has always been a useful tool to discuss the motion of a particle in space-time to avoid the tedious and nearly impossible nonlinear computations coming from the field equations of general relativity. Mukhopadhyay, in 2002, has introduced such a pseudo-Newtonian potential for rotating Kerr black hole which is efficient enough to replicate the scenario of the classical mechanics. But there was no such model to explain the dark energy realm. In 2016 S Ghosh introduced a Lagrangian for such rotating black hole embedded in quintessence. in this article we obtained a pseudo-Newtonian force for this new black hole solution embedded quintessence. This paper introduces a simple computational scheme to evaluate a pseudo-Newtonian force for any space-time metric. This model possesses at most $4.95 \%$ error corresponding to general relativistic results. Since we took a popular agent of dark energy, i.e., quintessence into account, this is a general form of pseudo-Newtonian force to explain late time accelerating universe. In this paper, it also has been discussed about the difference between the pseudo-Newtonian force with and without dark energy effect. This paper also explains the natures of our present universe and its fate(locally around a black hole when repulsive negative pressure of dark energy is taken into account).  


\end{abstract}

\myclassification{ 95.36.+x, 98.80.Cq, 98.80.-k, 97.60.Lf}
\section{Introduction}
		The properties and behaviors of particles orbiting around a central massive object were firstly studied by famous Kepler's law and Newtonian mechanics. To keep the corresponding motion in circular orbit, i.e., to make the eccentricity $e=0$, we required a better understanding of the fact how the energy corresponding to different orbits is kept fixed. Popularly, this did lead to the definition of effective potential. In classical regime, the force acting on an orbiting particle is not just the gravitational force in its own reference frame but there also exists a centrifugal force. Effective potential is taken to be the integration of the proper combination of the forces discussed above. While moving far from the central gravitating object, effective potential starts with very high magnitude, then reduces to a minimum and again starts to increase\footnote{Providing a particle the energy exactly equal to the minimum potential mentioned above leads the particle to a motion which should be bounded between two envelopes (i.e., between a lower and upper radii of orbits). This will however not allow the particle to fall in or go away from the system.}.
		
Introduction of general relativity(GR hereafter), especially from Mercury's perihelion precision observation, leads to the idea of open orbits. Another important prediction of GR is the existence of highly dense compact objects like black holes(BH hereafter) for which the radius turns to be $r_s=2M$ (with units $G=c=1$). Using the dimensionless parameters $x=\frac{r}{r_s}$ and assuming that one particle is extremely massive and is stationary in the center mass frame, we can introduce the Newtonian gravitational potential as 
\begin{equation}
U(x)=\frac{k}{r_s(x-1)}=-\frac{GmM}{2M(x-1)}=-\frac{m}{2(x-1)},
\end{equation}
here, $M$ is the mass of the stationary BH and $m$ is the mass of the orbiting particle. This potential is called the pseudo-Newtonian potential(PNP hereafter) because it is constituted of terms of both classical physics and GR\cite{Paczynsky Wiita PNP 1980}. To find the circular orbit we have 
\begin{equation}
\dot{\theta}=\left(\frac{1}{r}\frac{dU}{dr}\right)^{\frac{1}{2}}=\left(\frac{1}{xr_s}\frac{1}{r_s}\frac{dU}{dx}\right)^{\frac{1}{2}}=\frac{1}{\sqrt{2}r_s}\left\{\frac{1}{x(x-1)^2}\right\}^{\frac{1}{2}}.
\end{equation}
With the angular momentum $l=x^2\dot{\theta}$, we have $\frac{dl}{dx}=\frac{(x-3) x^{\frac{1}{2}}}{2\sqrt{2}(x-1)^2}$.  Here it is obvious that if $x<3$, $\frac{dl}{dx}$ is negative, i.e., the circular orbit is unstable whereas for $x>3$ it is stable. We call this marginally stable orbit, i.e., $x=3$ {\it the innermost stable circular orbit} or {\it last stable orbit}, it is the transition between stable and unstable orbits.

A. Qadir and M. Sharif\cite{Quadir sharif PNP 1992} have generalized the classical gravitational potential from a small variation of Minkowski space in 1992, using the $e\psi N$-formalism(here the authors used $e\psi N$ to denote extended pseudo-Newtonian and $\psi N$ to denote pseudo-Newtonian), for a static space-time. Though it was constructed for the static space-time, it can be extensible over any arbitrary space-time. They have also provided some applications in the de Sitter metric and in the Friedmann metric. They have included a new feature which is its zero-zeroth component of $g_{\mu\nu}$, i.e., the coefficient of $dt^2$. Since they have claimed a strong correlation between tidal force and $e\psi N$ potential, this new feature enabled them to approximate this $e\psi N$ potential differently from the tidal force, although this potential is an approximation for small variation from the Minkowski metric. For deSitter metric, i.e., 
\begin{equation}
ds^2=(1-\frac{r^2}{D^2})dt^2-(1-\frac{r^2}{D^2})^{-1}dr^2-r^2d\Omega^2,
\end{equation}
where $d\Omega^2=d\theta^2+sin^2\theta d\phi^2$ is the solid angle element and $D=\sqrt{\frac{3}{\Lambda}}$ is the radial distance to the event horizon and $\Lambda$ is the cosmological constant. They derived the potential as 
\begin{equation}
V=m~ln\sqrt{1-\frac{r^2}{D^2}}.
\end{equation}
Similarly for Friedmann metric 
\begin{equation}
ds^2=dt^2-a^2(t)\left[d\chi^2+\alpha^2(\chi) d\Omega^2\right],
\end{equation}
where $\chi$ is the hyperspherical angle and the extended $\psi N$ potential was in the form
\begin{equation}
U=\frac{m\left(1+ln\sqrt(2) \right)}{2a_0}.
\end{equation}

Since almost every astronomical objects rotate, it is more justified to consider rotating Kerr metric rather than non-rotating Schwarzschild metric. This metric is not spherically symmetric but also axially symmetric about the spinning axis. To consider the direction of the orbital trajectories relative to the spin direction, we have to focus on circular orbits in the equatorial plane. The Kerr metric expressed in the Boyer-Lindquist coordinates, for rotating BH with angular momentum $J$ and mass $M$ is
\begin{equation}
ds^2=-\left(1-\frac{2M}{\Sigma}\right)dt^2-\frac{4aMr ~sin^2\theta}{\Sigma}dt d\phi + \frac{\Sigma}{\Delta}dr^2 +\Sigma d\theta^2+\left(r^2+a^2+\frac{2Mra^2~sin^2\theta}{\Sigma}\right)sin^2\theta d \phi^2,
\end{equation}
here the BH rotates in the $\phi$ direction, the terms $a$, $\Sigma$ and $\Delta$ are defined as $a\equiv \frac{J}{M}$, $\Delta \equiv r^2-2Mr+a^2$, $\Sigma \equiv r^2+a^2~cos^2 \theta$. It is to be noted that when $a=0$ the Kerr Metric reduces to the Schwarzschild metric.

In 2002, Mukhopadhyay\cite{Mukhopadhyay B PNP 2002} introduced a PNP for accretion disks around a rotating Kerr BH, which was capable to replace the general relativistic effect on accretion disk. Even, to represent energy dissipation $\eta$ in the accretion disk, Mukhopadhyay's potential possessed at most $10\%$ error in comparison of general relativistic results, for counter-rotating BHs the errors are significantly less than those of the co-rotating one. Although the PNP formed by Mukhopadhyay was applicable close to the equatorial plane, i.e., $\theta=\frac{\pi}{2}$. The pseudo-Newtonian force(PNF hereafter) mentioned above was
\begin{equation}
F_x=\frac{(x^2-2a\sqrt{x}+a^2)^2}{x^3\left\{\sqrt{x}(x-2)+a^2\right\}^2}.
\end{equation}		 
By choosing $a=0$, this potential can be reduced to the Paczynski-Wiita potential. 
 
In 2004, Jan Fukue\cite{Fukue J PNP 2004} proposed a metric correction as well as Lorentz correction on velocities calculated from the Paczynski-Wiita PNP to treat the quantities without divergence to infinity, for models of slim and accretion flow which is advection dominated. 

A similar study for a rotating charged BH in the Kerr-Newman geometry done by Ivanov and Prodanov\cite{Ivanov and Prodanov PNP 2005} in 2005 has considered the equatorial circular motion of a particle with specific charge $q \ll m$, where $m$ is the mass of the accreting particle. this leads them toward a PNF(PNF hereafter) given as
\begin{equation}
F_x=\frac{1}{x^3}\frac{J^2}{E^2}=\frac{1}{x^3}\left(\frac{J_0}{E_0}\right)^2\left(\frac{1+\frac{q}{m}\frac{J_1}{J_0}}{1+\frac{q}{m}\frac{E_1}{E_0}}\right)=F_0\left\{1+\frac{2q}{m}\left(\frac{J_1}{J_0}-\frac{E_1}{E_0}\right)+O\left[\left(\frac{q}{m}\right)^2\right]\right\},
\end{equation}
where $E$ is the conserved energy and $J$ is the angular momentum of the projected particle, $E_0$ and $J_0$ are the similar terms for a neutral particle and $E_1$ \& $J_1$ are for charged particles. This PNP can mimic the corresponding general relativistic problem.

Later on, Stuchlik and Kovar\cite{Stuchlik and Kovar PNP 2008} introduced a pseudo-Newtonian gravitational potential which described the gravitational field of static and spherically symmetric BHs in the universe with repulsive cosmological constant. The potential demonstrated by them had left just 12\% error compared with general relativistic counterpart given by Schwarzschild-deSitter geometry with the cosmological parameter $y=\frac{\Lambda M^2 }{3}~\leq 10^{-6}$. The pseudo-Newtonian gravitational force produced by them was in the form
\begin{equation}
F(x,y)=\frac{x^3y-3xy^{\frac{1}{3}}+2}{2\left(1-3y^{\frac{1}{3}}\right)(2-x+x^3y)}.
\end{equation}

In 2011, M. Sharif used the formula $v=\frac{1}{2}(kk-1)$, where $V$ is the PNP and $k$ is the killing vector for the time like isometry. Using this result a new formula was developed earlier, i.e., the approximated value of PNP as 
\begin{equation}
V\approx \frac{1}{2}Mln~(g_{00}-1),
\end{equation}
where $M$ is the mass of the central object and $g_{00}$ is the coefficient of $dt^2$. He precisely used this formula for two matrices directly drawn from String theory by Gary T. Horowitz\cite{G T Horowitz SBH},
i.e., 
\begin{equation}
ds^2=\frac{1-\frac{2m}{r}}{\left(1+\frac{2msinh^2\alpha}{r}\right)^2}dt^2+\frac{dr^2}{1-\frac{2m}{r}}+r^2d\Omega^2,
\end{equation}
where $d\Omega^2=d\theta^2+sin^2\theta d\phi^2$, and $\alpha$ is the parameter which is related to the charge of the BH with the relation $tanh^2 \alpha=\frac{Q^2}{M^2}$, where $Q$ being the charge and $m$ being the parameter related to the mass of the BH. With the help of this metric, his derived potential was 
\begin{equation}
V=-\frac{\frac{m}{r}\left(1+\frac{2msinh^4\alpha}{r}+2sinh^2\alpha\right)}{2\left(1+\frac{2msinh^2\alpha}{r}\right)}.
\end{equation}
Although there was a significant amount of similarities with a charged Kerr metric, Sharif's metric of consideration was far away from the well known Kerr metric in Boyer-Lindquist coordinates, since the metric was directly picked up for stringy BH. This metric doesn’t contain the off-diagonal element, i.e., the coefficient of  $dtd\phi$ as mentioned in the regular one. Since the key feature of this Kerr metric in Boyer-Lindquist coordinate is the cross product term of $dt$ and $d\phi$, which signifies the coupling between time and motion in the plane of rotation. As we have considered steady state Kerr metric, the Lagrangian should be independent of $t$ and $\phi$, which will give us the conjugate momenta. This is why the presence of the non-vanishing coefficient of $dtd\phi$ in the corresponding metric will be required. Along with this, the rotational parameter should be included in the metric as well as in the Lagrangian such that we can speculate its effects on these physical quantities. This single off-diagonal element is the only link to understand the asymptotic nature of the event horizon and the difference between Ergosphere.

Observational evidence from the supernova $(SNe~Ia)$ suggests that we are going through a late-time cosmic acceleration\cite{Riess Supernovae observation 1998, Perlmutter Supernovae observation 1999}. Huterer
and Turner\cite{Huterer Turner DE 1999} had suggested a new concept, dark energy(DE hereafter), which occupies nearly $70\%$ of the energy density of today's universe, apart from the topological defect\cite{re1} and a scalar field\cite{re2,re3} which revolving, called quintessence\cite{Caldwell R R et al Quint. 1998} to explain this fact.

Baking up for a moment, the classification scheme for a supernova, a late-time evolutionary stage of a star or a white dwarf ends up with a massive explosion, recently encountered in 17th October 1604, first devised by Minkowski and Zwicky. Type-Ia supernova was introduced where participants of a stellar binary are the white dwarf and a red giant, which dumped on the white dwarf making increase of mass of the smaller star up to a certain limit, i.e., $0.44M_\odot$, which is well known Chandrasekhar limit. At that point, it starts to burn its nuclear fuel as it can't support its own mass and then suddenly explodes, this causes a standard candle with an absolute magnitude of luminosity $M$. If $m$ be the apparent magnitude of luminosity and $D$ is the distance then $m=M-97.5+5logD$, which is the proper scale to measure the distance of the light source. This leads to measuring Hubble's constant $H(z)$ as a function of cosmological red-shift, used to ensure the evidence of cosmic acceleration. The existence of DE mentioned above has been confirmed independently from the observational data support of cosmic microwave background(CMB)\cite{Spergel mo 2003,Hinshaw mo 2013,Ade mo 2014,Ade mo 2016,Aghanim mo 2018} and also baryon acoustic oscillation(BAO)\cite{Eisenstein mo 2005}. Though the origin of this exotic energy, i.e., DE has not been classified yet, we have been able to obtain a key quantity, i.e., its equation of State(EOS). If a change in volume is $dV$ which requires $(-PdV)$ amount of change in energy, where $P$ is the pressure considering a vacuum container of energy increase via increment of volume which indicates $P=-\rho$, where $\rho$ is the energy density. This popular model is known as $\lambda-CDM$ model.          

If we dig further to find the DE agents, the simplest one is for $\omega=-1$, i.e., the cosmological constant $\Lambda$. This was introduced from the perspective of particle physics, but the major drawback with this approach is that the energy scale of this cosmological constant $\Lambda$ differs from the energy scale of DE with an enormously high value\cite{Winberg DE 1989}. Therefore to explain the origin of DE we needed an alternative mechanism. So classification of DE was started. The very first one is quintessence\cite{Fujii Y Quint. 1982, Ford L H Quint. 1987,Wetterich C Quint. 1988,Ratra B Quint. 1988, Chiba T et al Quint. 1997,Ferreira P G Quint. 1997, Copeland E J Quint. 1998, Caldwell R R et al Quint. 1998, Zlatev I Quint. 1999}, k-essence\cite{Chiba T et al K-ess 2000, Armendariz-Picon C K-ess 2000} and Chaplygin gas\cite{Kamenshchik A Y Chap 2001}. On the other hand, modified gravity at large distance is the second one. Quintessence is the scalar field scenario which is free from theoretical problems like Lagrange's instabilities and appearance of the ghost. Simply quintessence is a canonical scalar field which is coupled with gravity. the basic idea of quintessence is to describe a slowly varying scalar field $Q$ along with a potential $V(Q)$ which can explain the accelerated universe with equation of state $p_q=\omega_q \rho_q$, $(-1<\omega_q<\frac{1}{3})$ where $p_q$ is the quintessence pressure and $\rho_q$ is the quintessence density. The quintessence EOS parameter $\omega_q$ can be expressed as 
\begin{equation}
\omega_q=\frac{p_q}{\rho_q}=\frac{\frac{1}{2}\dot{Q}^2-V(Q)}{\frac{1}{2}\dot{Q}^2-V(Q)},
\end{equation}
where $\dot{Q}$ represents first order derivative of $Q$ with respect to proper cosmic time $t$.
Observational value suggests the values of this parameter to be $\omega_q<-0.51$ examined by Sereno M. et al.\cite{Sereno M. quint 2002} sample of 112 GRBs from the BATSE catalogue and $\omega_q>-1$ by Mortsell et al.\cite{Mortsell E. Quint 2001} From SNe Ia. Furthermore, there are two types of quintessence model, one is thawing model where Hubble friction during early cosmological epoch nearly freezes the field and another is freezing model where potential tends to be flimsy at late times causing gradual slow down of the fields.

Now we will briefly state our motivation towards this article. There are many pieces of evidence proving that an isolated object does not contain free charge as a whole. Even if we consider charge(as at the time of supernova explosions the electron of the outer layers flashes out with shocks leaving behind the preferably positively charged core which is collapsed to form a compact object) as we see in the Kerr-Newmann black hole, previously done by Ivanov and Prodanov\cite{Ivanov and Prodanov PNP 2005} and they have considered the ratio of $q$ and $m$ for further comment but, it is difficult to analyze quantitatively the charge, due to lack of observational evidences,  which makes the case with charge complicated. But it affects the surroundings when it is rotating. Indeed when a protostellar dust cloud collapses to form a stable star due to the conservation of angular momentum, the rotation is obvious to come on the stage. Stellar spots support this incident when a post-main-sequence star forms a compact object through a catastrophic collapse, in general, the rotation should be sustained. This is the reason why we have considered a rotating black hole embedded in quintessence.

In the next section, we will rebuild a PNF for regular Kerr metric in Boyer-Lindquist coordinate and then calculated the same for a Kerr metric embedded in quintessence given by S. Ghosh. In section 3 we will analyze our results graphically and will make a detailed comparison with that of B. Mukhopadhyay's result. In section 4 we will make a tabular analysis of our PNP. In the final section we will discuss our derived results in details.

\section{Basic Calculations}
Mukhopadhyay started by considering the Lagrangian density for a particle moving in equatorial plane ($\theta=\frac{\pi}{2}$) in Kerr spacetime. Then he derived the geodesic equation of motion ($E=$constant and $\lambda=$constant) to obtain an effective potential for the radial geodesic motion, i.e., $\psi$. Hereafter he has used two equations previously derived by Bardeen\cite{Bardeen BH 1973} in 1973 and hence derived the Keplerian angular momentum distribution $\lambda_k=\frac{\lambda}{k}$ to get the desired value of PNF. Whereas we have replaced the very old equation used by Bardeen and introduced new and easy equations.

In this section, we have tried to rebuild the expression of PNP with a different approach. We focused on a rotating BH solution mentioned by Mukhopadhyay\cite{Mukhopadhyay B PNP 2002}, i.e., Kerr metric in Boyer-Lindquist coordinates in the equatorial plane($\theta=\pi/2$). 
\begin{equation}
ds^2=-\left(1-\frac{2GM}{c^2r}\right)\dot{t}^2+\frac{r^2}{\Delta}\dot{r}^2-\frac{4GMa}{c^3r}\dot{t} \dot{\phi}+\left(r^2+\frac{a^2}{r^2}+\frac{2GMa^2}{c^4r}\right)\dot{\phi}^2,
\end{equation}
where $\Delta=r^2+\frac{a^2}{c^2}-\frac{2GMr}{c^2}$.\\
Hence, the Lagrangian $\mathcal{L}$\footnote{$\mathcal{L}=\frac{1}{2} g_{\mu \nu}\frac{dx^{\mu}}{d\tau}\frac{dx^{\nu}}{d\tau}$, $\tau$ is the proper time} can be treated as
\begin{equation}
2\mathcal{L}=-\left(1-\frac{2GM}{c^2r}\right)\dot{t}^2+\frac{r^2}{\Delta}\dot{r}^2-\frac{4GMa}{c^3r}\dot{t} \dot{\phi}+\left(r^2+\frac{a^2}{r^2}+\frac{2GMa^2}{c^4r}\right)\dot{\phi}^2
\end{equation}
Now from $E=-\frac{\partial\mathcal{L}}{\partial\dot{t}}$, $\lambda=\frac{\partial\mathcal{L}}{\partial\dot{\phi}}$ and the expression of $\mathcal{L}$, we can easily establish that
\begin{equation}
-m^2=-\frac{1}{\Delta}\left(r^2+\frac{a^2}{r^2}+\frac{2GMa^2}{c^4r}\right)E^2+\frac{4GMa}{c^3r\Delta}E\lambda+\frac{1}{\Delta}\left(1-\frac{2GM}{c^2r}\right)\lambda^2+\frac{r^2}{\Delta}\dot{r}^2
\end{equation}
where $\mathcal{L}$ is replaced by $-m^2$,for photon of mass $m$ to define a function $\psi$ as 
\begin{equation}
\dot{r}^2=-m^2=\frac{1}{r^2}\left(r^2+\frac{a^2}{r^2}+\frac{2GMa^2}{c^4r}\right)E^2-\frac{4GMa}{c^3r^3}E\lambda-\frac{1}{r^2}\left(1-\frac{2GM}{c^2r}\right)\lambda^2+\frac{r^2}{\Delta}\dot{r}^2-m^2\frac{\Delta}{r^2}=\psi
\end{equation}
To solve $\psi=0$ and $\frac{d\psi}{dr}=0$, we have just eliminated $m^2$ to get a quadratic equation of $E$ and $\lambda$ as
\begin{equation}
A~\lambda^2+B~E\lambda+C~E^2=0
\end{equation}
where $A=\frac{2 a^2 G M-2 r \left(c^2 r-2 G M\right)^2}{c^4 r^6}$, $B=-\frac{4 a G M \left\{a^2+r \left(3 c^2 r-4 G M\right)\right\}}{c^5 r^6}$ and $C=\frac{2 G M \left\{a^4+2 a^2 r \left(c^2 r-2 G M\right)+c^4 r^4\right\}}{c^6 r^6}$. 
\\Solving this, we can derive the value of $\lambda_k$\footnote{$\lambda_k=\frac{\lambda}{E}$}(ratio of $\lambda$ and $E$ in Keplerian orbit) as
\begin{equation}
\lambda_k=\frac{16 a^2 G M \left[a^4 \left\{G M \left(r^2+1\right)-c^2 r^3\right\}-2 a^2 r \left\{c^4 r^4-3 c^2 G M \left(r^2+1\right) r+2 G^2 M^2 \left(r^2+2\right)\right\}\right]}{c^{10} r^{12}}
\end{equation}
$$+\frac{16 a^2 G M r^2 \left[-c^6 r^5+c^4 G M r^2 \left(r^2+9\right)-24 c^2 G^2 M^2 r+16 G^3 M^3\right]}{c^{10} r^{12}}$$
For preudo Newtonian force $F_x$, we get, 
\begin{equation}
F_x=\frac{\left\{a^3+x^6 \sqrt{\frac{\left(a^2-2x+x^2\right)^2}{x^9}}+a (3 x-4) x\right\}^2}{x^3 \left\{a^2-(x-2)^2 x\right\}^2}.
\end{equation}
Now, we will start with the spacetime metric for a rotating BH embedded in quintessence which was given by Sushanth G. Ghosh\cite{Ghosh S G PNP 2016} in 2016 as
\begin{equation}
ds^2=-\frac{\Delta-a^2sin^2\theta}{\Sigma}dt^2+\frac{\Sigma}{\Delta}dr^2-2asin^2\theta\left(1-\frac{\Delta-a^2sin^2\theta}{\Sigma}\right)dt d\phi
\end{equation}
$$+\Sigma d\theta^2+sin^2\theta\left[\Sigma+a^2sin^2\theta\left(2-\frac{\Delta-a^2sin^2\theta}{\Sigma}\right)\right]d\phi^2$$
where $\Sigma=r^2+a^2cos^2\theta$ and $\Delta=r^2+a^2+2Mr-\frac{{\cal A}_q}{r^{3\omega_q-1}}$.\\
Using this for $\theta=\pi/2$ we can express the Lagrangian $\mathcal{L}$ as
\begin{equation}
2\mathcal{L}=-\frac{\Delta-a^2}{\Sigma}dt^2+\frac{\Sigma}{\Delta}dr^2-2a\left(1-\frac{\Delta-a^2}{\Sigma}\right)dt d\phi+\left[\Sigma+a^2\left(2-\frac{\Delta-a^2}{\Sigma}\right)\right]d\phi^2
\end{equation}
where $\Sigma=r^2$ and $\Delta=r^2+a^2+2Mr-\frac{{\cal A}_q}{r^{3\omega_q-1}}$.\\
Now from $E=-\frac{\partial\mathcal{L}}{\partial\dot{t}}$ and $\lambda=\frac{\partial\mathcal{L}}{\partial\dot{\phi}}$ and the expression of $\mathcal{L}$ we can easily establish that
\begin{equation}
-m^2=-\frac{1}{\Delta}\left[r^2+a^2\left(2-\frac{\Delta-a^2}{r^2}\right)\right]E^2+\frac{2a}{\Delta}\left(1-\frac{\Delta-a^2}{r^2}\right)E\lambda+\frac{1}{\Delta}\frac{\Delta-a^2}{r^2}\lambda^2+\frac{r^2}{\Delta}\dot{r}^2
\end{equation}
where $\mathcal{L}$ is replaced by $-m^2$ to define a function $\psi$ as 
\begin{equation}
\dot{r}^2=\frac{1}{r^2}\left[r^2+a^2\left(2-\frac{\Delta-a^2}{r^2}\right)\right]E^2-\frac{2a}{r^2}\left(1-\frac{\Delta-a^2}{r^2}\right)E\lambda-\frac{1}{r^2}\frac{\Delta-a^2}{r^2}\lambda^2-m^2\frac{\Delta}{r^2}=\psi
\end{equation}
To solve $\psi=0$ and $\frac{d\psi}{dr}=0$ we just eliminated $m^2$ to get a quadratic equation of $E$ and $\lambda$ as
\begin{equation}
A~\lambda^2+B~E\lambda+C~E^2=0
\end{equation}
where $A=r^{-6 (\omega_q+1)} \left[a^2 r^{3 \omega_q} \left({\cal A}_q +2 M r^{3 \omega_q}+3 {\cal A}_q  \omega_q\right)-2 r \left({\cal A}_q +2 M r^{3 \omega_q}-r^{3 \omega_q+1}\right)^2\right]$,
\\$B=-2 a r^{-6 (\omega_q+1)} \left[a^2 r^{3 \omega_q} \left({\cal A}_q +2 M r^{3 \omega_q}+3 {\cal A}_q  \omega_q\right)+r \left\{-8 M^2 r^{6 \omega_q}+M \left(6 r^{6 \omega_q+1}-8 {\cal A}_q r^{3 \omega_q}\right)\right. \right.$
\\$~~~~~~~~~~~~~~~~~~~~~~~~~~~~~~~~~~~~~~~~~~~~~~~~~~~~~~~~~~~~~~~~~~~~~~~~~~~~-\left. \left. 3 {\cal A}_q (\omega_q+1) r^{3 \omega_q+1}-2 {\cal A}_q^2\right\}\right] $ \\and $
C=r^{-6 (\omega_q+1)} \left[a^4 r^{3 \omega_q} \left({\cal A}_q +2 M r^{3 \omega_q}+3 {\cal A}_q  \omega_q\right)+2 a^2 r \left\{-4 M^2 r^{6 \omega_q}+2 M r^{3 \omega_q} \left(r^{3 \omega_q+1}-2 {\cal A}_q \right)\right.\right.$ 
\\$~~~~~~~~~~~~~~~~~~~~~~~~~~~~~~~~~~~~~~~~~~~~~~~+\left. \left. {\cal A}_q (3 \omega_q+1) r^{3 \omega_q+1}-{\cal A}_q^2\right\} +r^{3 \omega_q+4} \left({\cal A}_q +2 M r^{3 \omega_q}+3 {\cal A}_q  \omega_q \right)\right]$
\\Solving this we can derive the value of $\lambda_k$ as
\begin{equation}
\lambda=\frac{-a {\cal A}_q  r^{3 \omega_q} \left[a^2 (3 \omega_q+1)+r \left\{3 r (\omega_q+1)-8 M\right\}\right]-2 a M r^{6 \omega_q} \left\{a^2+r (3 r-4 M)\right\}+2 a {\cal A}_q ^2 r}{2 r \left\{{\cal A}_q +(2 M-r) r^{3 \omega_q}\right\}^2-a^2 r^{3 \omega_q} \left\{{\cal A}_q +2 M r^{3 \omega_q}+3 {\cal A}_q  \omega_q \right\}}
\end{equation}
$$+\frac{\sqrt{2} r^{6 \omega_q+6} \sqrt{r^{-9 (\omega_q+1)} \left({\cal A}_q +2 M r^{3 \omega_q}+3{\cal A}_q  \omega_q \right) \left[r^{3 \omega_q} \left\{a^2+r (r-2 M)\right\}-{\cal A}_q  r\right]^2}}{2 r \left\{{\cal A}_q +(2 M-r) r^{3 \omega_q}\right\}^2-a^2 r^{3 \omega_q} \left({\cal A}_q +2 M r^{3 \omega_q}+3 {\cal A}_q  \omega_q\right)}$$
Using $F_x=\frac{\lambda_k^2}{x^3}$ we have obtained PNF as
$$F_x=\left[ a {\cal A}_q  x^{3 \omega_q} \left\{ a^2 (3 \omega_q+1)+ 3x^2 (\omega_q+1)-8x\right\}+2 a \left\{a^2+x (3 x-4)\right\} x^{6 \omega_q}-2 a {\cal A}_q ^2 x\right.$$
\begin{equation}\label{PNP_quint}
\frac{\left. -\sqrt{2} x^{6 \omega_q+6}  \left\{ \left(a^2+x^2-2x\right) x^{3 \omega_q}-{\cal A}_q x\right\}\sqrt{x^{-9 (\omega_q+1)} \left({\cal A}_q +3 {\cal A}_q  \omega_q+2 x^{3 \omega_q}\right)}\right]^2}{x^3 \left[a^2 x^{3 \omega_q} \left({\cal A}_q +3 {\cal A}_q  \omega_q +2 x^{3 \omega_q}\right)-2 x \left\{{\cal A}_q +(2-x) x^{3 \omega_q}\right\}^2\right]^2}
\end{equation}

\section{Graphical Interpretations}
\begin{figure}
\centering
Fig $1.1.a$ \hspace{1.6 cm} Fig $1.1.b$ \hspace{1.6 cm} Fig $1.1.c$ \hspace{1.6 cm} Fig $1.1.d$
\includegraphics*[scale=0.3]{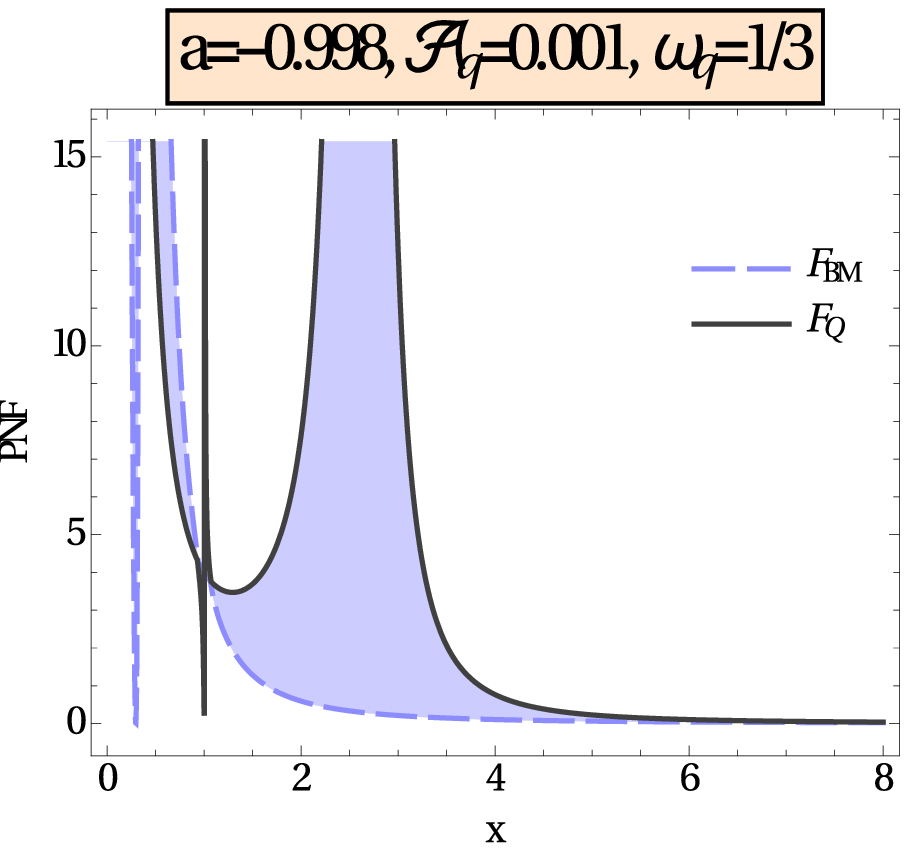}~~
\hspace*{0.5 cm}
\includegraphics*[scale=0.3]{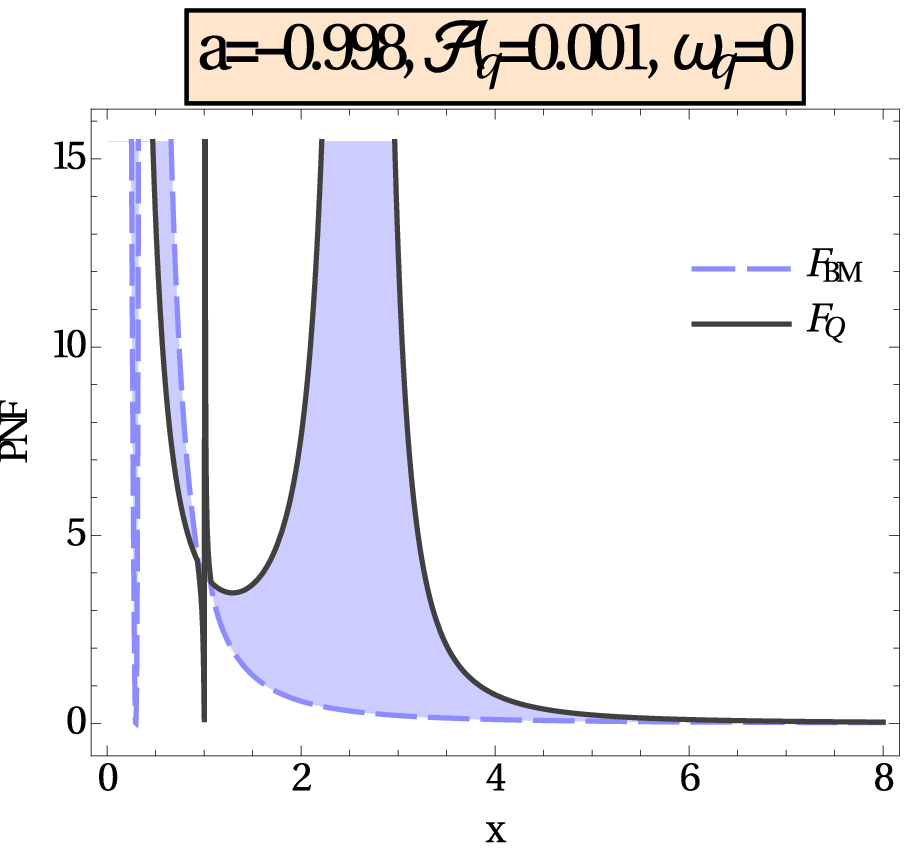}~~
\hspace*{0.5 cm}
\includegraphics*[scale=0.3]{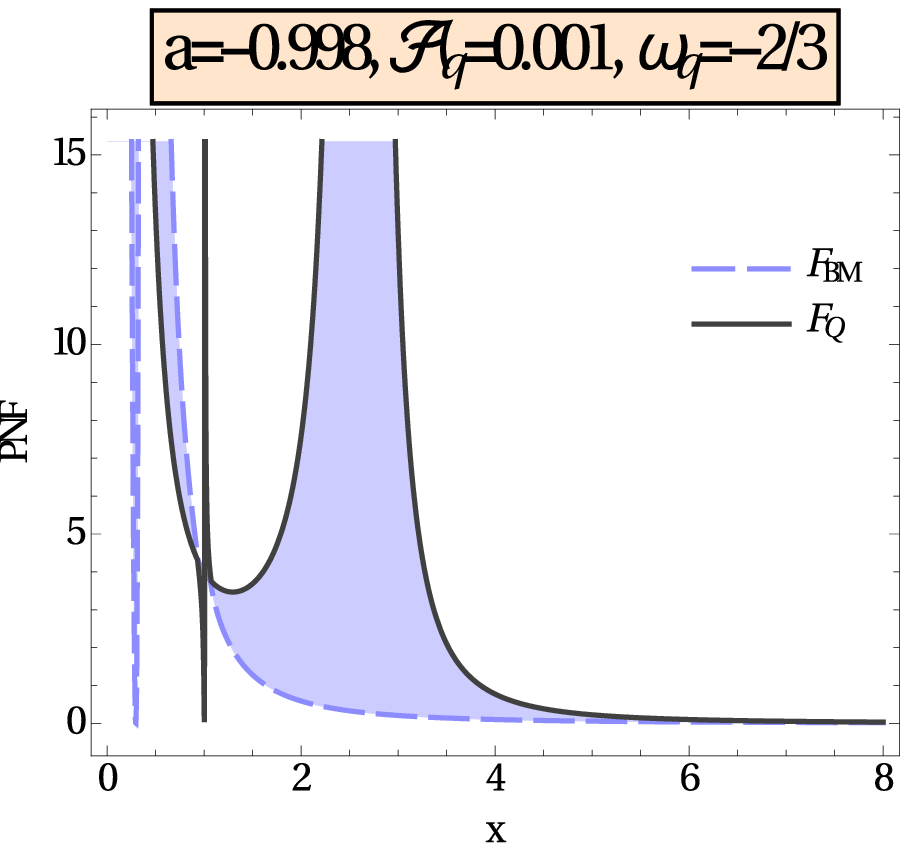}~~
\hspace*{0.5 cm}
\includegraphics*[scale=0.3]{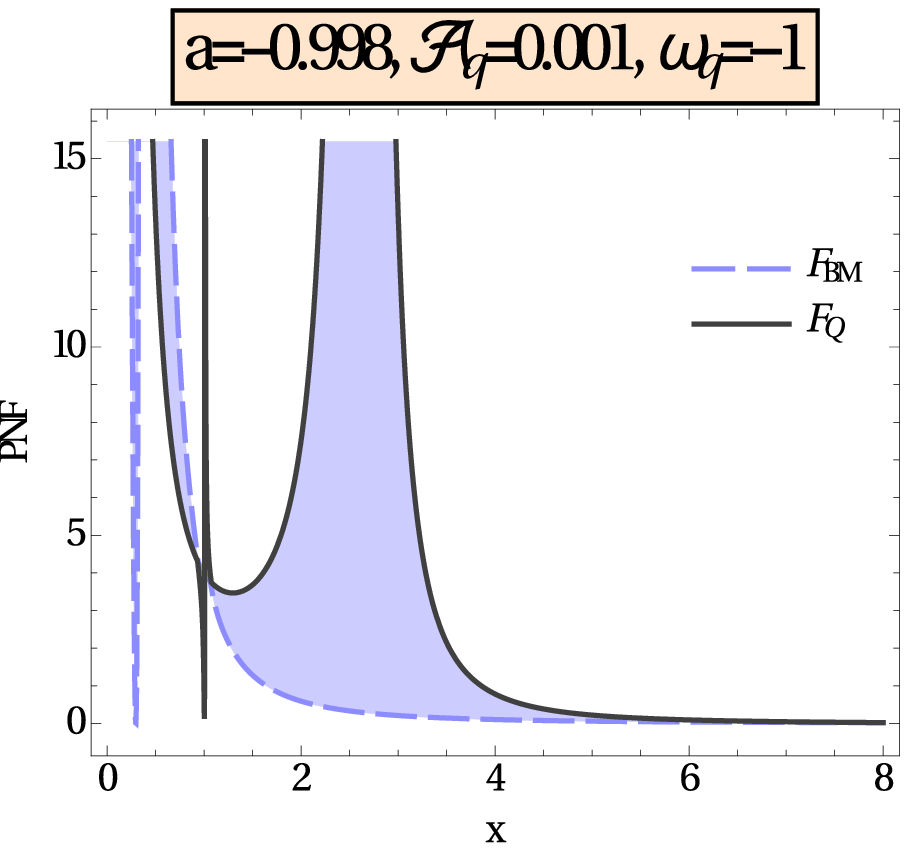}~~\\

Fig $1.2.a$ \hspace{1.6 cm} Fig $1.2.b$ \hspace{1.6 cm} Fig $1.2.c$ \hspace{1.6 cm} Fig $1.2.d$
\includegraphics*[scale=0.3]{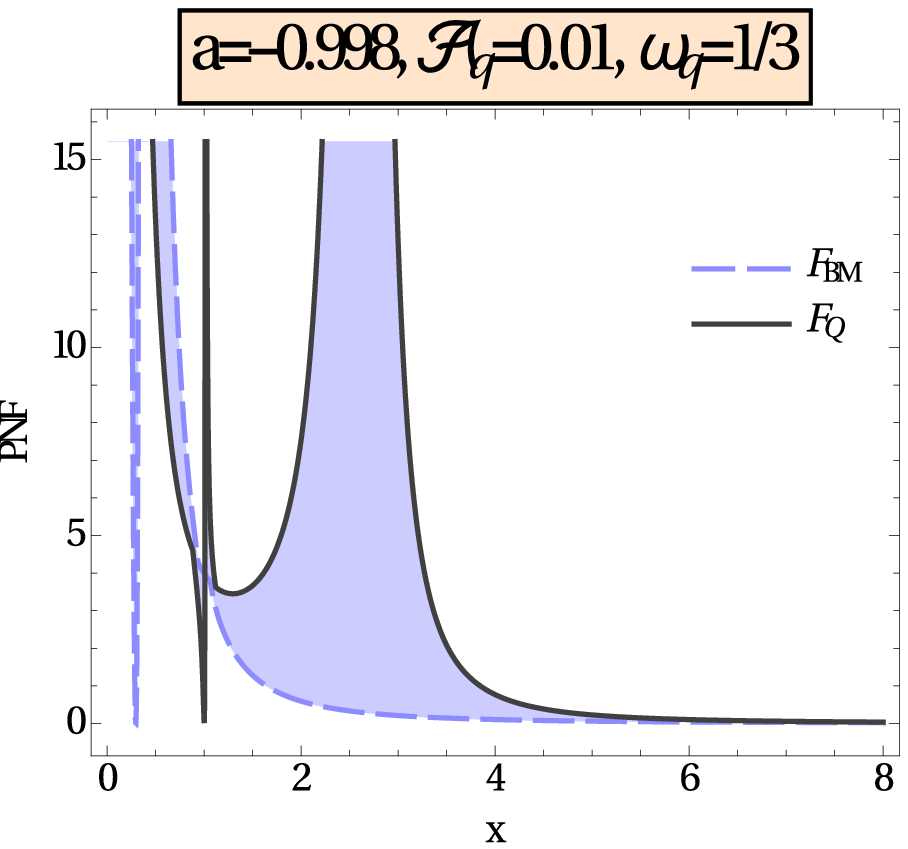}~~
\hspace*{0.5 cm}
\includegraphics*[scale=0.3]{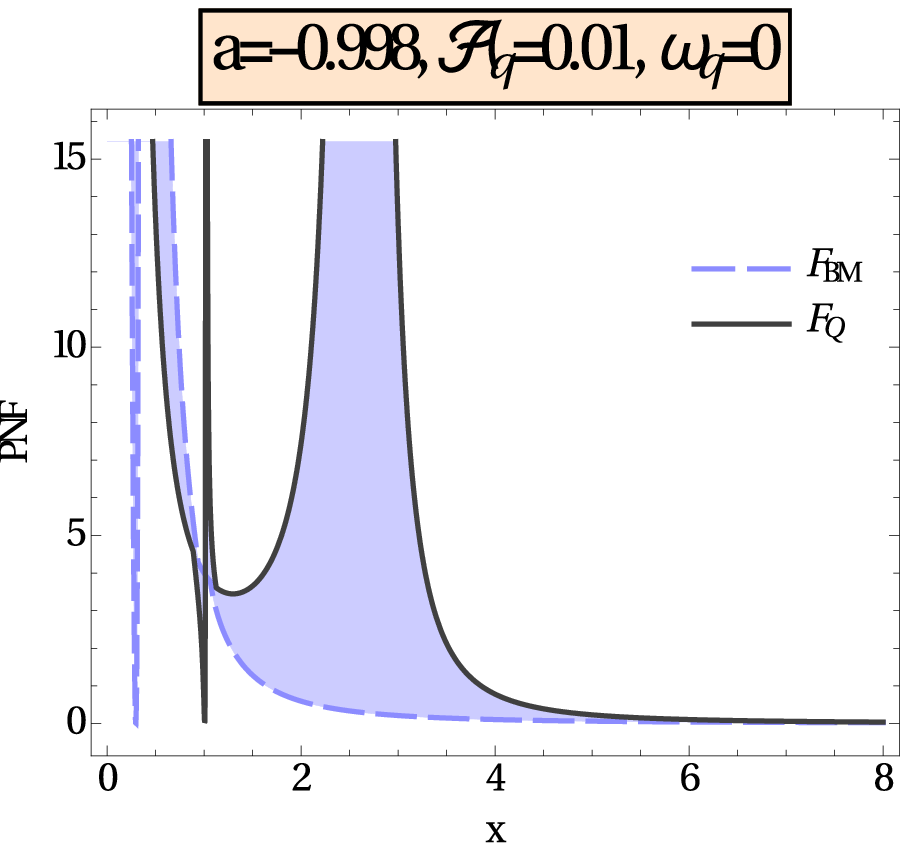}~~
\hspace*{0.5 cm}
\includegraphics*[scale=0.3]{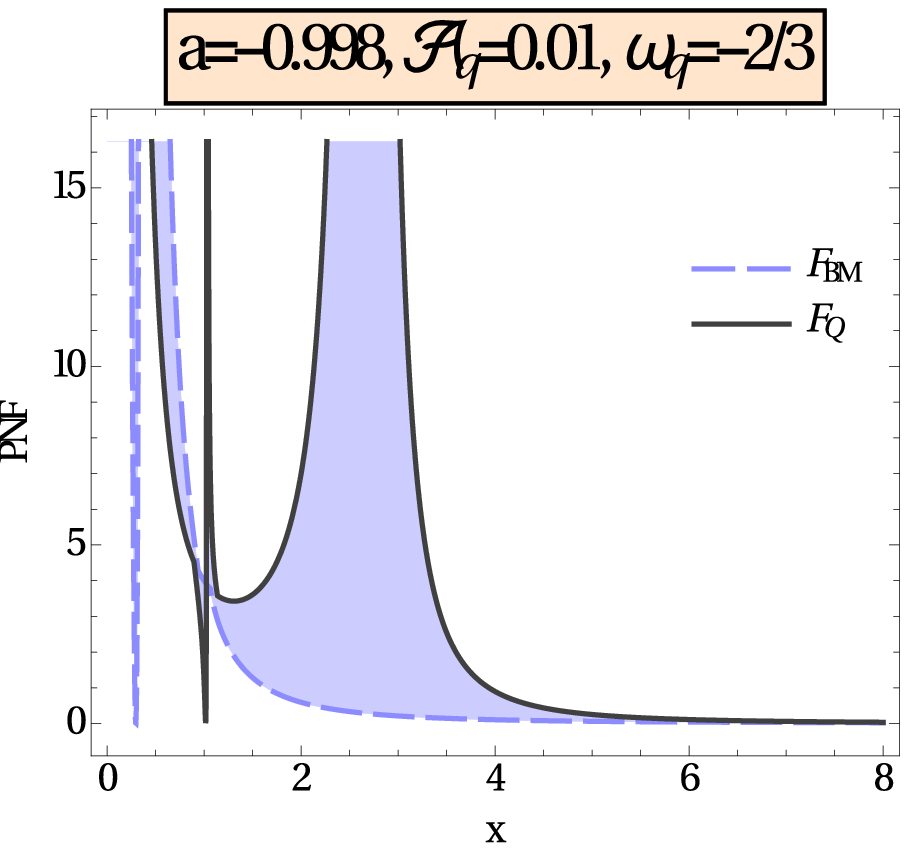}~~
\hspace*{0.5 cm}
\includegraphics*[scale=0.3]{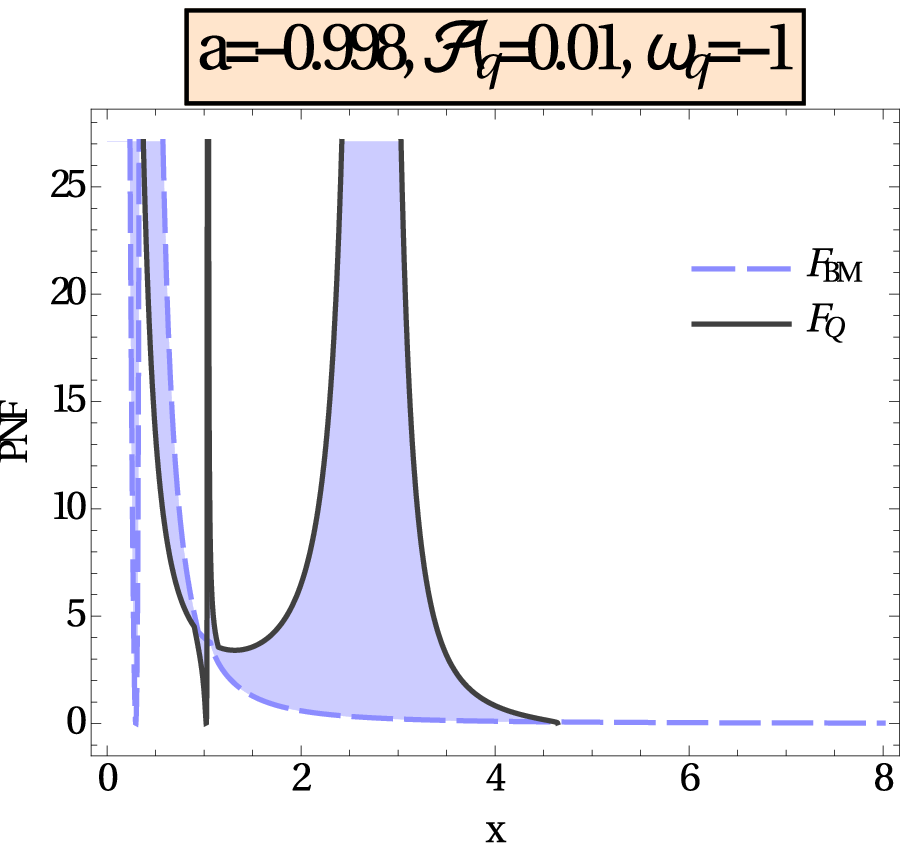}~~\\

Fig $1.3.a$ \hspace{1.6 cm} Fig $1.3.b$ \hspace{1.6 cm} Fig $1.3.c$ \hspace{1.6 cm} Fig $1.3.d$
\includegraphics*[scale=0.3]{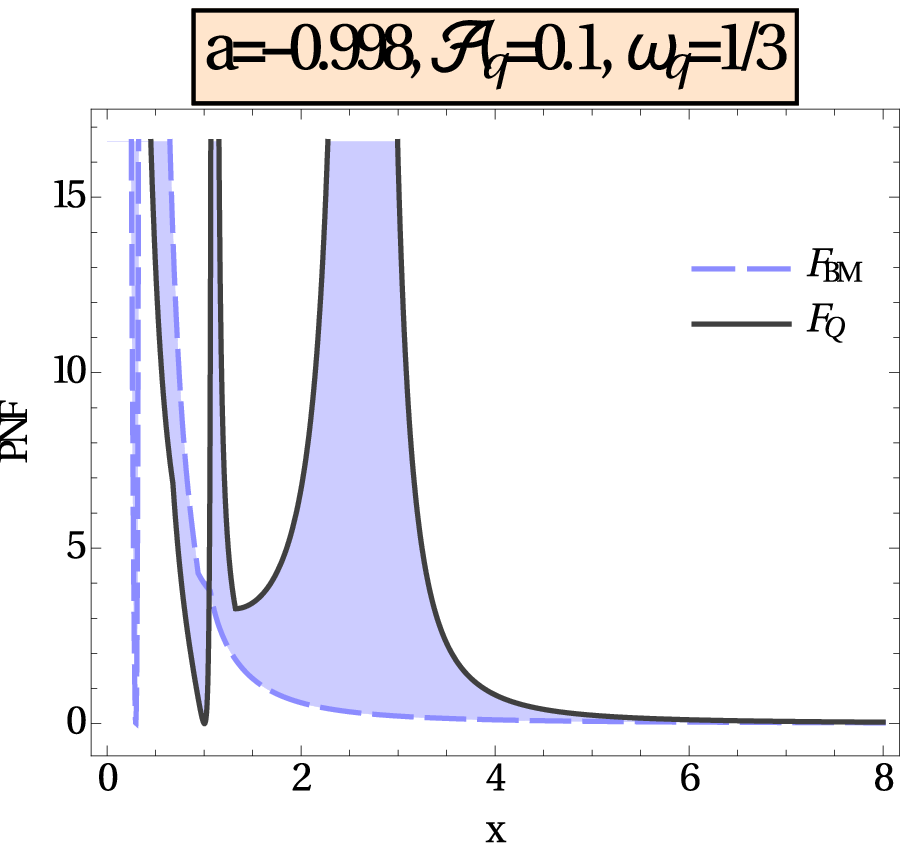}~~
\hspace*{0.5 cm}
\includegraphics*[scale=0.3]{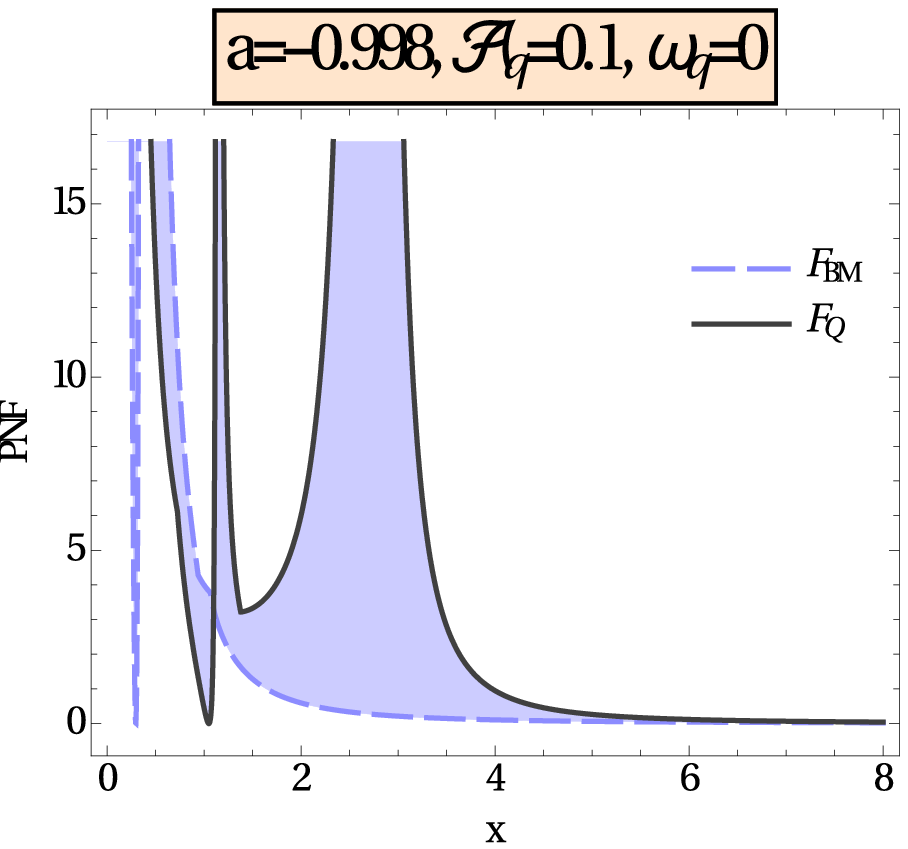}~~
\hspace*{0.5 cm}
\includegraphics*[scale=0.3]{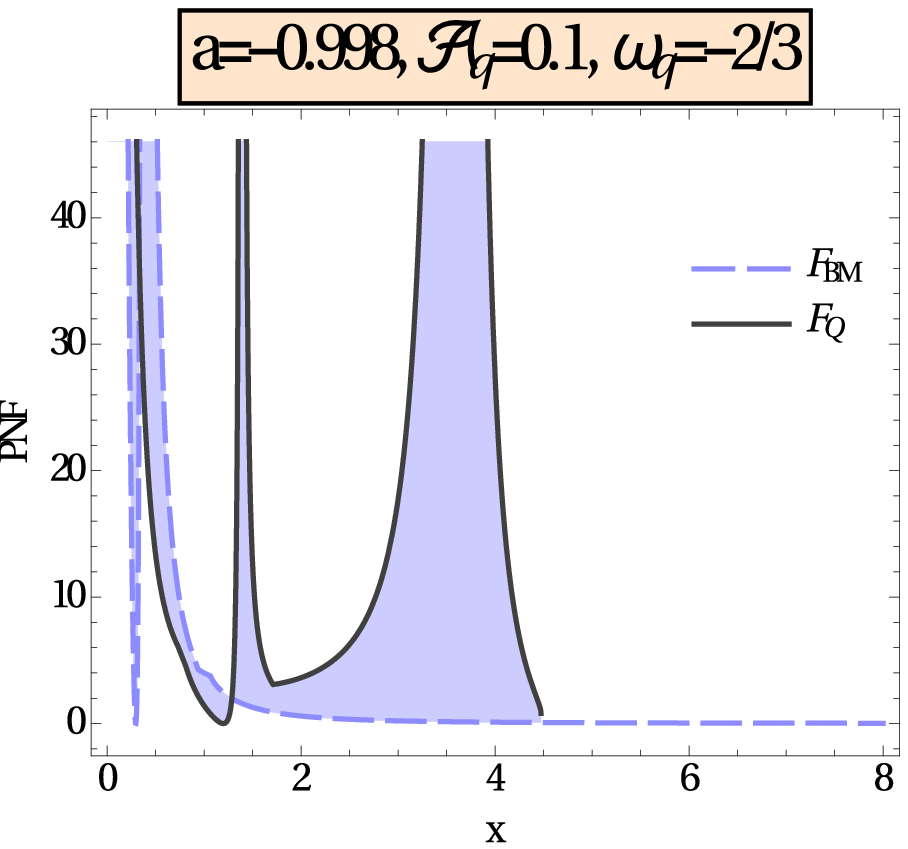}~~
\hspace*{0.5 cm}
\includegraphics*[scale=0.3]{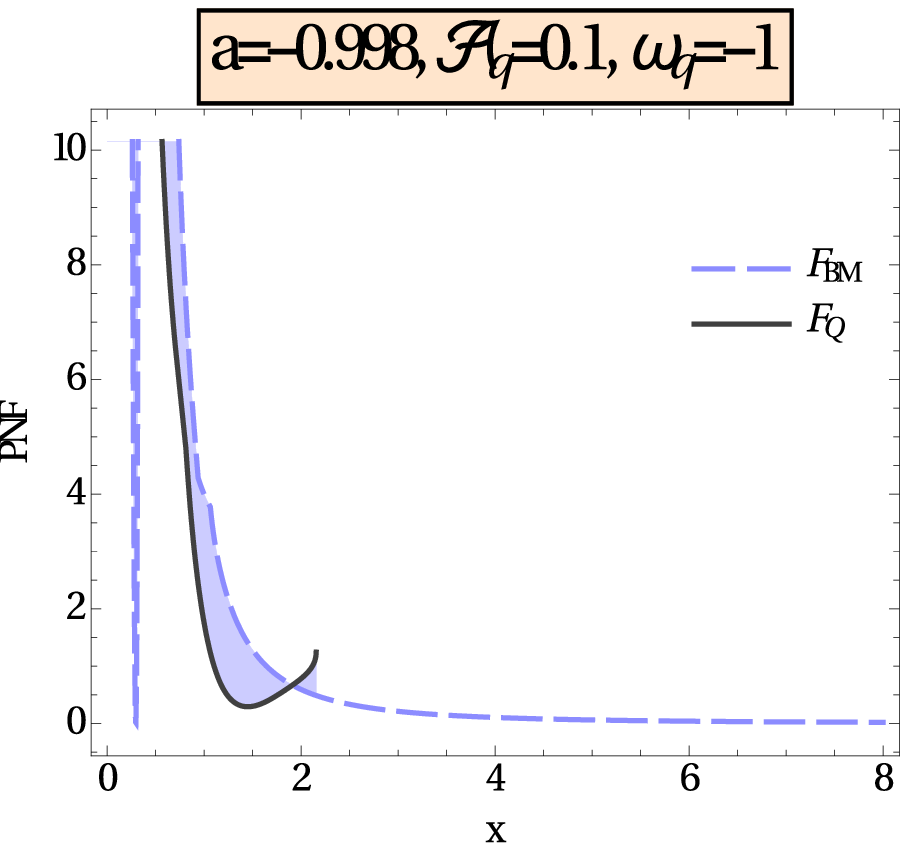}~~\\

\it{Here $a=-0.998$, for each values of ${\cal A}_q$, i.e., $10^{-3}$, $10^{-2}$ and $10^{-1}$ we have drawn a relative results for $\omega_q=\frac{1}{3}$ for radiation, $\omega_q=0$ for dust, $\omega_q=-\frac{2}{3}$ for quintessence and $\omega_q=-1$ for phantom barrier.}\\

Fig $1.4.a$ \hspace{1.6 cm} Fig $1.4.b$ \hspace{1.6 cm} Fig $1.4.c$ \hspace{1.6 cm} Fig $1.4.d$
\includegraphics*[scale=0.3]{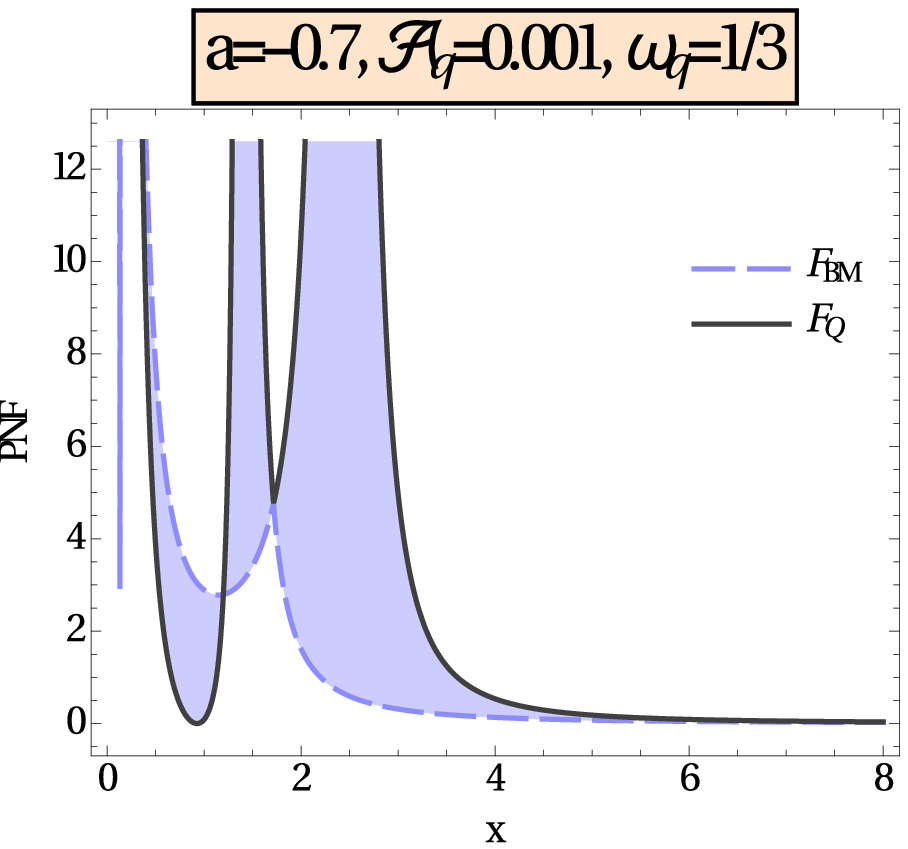}~~
\hspace*{0.5 cm}
\includegraphics*[scale=0.3]{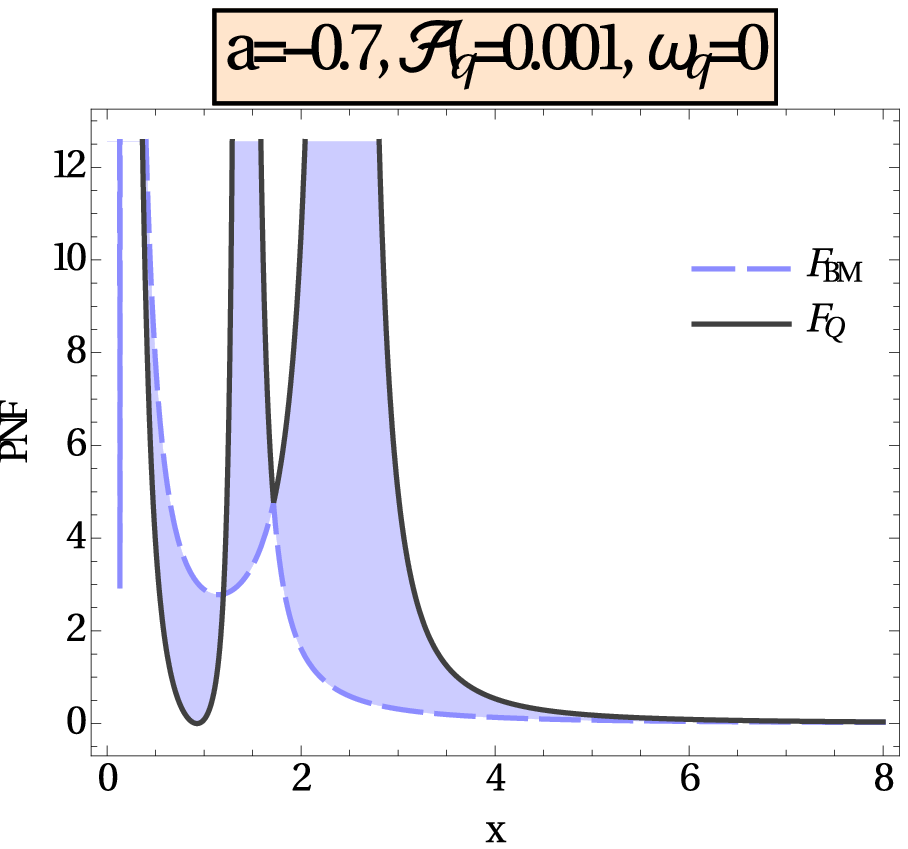}~~
\hspace*{0.5 cm}
\includegraphics*[scale=0.3]{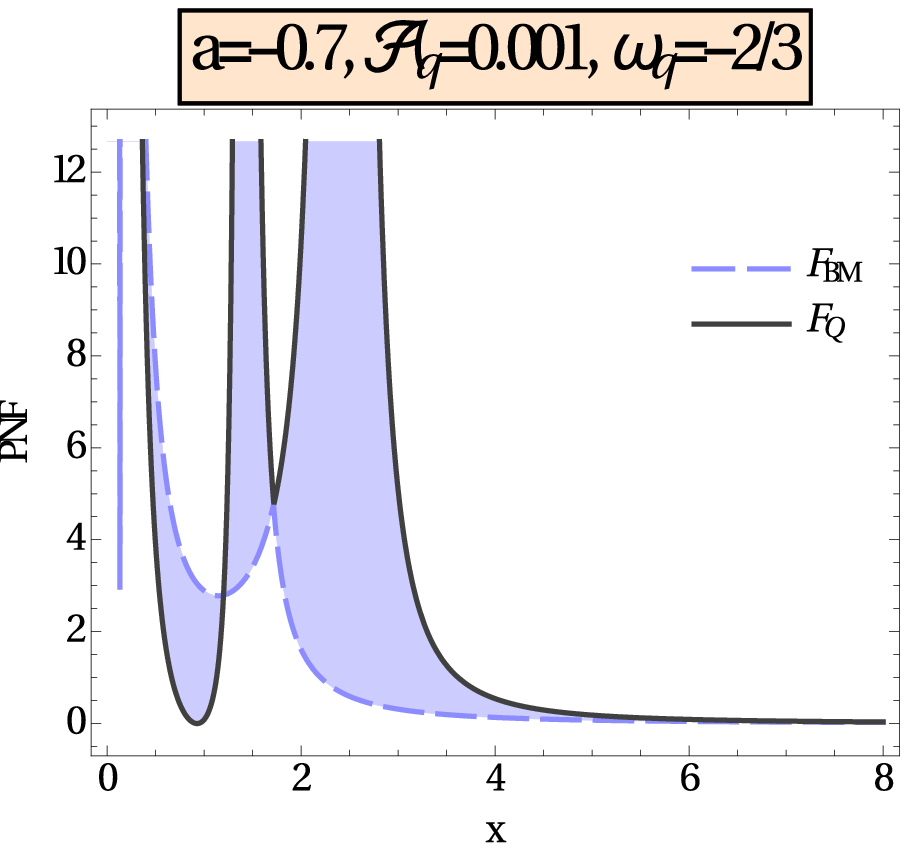}~~
\hspace*{0.5 cm}
\includegraphics*[scale=0.3]{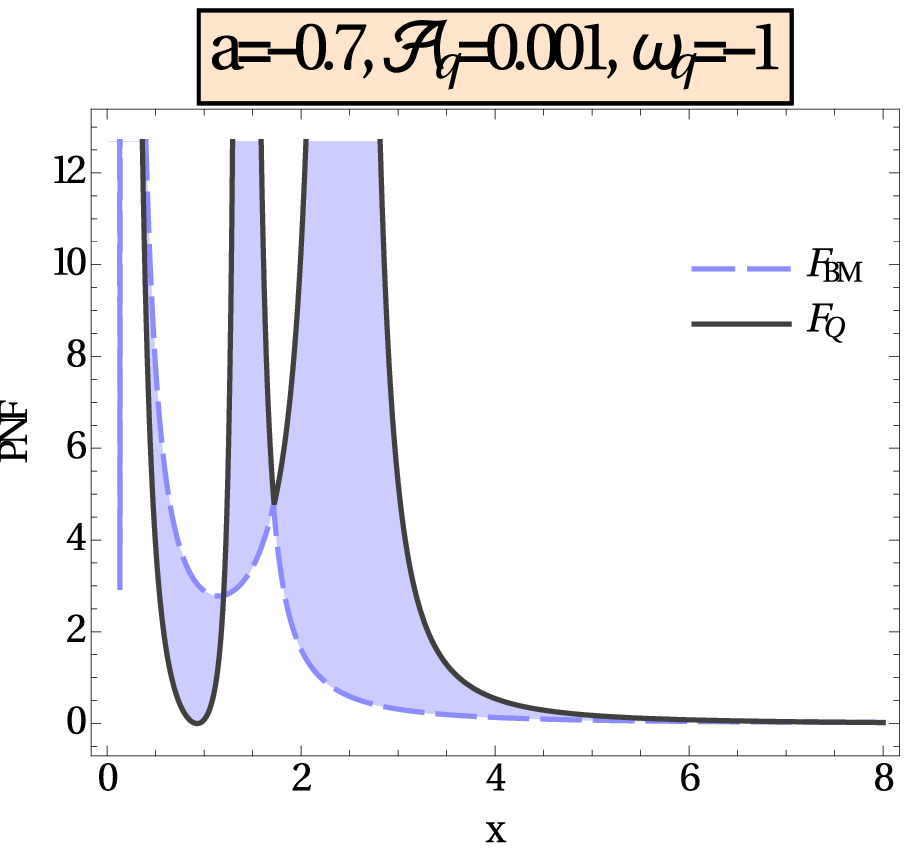}~~\\

Fig $1.5.a$ \hspace{1.6 cm} Fig $1.5.b$ \hspace{1.6 cm} Fig $1.5.c$ \hspace{1.6 cm} Fig $1.5.d$
\includegraphics*[scale=0.3]{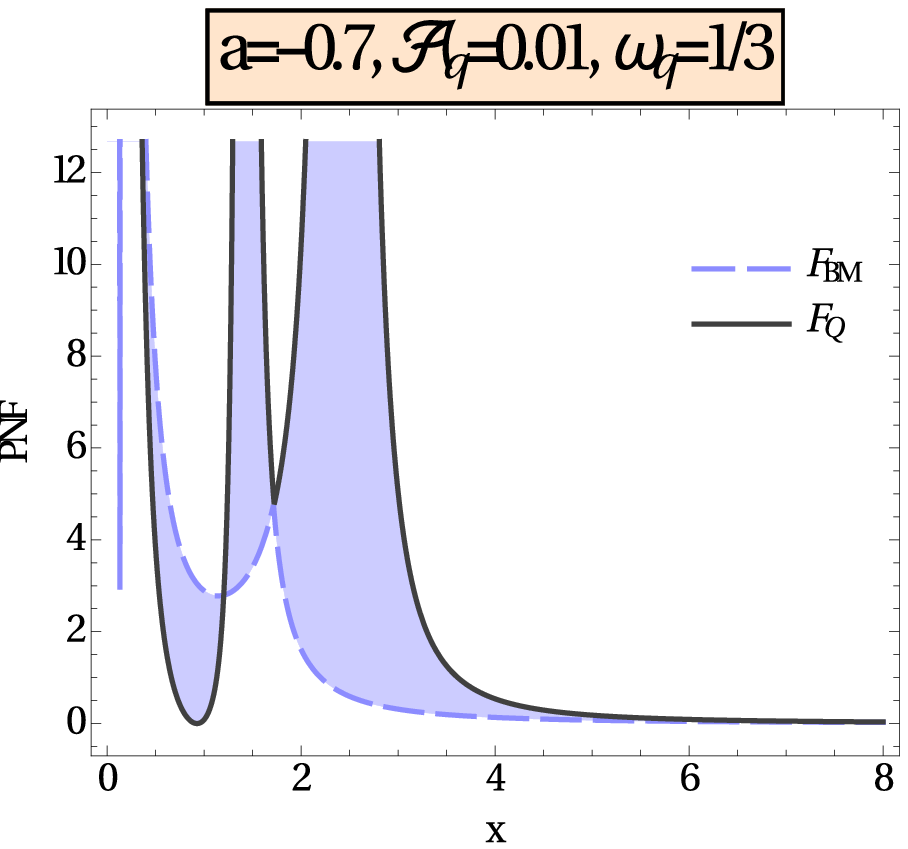}~~
\hspace*{0.5 cm}
\includegraphics*[scale=0.3]{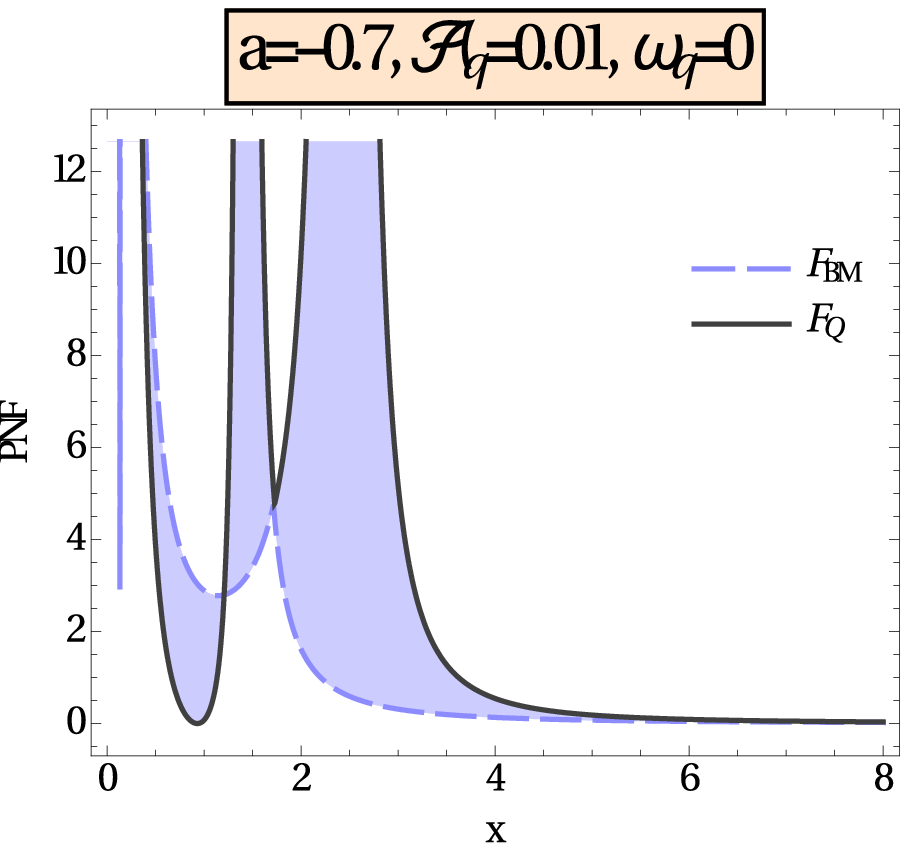}~~
\hspace*{0.5 cm}
\includegraphics*[scale=0.3]{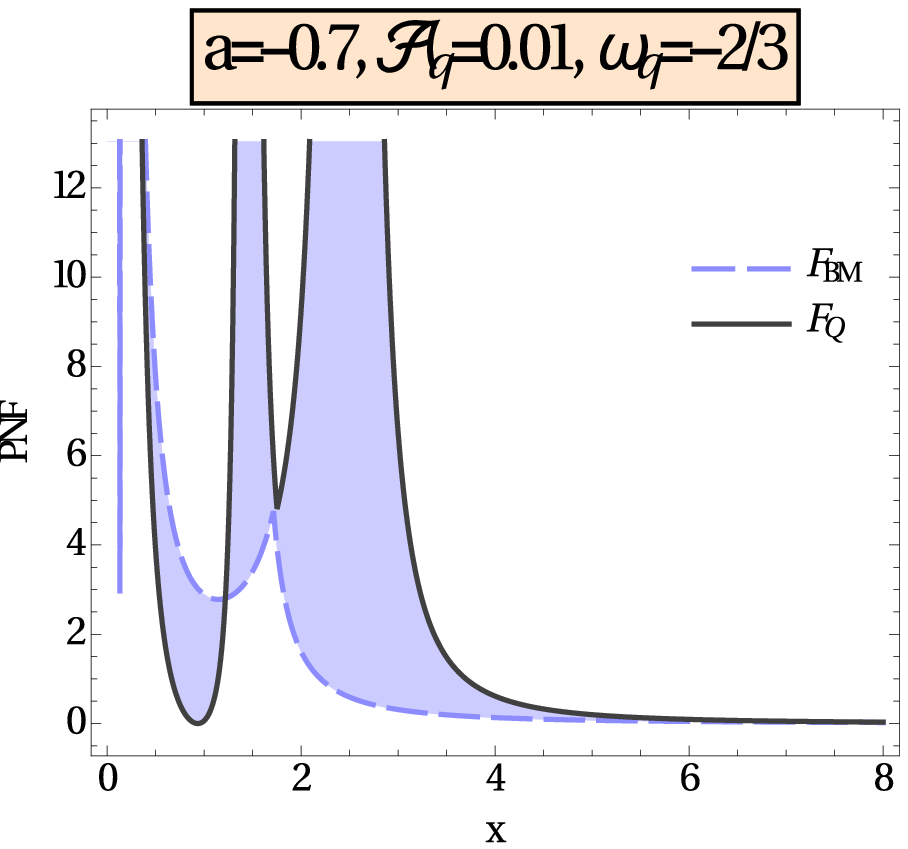}~~
\hspace*{0.5 cm}
\includegraphics*[scale=0.3]{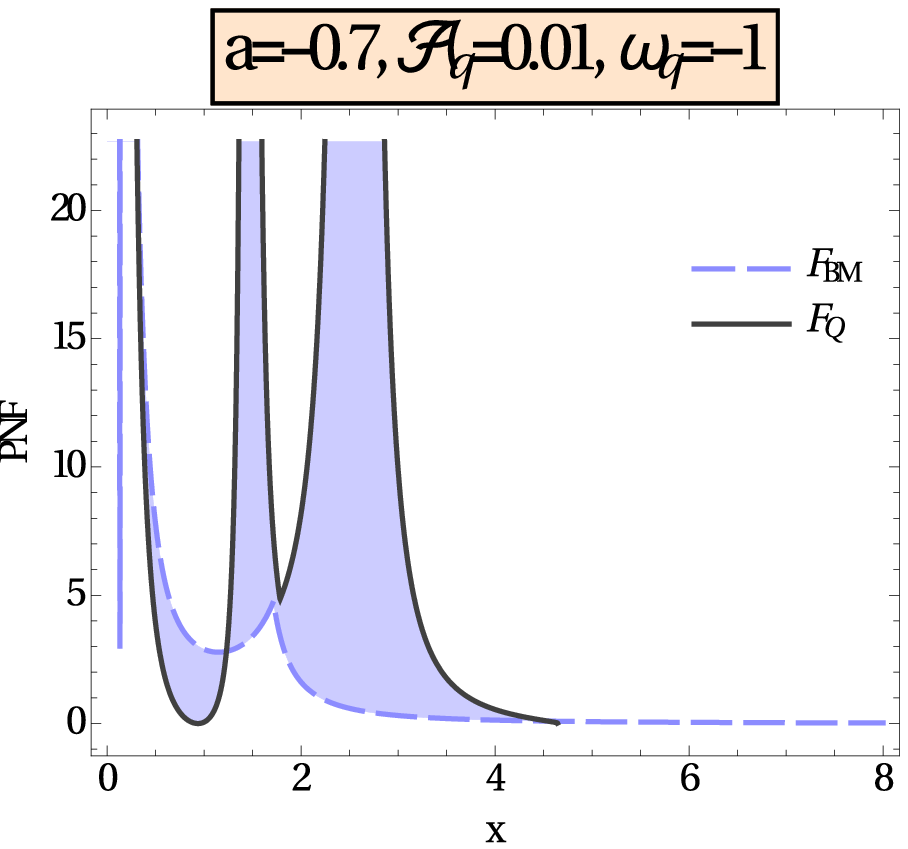}~~\\

Fig $1.6.a$ \hspace{1.6 cm} Fig $1.6.b$ \hspace{1.6 cm} Fig $1.6.c$ \hspace{1.6 cm} Fig $1.6.d$
\includegraphics*[scale=0.3]{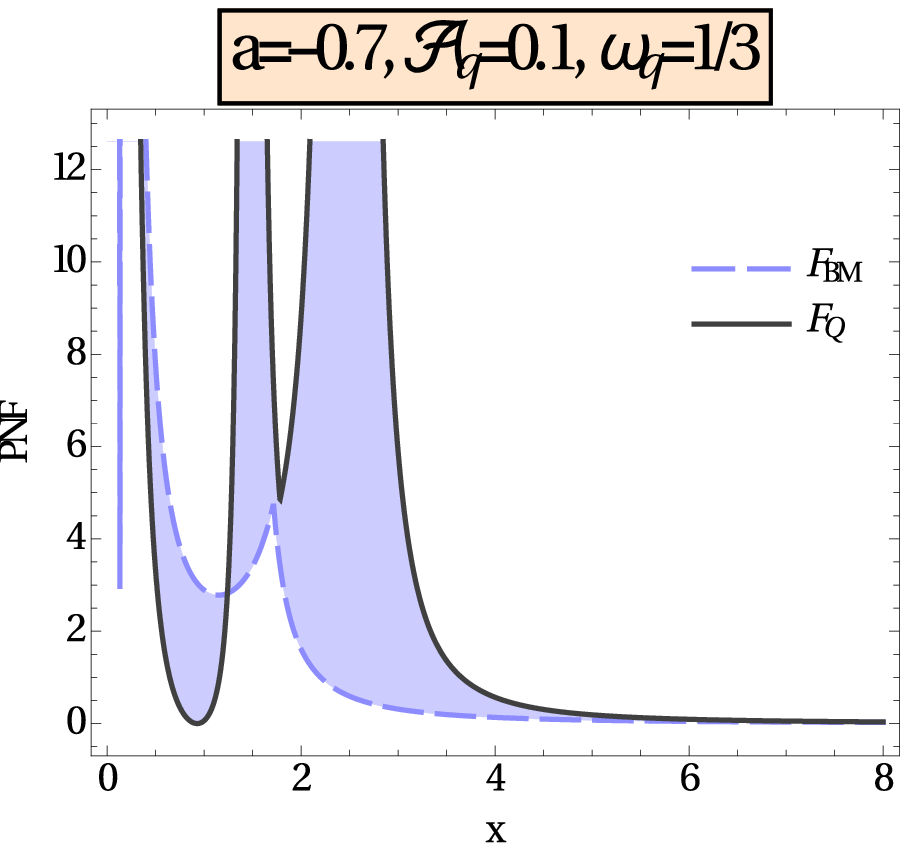}~~
\hspace*{0.5 cm}
\includegraphics*[scale=0.3]{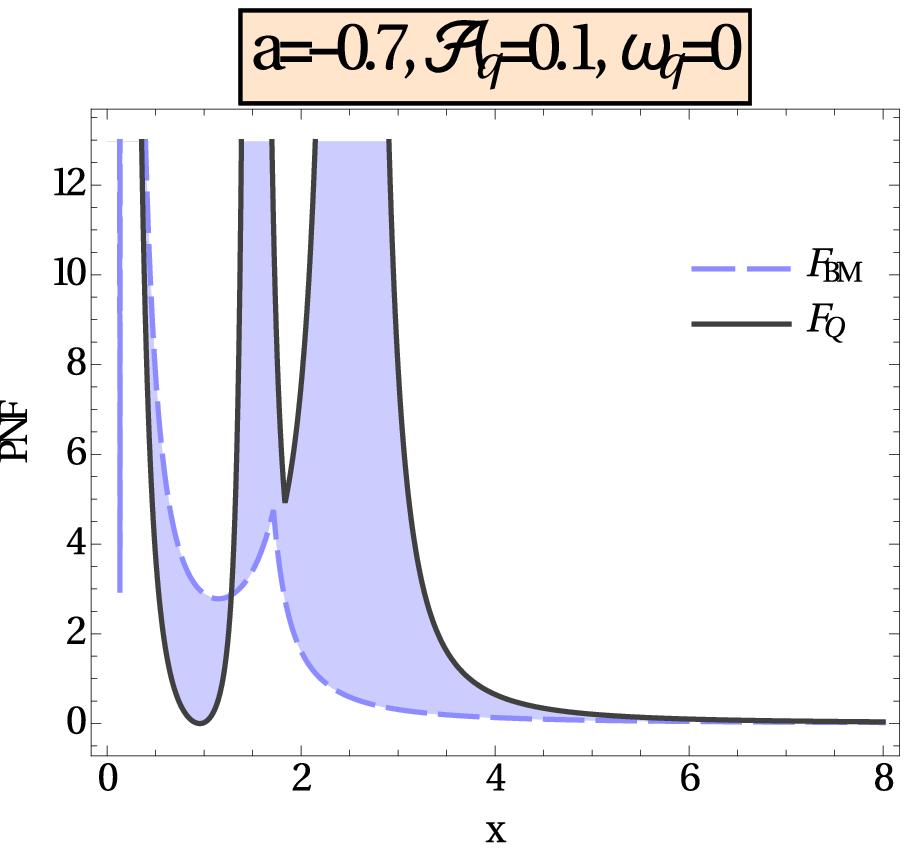}~~
\hspace*{0.5 cm}
\includegraphics*[scale=0.3]{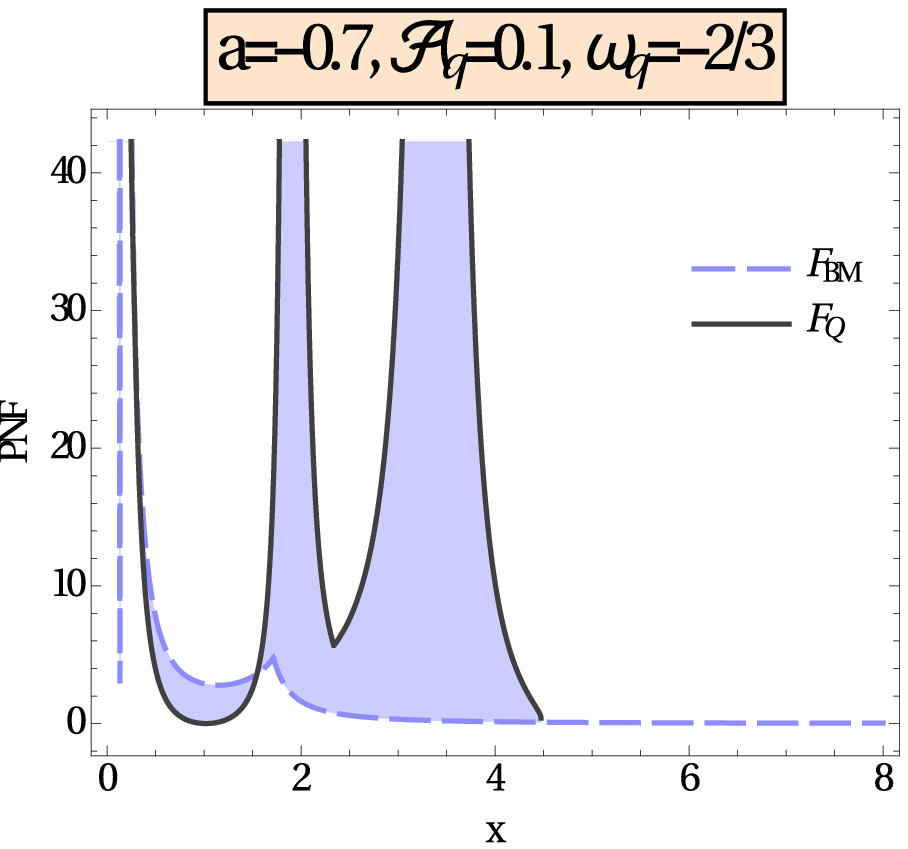}~~
\hspace*{0.5 cm}
\includegraphics*[scale=0.3]{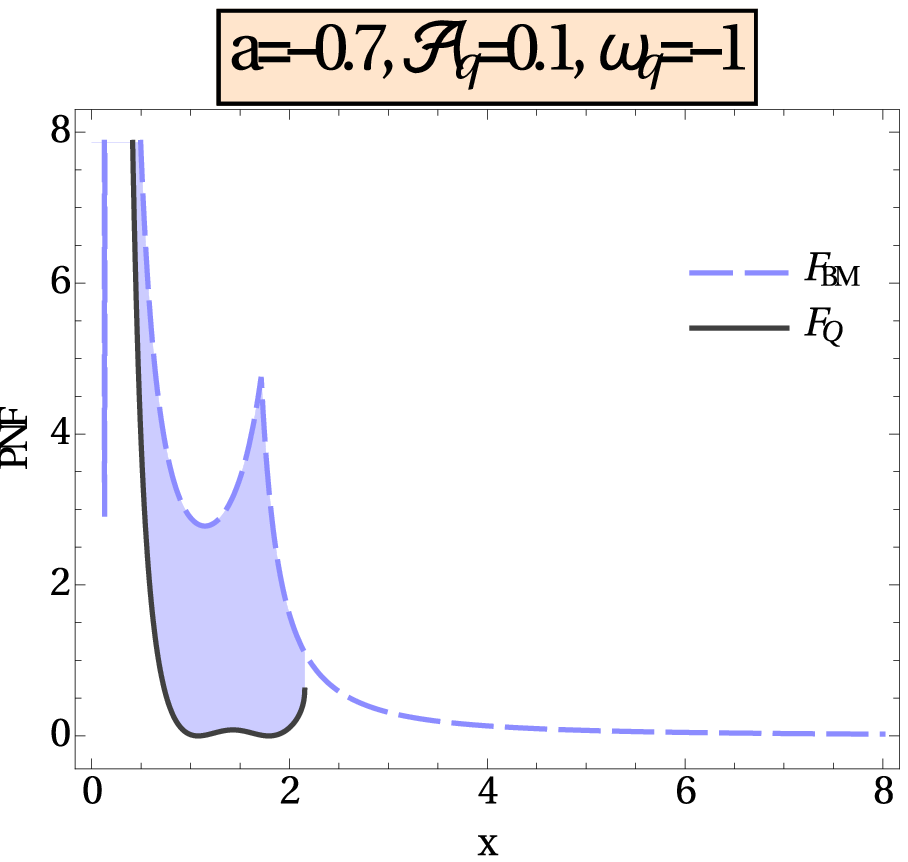}~~\\

\it{Here $a=-0.7$, for each value of ${\cal A}_q$, i.e., $10^{-3}$, $10^{-2}$ and $10^{-1}$ we have drawn relative results for $\omega_q=\frac{1}{3}$ for radiation, $\omega_q=0$ for dust, $\omega_q=-\frac{2}{3}$ for quintessence and $\omega_q=-1$ for phantom barrier.}
\end{figure}

In this section, we have plotted both the PNFs derived by equation (\ref{PNP_quint}) and Mukhopadhyay with respect to $x$. We have followed a specific pattern for every value of $a$. We have plotted twelve sets of graphs, i.e., for ${\cal A}_q$ is taken to be very low varying from $10^{-3}$ to $10^{-1}$ and $\omega_q$ from $\frac{1}{3}$ to $-1$, i.e., form radiation era to the phantom barrier. For the first set of graphs, i.e., from figure $1.1. a$ to $1.3.d$ we have taken the central object to be counter rotating with a high angular momentum $a=-0.998$. In figures $1.1.a$ to $1.1.d$ ${\cal A}_q$ is taken to be $10^{-3}$ and $\omega_q$ varies as a predefined sequence. For low $x$, the forces derived by Mukhopadhyay is dominating. For high $x$, these two are almost similar. But in a region $1<x<5$, we find forces derived by us to blow up and makes a clear difference with the forces derived by Mukhopadhyay.

In figures $1.2.a~-~1.2.d$, we have increased ${\cal A}_q$ to 0.01. This increases the exotic fluid's impact. The basic natures of the graphs are equivalent to the previous one. In $1.3.a~-~1.3.d$ ${\cal A}_q$ is increased to 0.1. Here we have found two places where our force blows up. The only explanation for the bizarre behavior of the PNF produced by us is due to the DE effect. Especially for $a=-0.998,~ {\cal A}_q=0.1~ \&~\omega_q=-2/3$, i.e., for a BH embedded in quintessence, we have come up with the fact that our PNF have just vanished near $x=4.5$ but up to that range it is relevant to its predecessors but surprisingly for $a=-0.998,~{\cal A}_q=0.1~ \&~ \omega_q=-1$, i.e., for phantom barrier, we got no trace of PNF after $x=2$. So for effect of quintessence, our force retains its behavior but vanishes after a certain value of $x$ and for the phantom barrier, this force has different behavior and then vanishes too early.

This may be physically interpreted as a rotating BH with a high amount of counter rotation which initially attracts particles towards its center but when its radius increases gradually the attraction force decreases rapidly. But in case of highly counter rotating BH embedded in quintessence, the central object can attract with more impulse if the radius is in the region $2<x<4$, i.e., for specific radii, BHs with DE attract with a force extremely high. On the other hand for small radius, BHs with DE attract with magnitude of more force than that with regular counter rotating one and for a high radius, DE effect nearly tends to zero. Finally, as we increase ${\cal A}_q$, i.e, the effect of DE one extra shatter region shows up and gradually it becomes prominent.

For the second set of graphs, we have decreased the angular momentum of rotation to $a=-0.7$ maintaining its direction for rotation unaltered. Like previous results, when $x$ is very low and very high, these two forces have same behaviors but with a special framework, i.e, Mukhopadhyay's force is dominating for very low $x$ and our force is dominating for high $x$. Like previous results, our force blows up exactly twice between 1 to 4 for figures $1.4.a$ to $1.7.c$ whereas Mukhopadhyay's force shows a regular curve with one double point. Here we have got an interesting fact, i.e., Mukhopadhyay's PNF has a double point directing vertically upward and our PNF also has a double point but directed vertically downwards. For $1<x<4$ both the forces generate a curve exactly reciprocal to each other. In the last two cases where $a=-0.7,~{\cal A}_q=0.1,~\omega_q=-2/3~ \&~ \omega_q=-1$ our forces vanishes too quickly, for first one it vanishes near $x=4.5$ and for second one, i.e., for phantom barrier it vanishes near $x=2$.
 
Here we have got a relatively strong region where the effect of quintessence is relatively high. For a mediocre counter rotating BH, the impact of DE accreted into a rotating BH is remarkable. A double point occurs in both the cases but with an opposite nature. For Mukhopadhyay's PNF it is pointing upwards, whereas our PNF is pointing upside down. So before that very point, Mukhopadhyay's force is increasing sharply and after that, it decreases rapidly, i.e., it has a critical radius where attracting force has absurd value. Similarly, in that point, our force also obtains a double point such that, before that point, it decreases sharply and after that, it has a highly increased value.
\begin{figure}
\centering
Fig $1.7.a$ \hspace{1.6 cm} Fig $1.7.b$ \hspace{1.6 cm} Fig $1.7.c$ \hspace{1.6 cm} Fig $1.7.d$
\includegraphics*[scale=0.3]{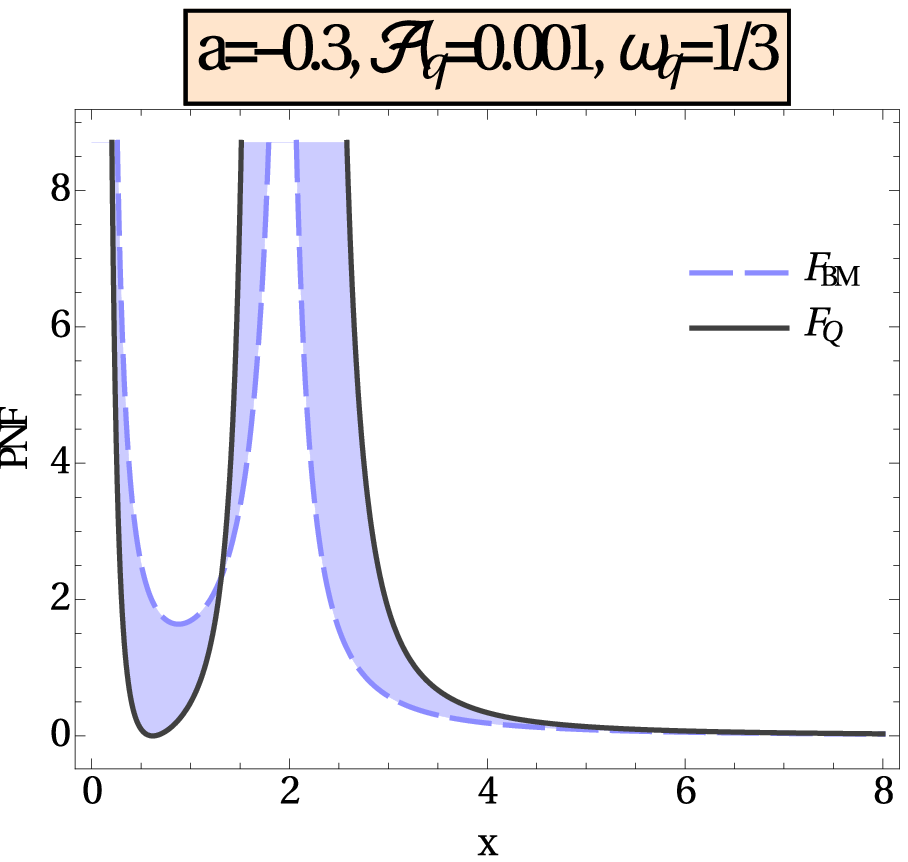}~~
\hspace*{0.5 cm}
\includegraphics*[scale=0.3]{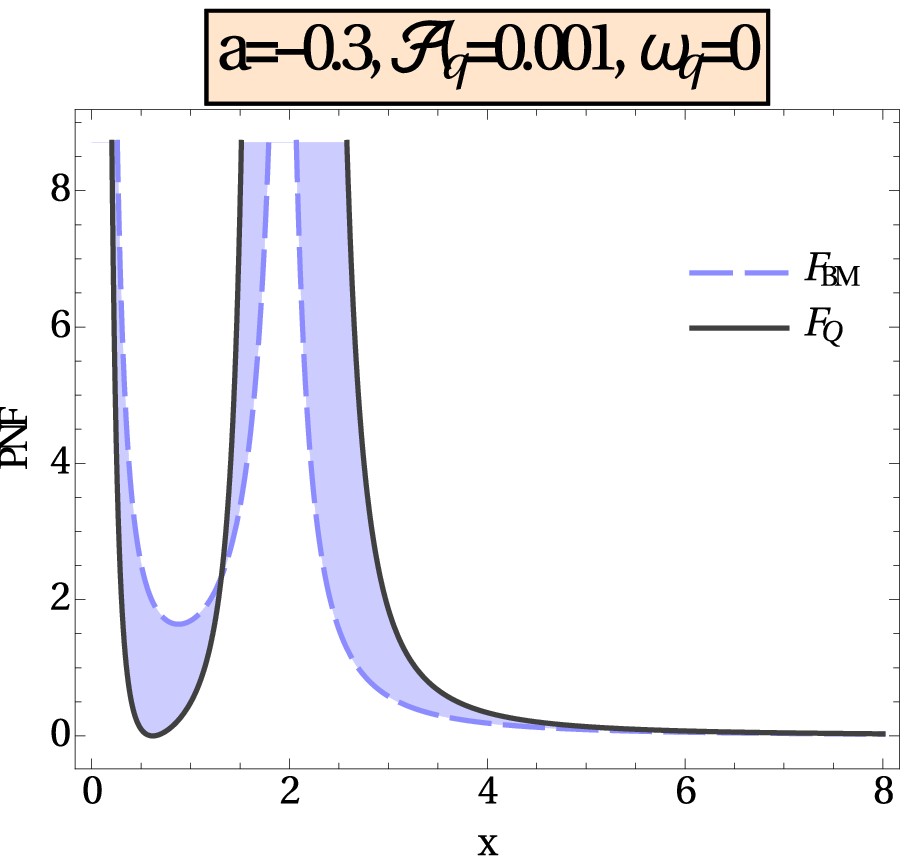}~~
\hspace*{0.5 cm}
\includegraphics*[scale=0.3]{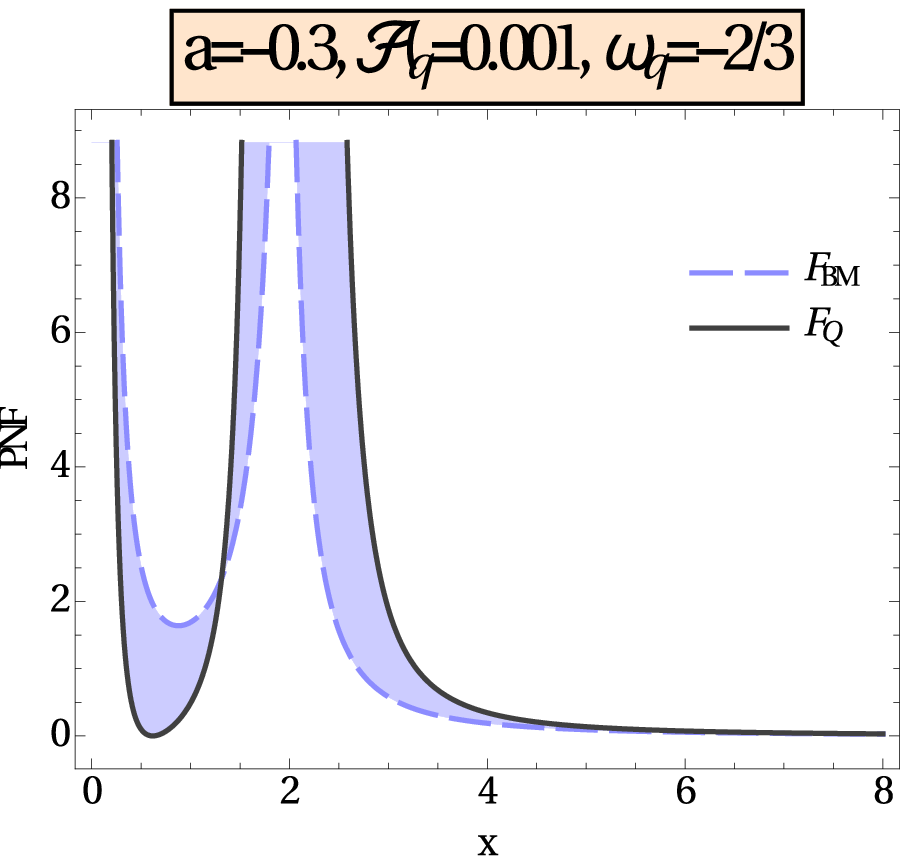}~~
\hspace*{0.5 cm}
\includegraphics*[scale=0.3]{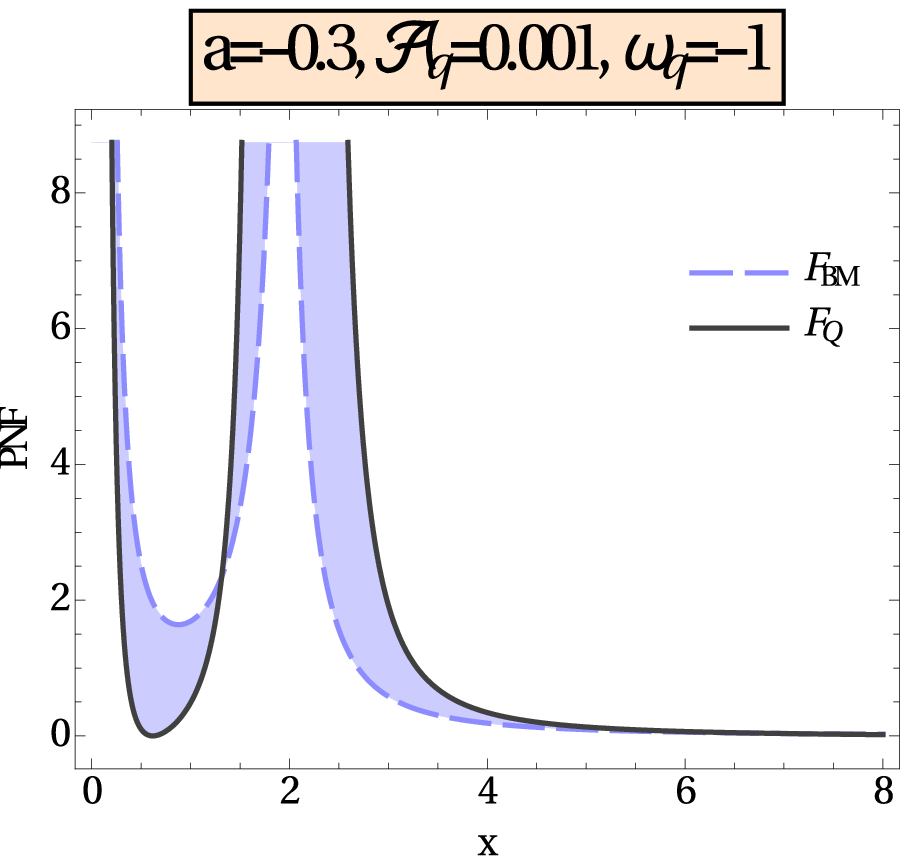}~~\\

Fig $1.8.a$ \hspace{1.6 cm} Fig $1.8.b$ \hspace{1.6 cm} Fig $1.8.c$ \hspace{1.6 cm} Fig $1.8.d$
\includegraphics*[scale=0.3]{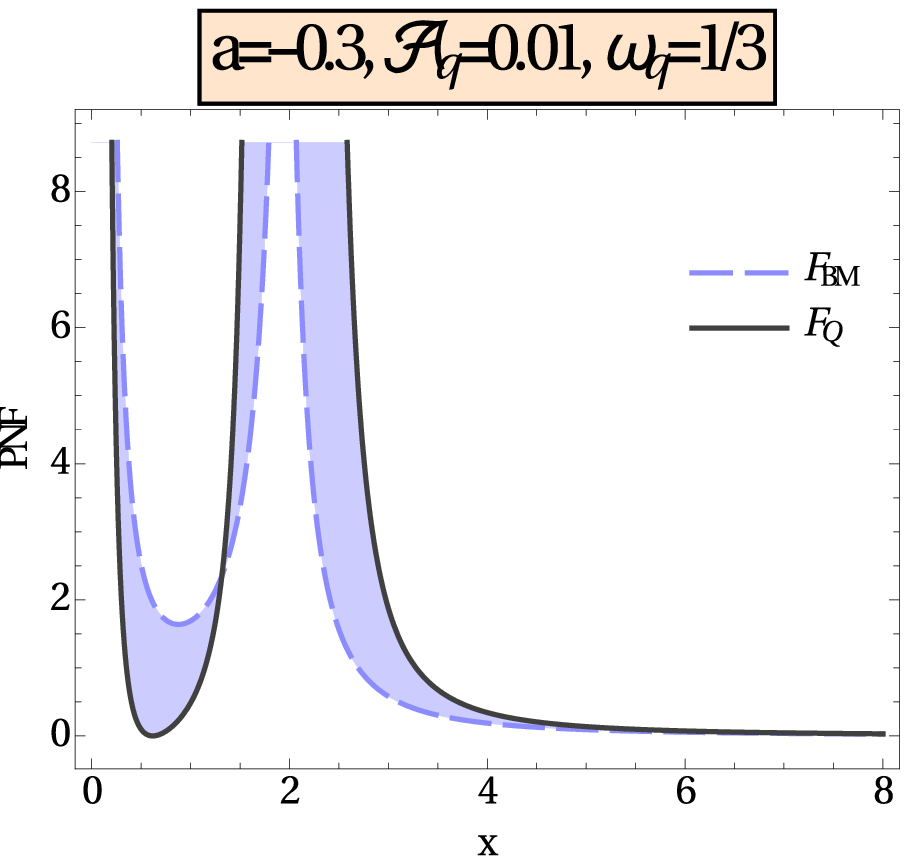}~~
\hspace*{0.5 cm}
\includegraphics*[scale=0.3]{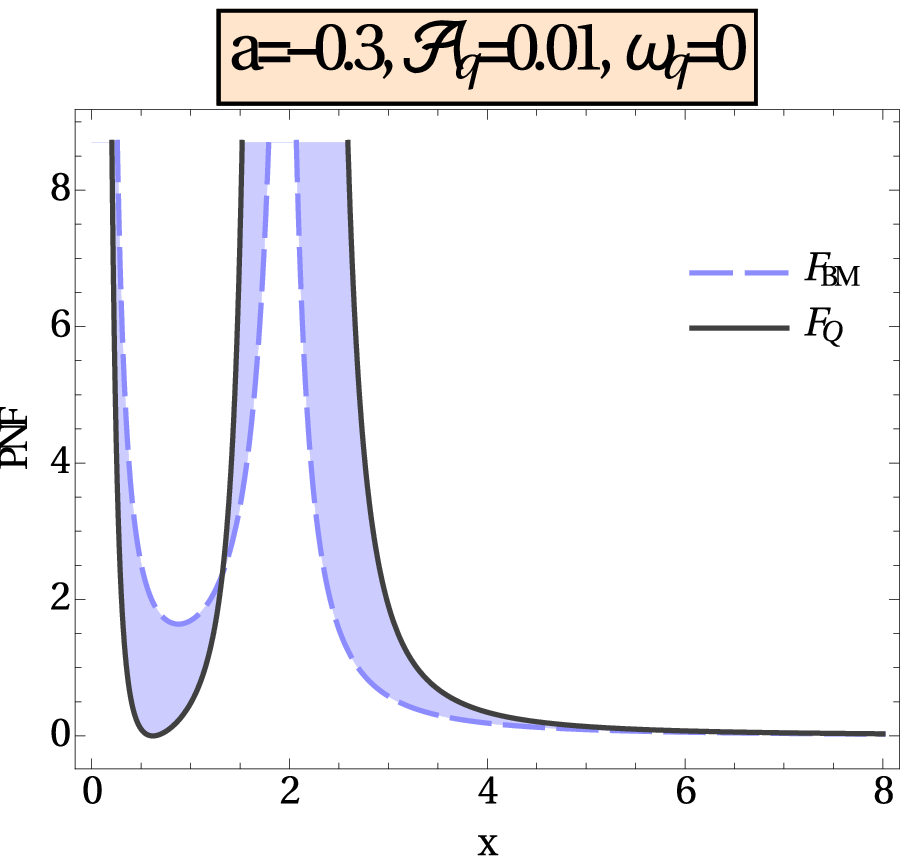}~~
\hspace*{0.5 cm}
\includegraphics*[scale=0.3]{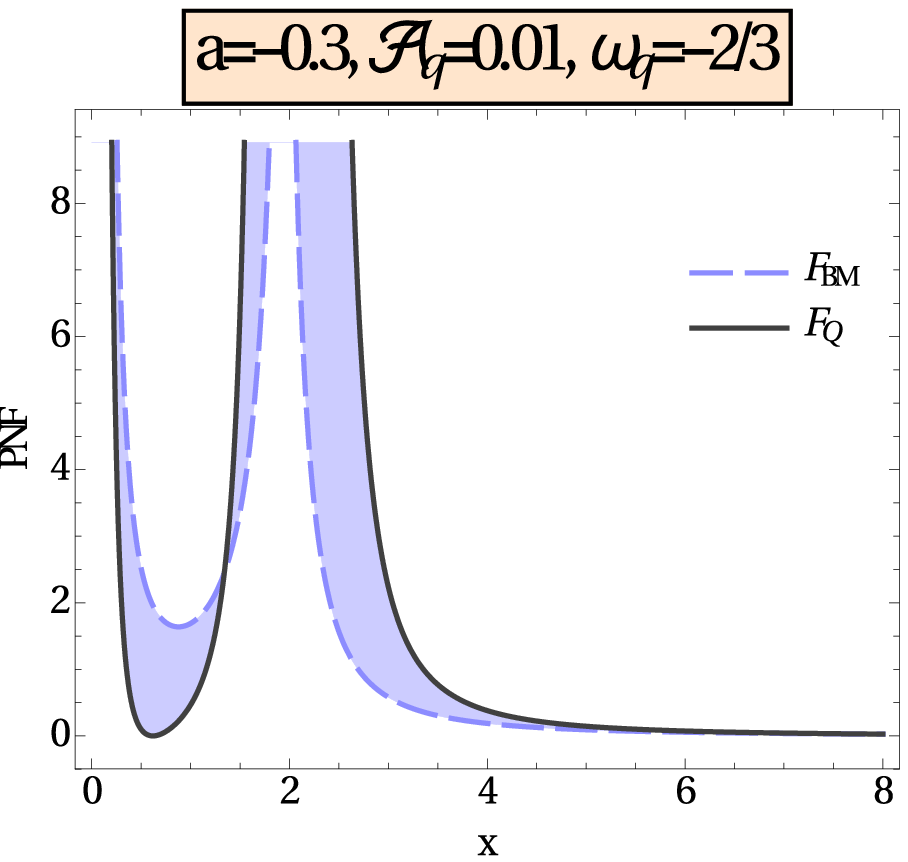}~~
\hspace*{0.5 cm}
\includegraphics*[scale=0.3]{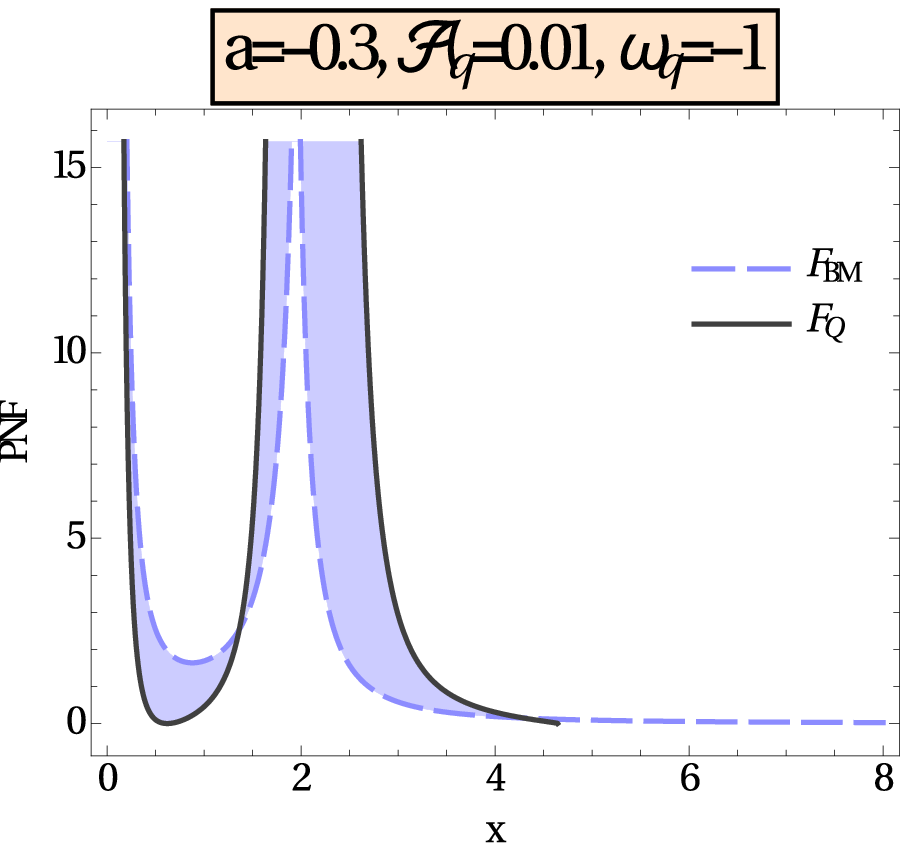}~~\\

Fig $1.9.a$ \hspace{1.6 cm} Fig $1.9.b$ \hspace{1.6 cm} Fig $1.9.c$ \hspace{1.6 cm} Fig $1.9.d$
\includegraphics*[scale=0.3]{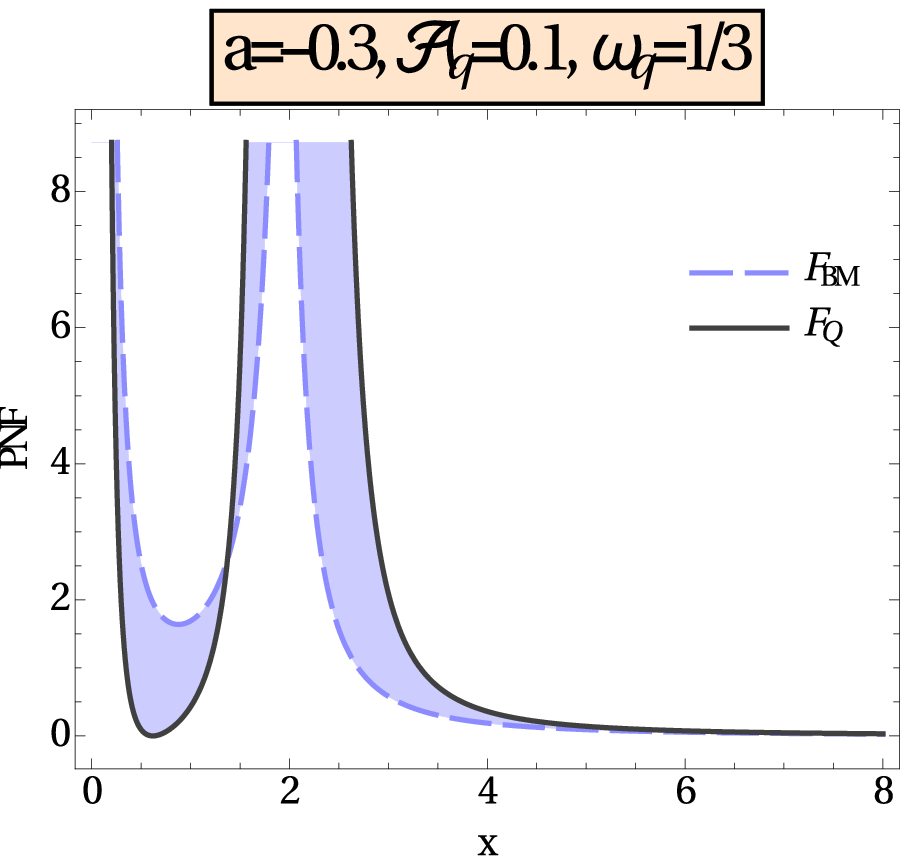}~~
\hspace*{0.5 cm}
\includegraphics*[scale=0.3]{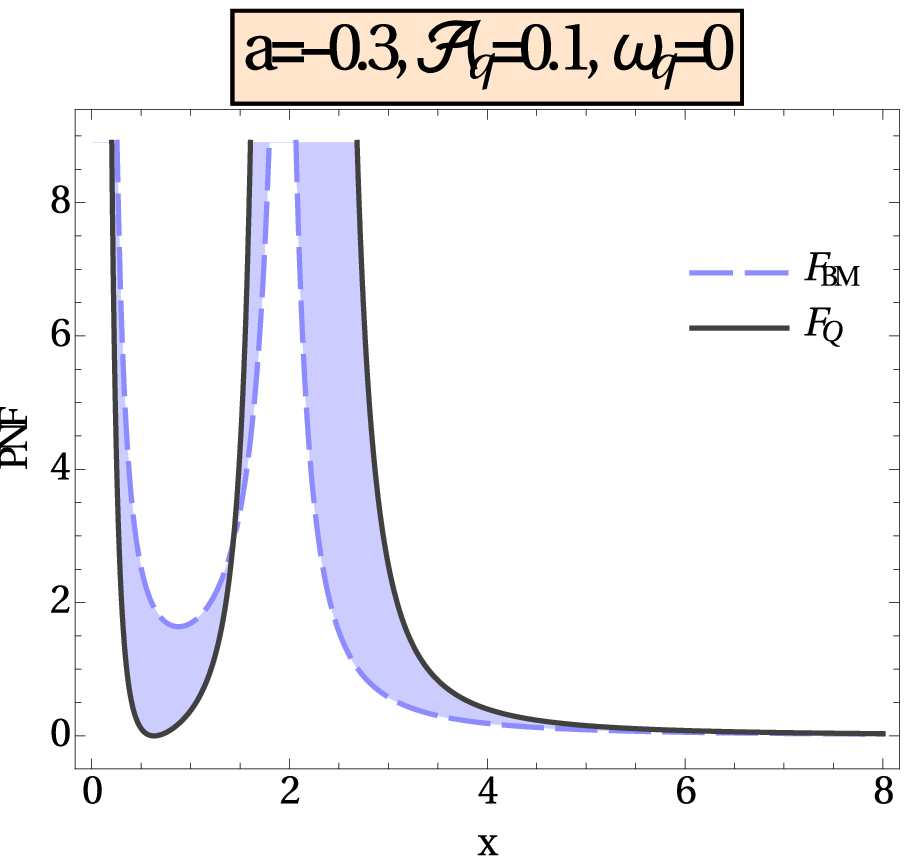}~~
\hspace*{0.5 cm}
\includegraphics*[scale=0.3]{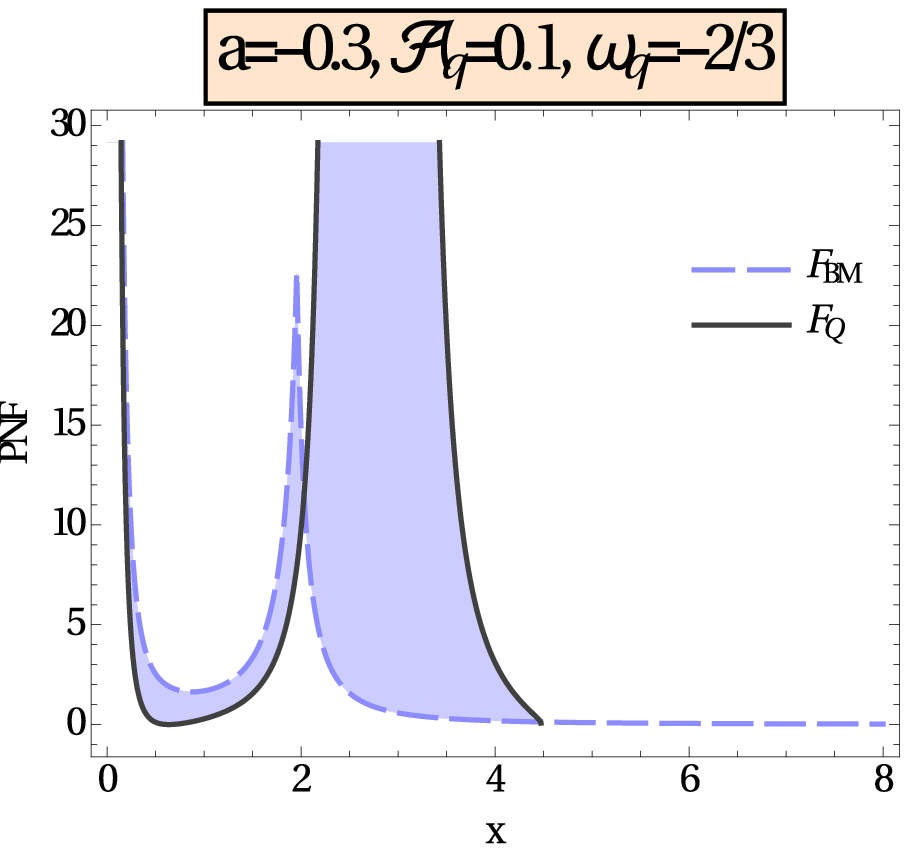}~~
\hspace*{0.5 cm}
\includegraphics*[scale=0.3]{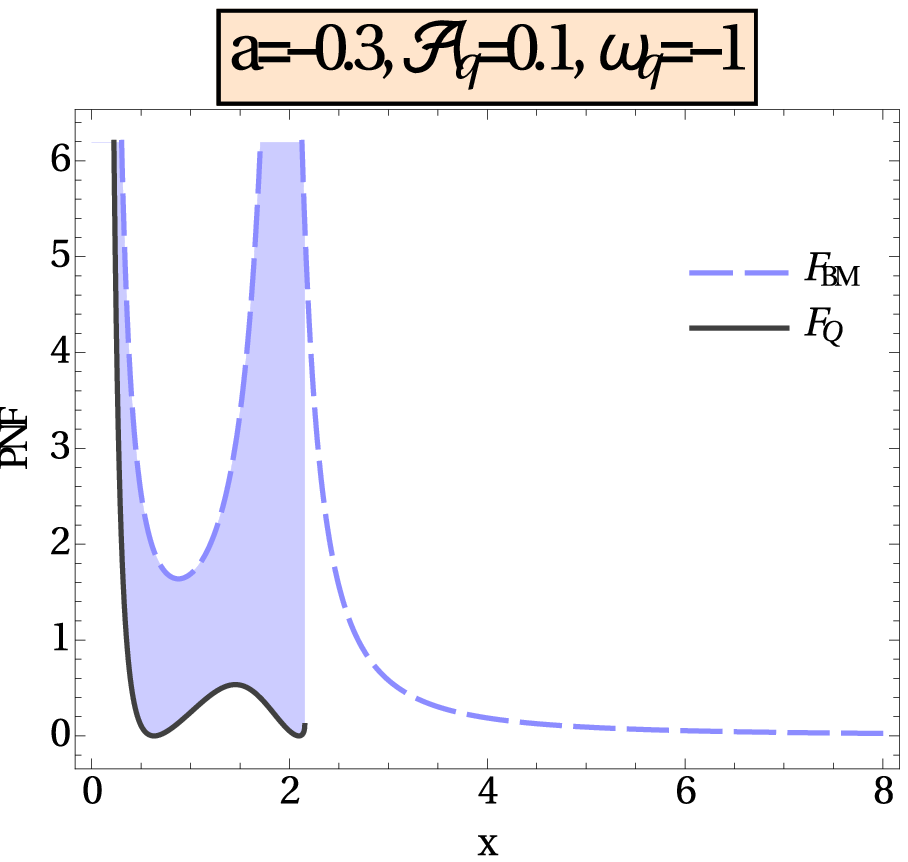}~~\\

\it{Here $a=-0.3$, for each values of ${\cal A}_q$, i.e., $10^{-3}$, $10^{-2}$ and $10^{-1}$ we have drawn a relative results for $\omega_q=\frac{1}{3}$ for radiation, $\omega_q=0$ for dust, $\omega_q=-\frac{2}{3}$ for quintessence and $\omega_q=-1$ for phantom barrier.}\\

Fig $1.10.a$ \hspace{1.6 cm} Fig $1.10.b$ \hspace{1.6 cm} Fig $1.10.c$ \hspace{1.6 cm} Fig $1.10.d$
\includegraphics*[scale=0.3]{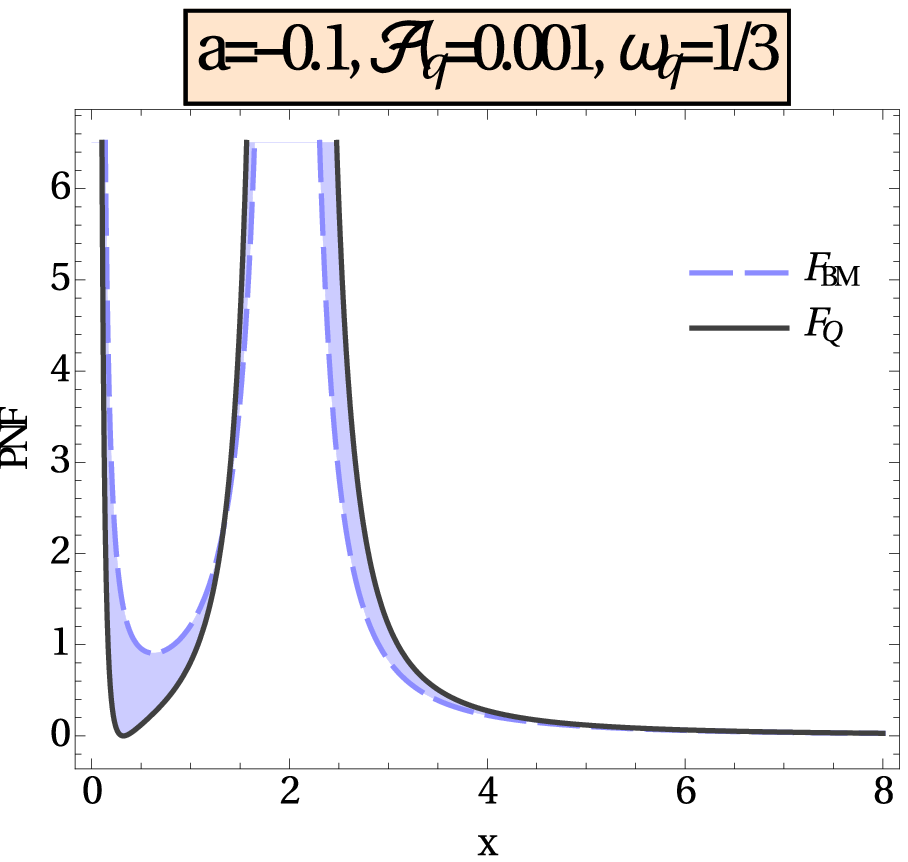}~~
\hspace*{0.5 cm}
\includegraphics*[scale=0.3]{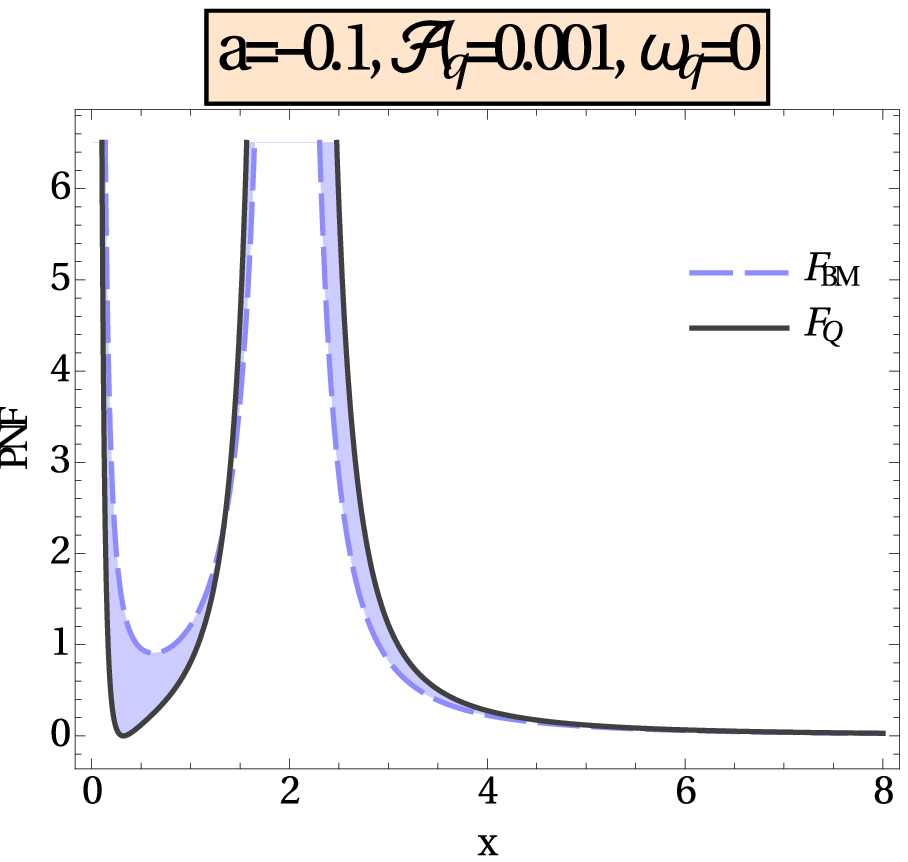}~~
\hspace*{0.5 cm}
\includegraphics*[scale=0.3]{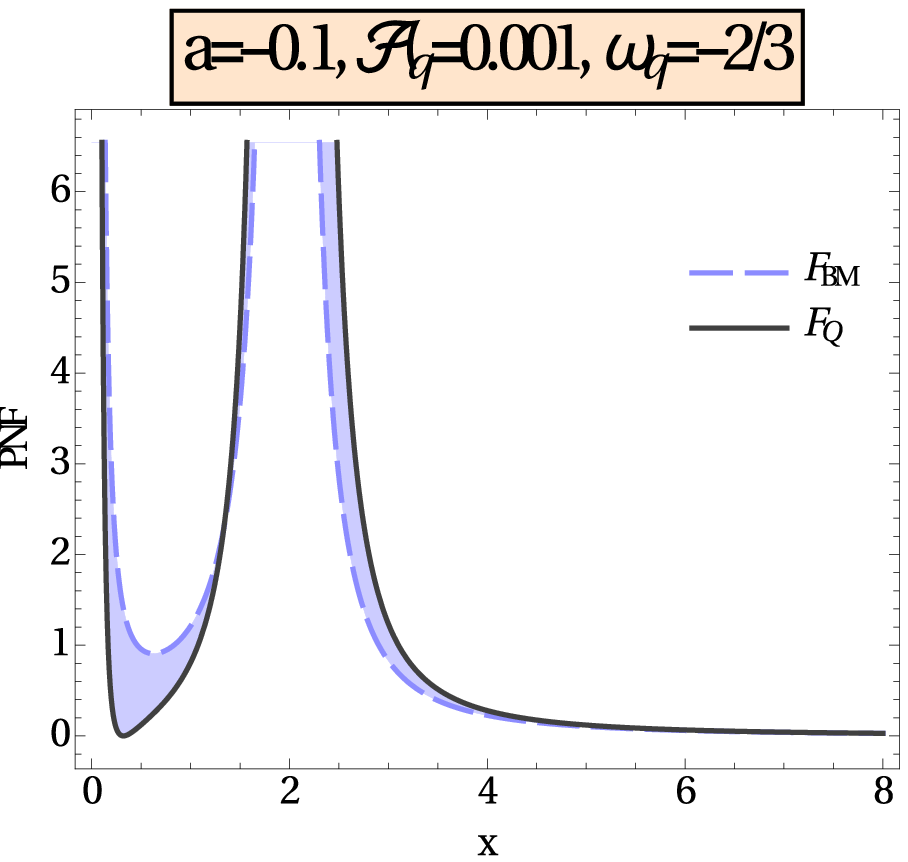}~~
\hspace*{0.5 cm}
\includegraphics*[scale=0.3]{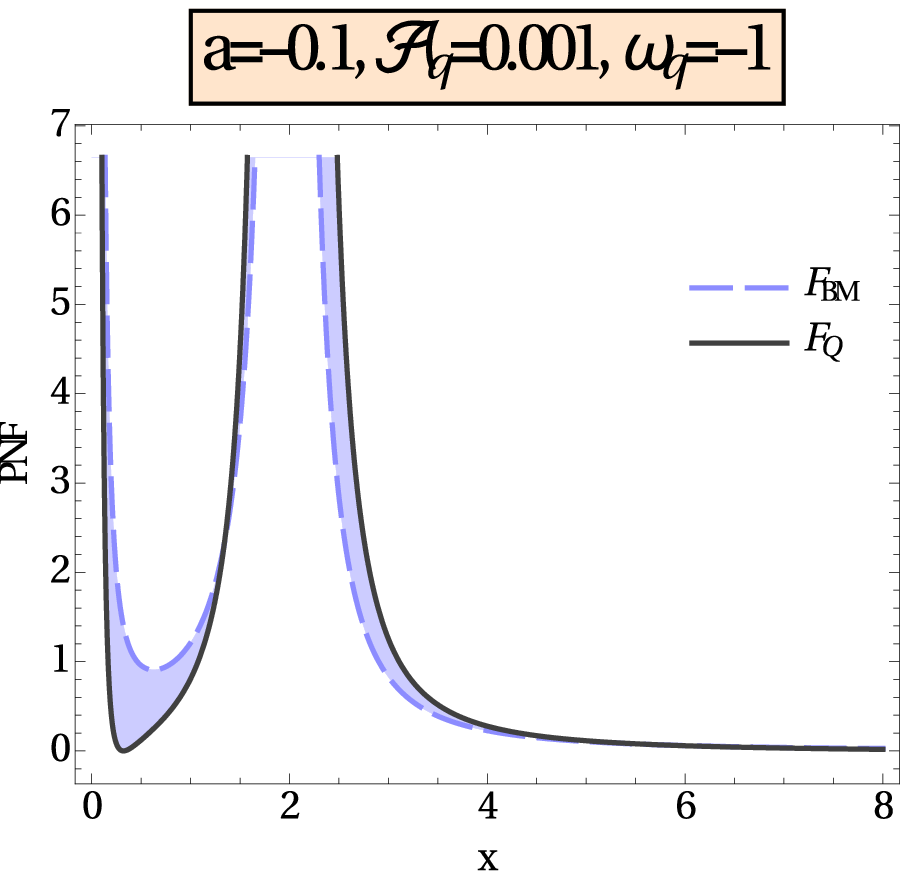}~~\\

Fig $1.11.a$ \hspace{1.6 cm} Fig $1.11.b$ \hspace{1.6 cm} Fig $1.11.c$ \hspace{1.6 cm} Fig $1.11.d$
\includegraphics*[scale=0.3]{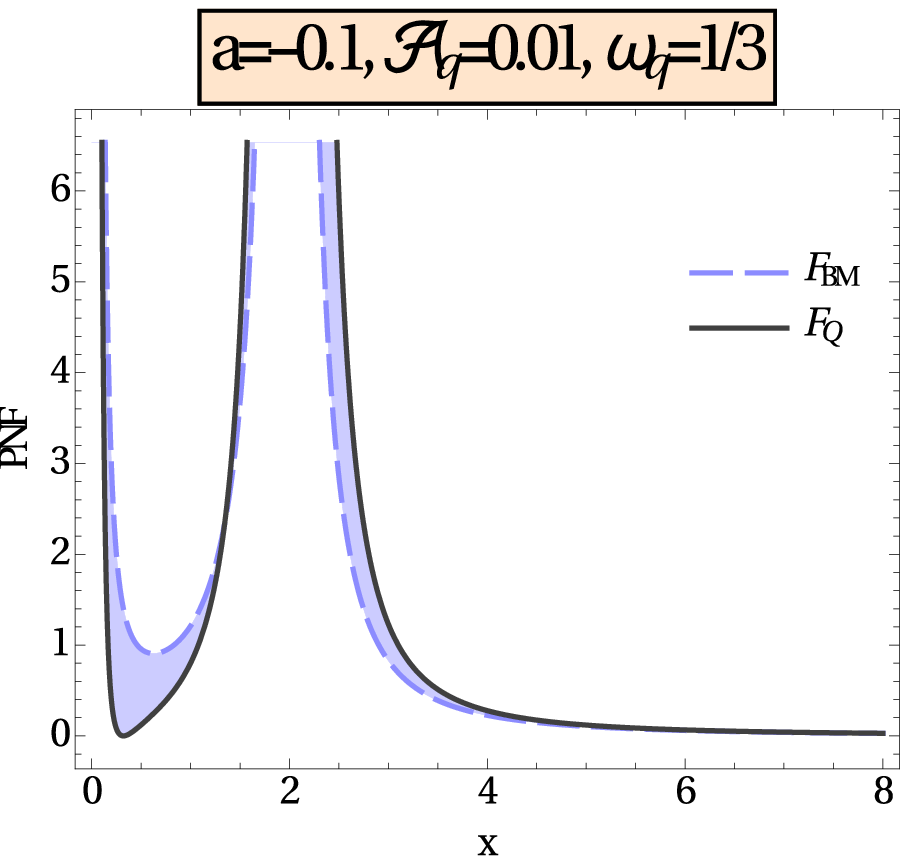}~~
\hspace*{0.5 cm}
\includegraphics*[scale=0.3]{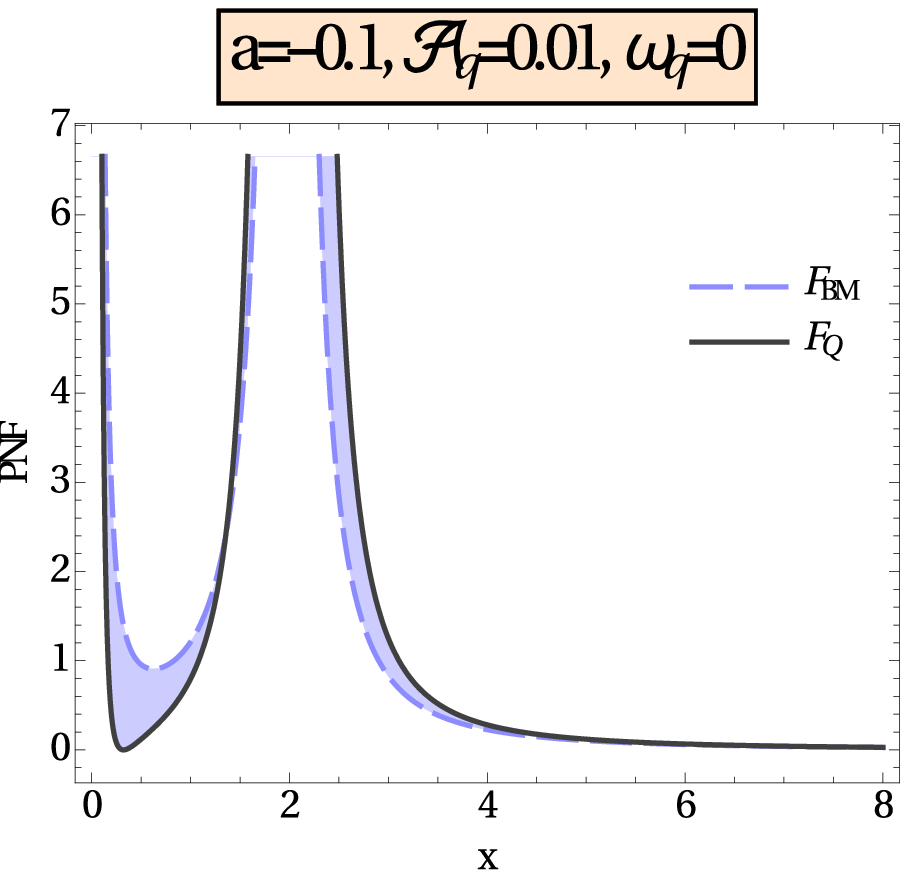}~~
\hspace*{0.5 cm}
\includegraphics*[scale=0.3]{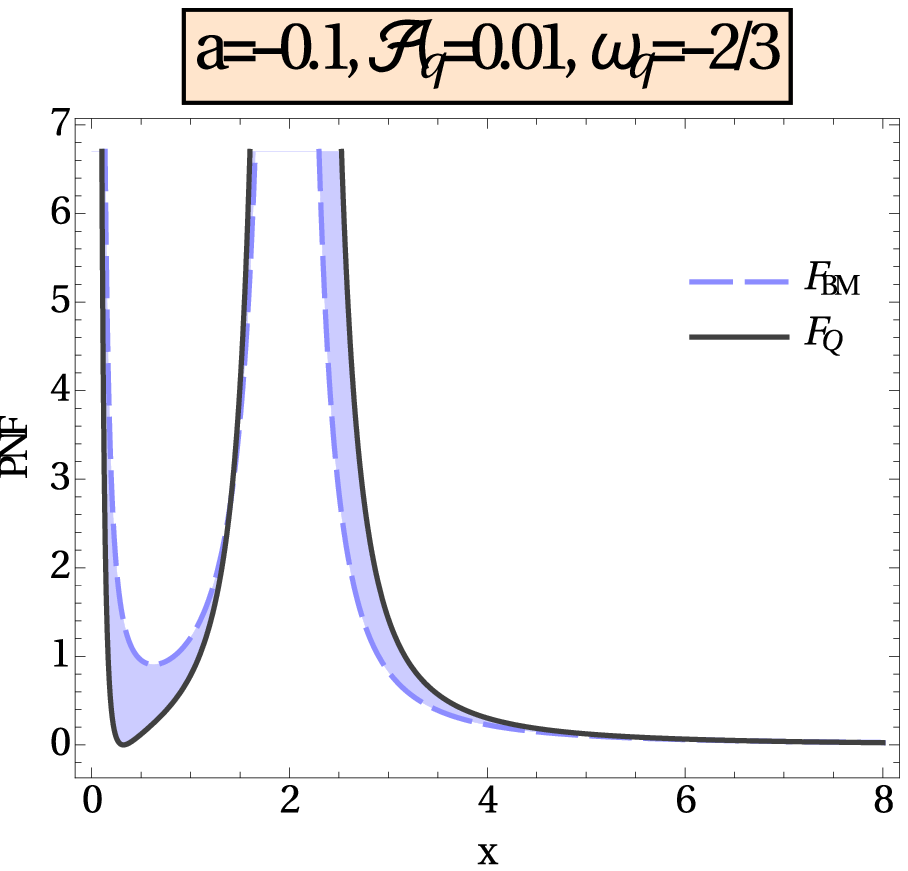}~~
\hspace*{0.5 cm}
\includegraphics*[scale=0.3]{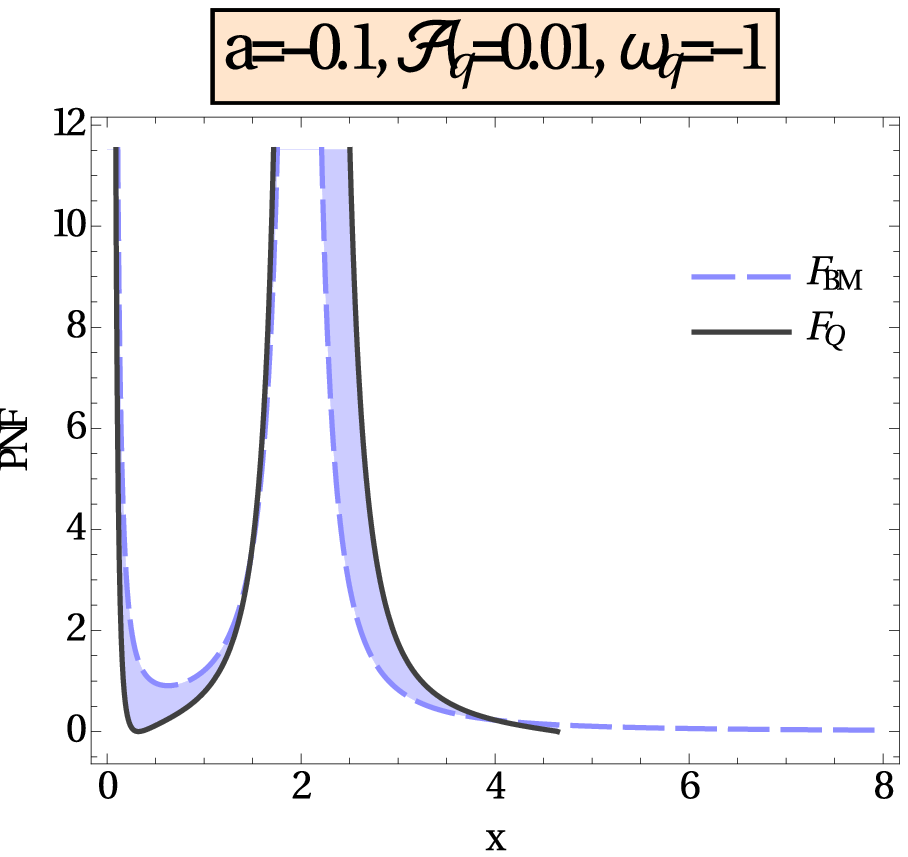}~~\\

Fig $1.12.a$ \hspace{1.6 cm} Fig $1.12.b$ \hspace{1.6 cm} Fig $1.12.c$ \hspace{1.6 cm} Fig $1.12.d$
\includegraphics*[scale=0.3]{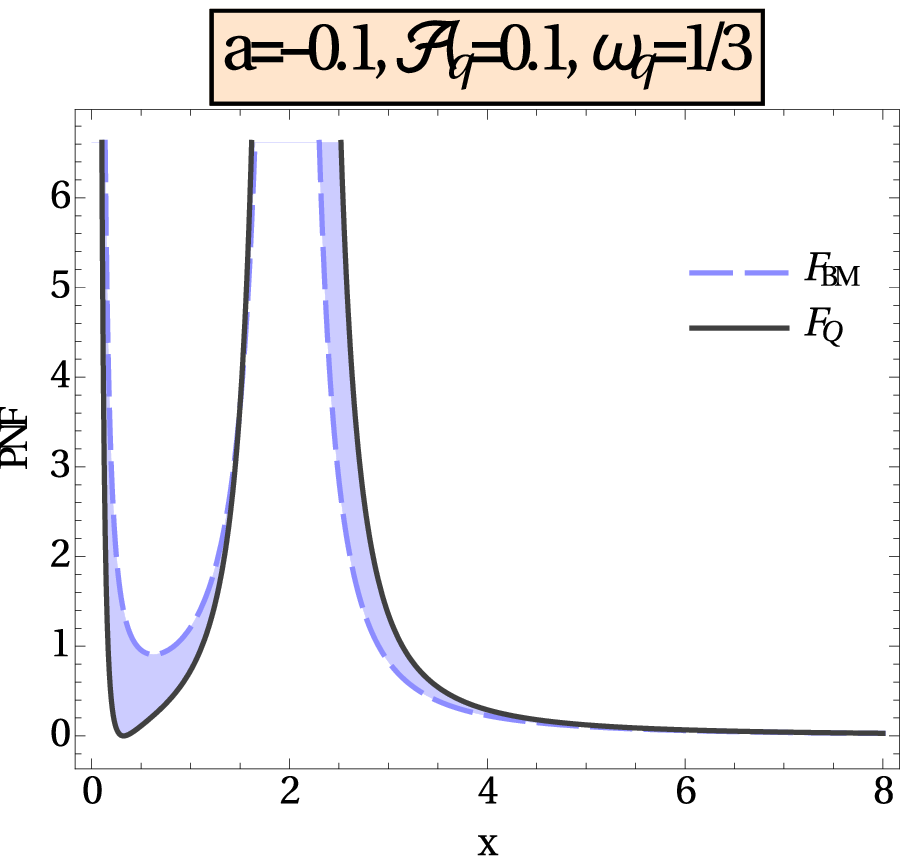}~~
\hspace*{0.5 cm}
\includegraphics*[scale=0.3]{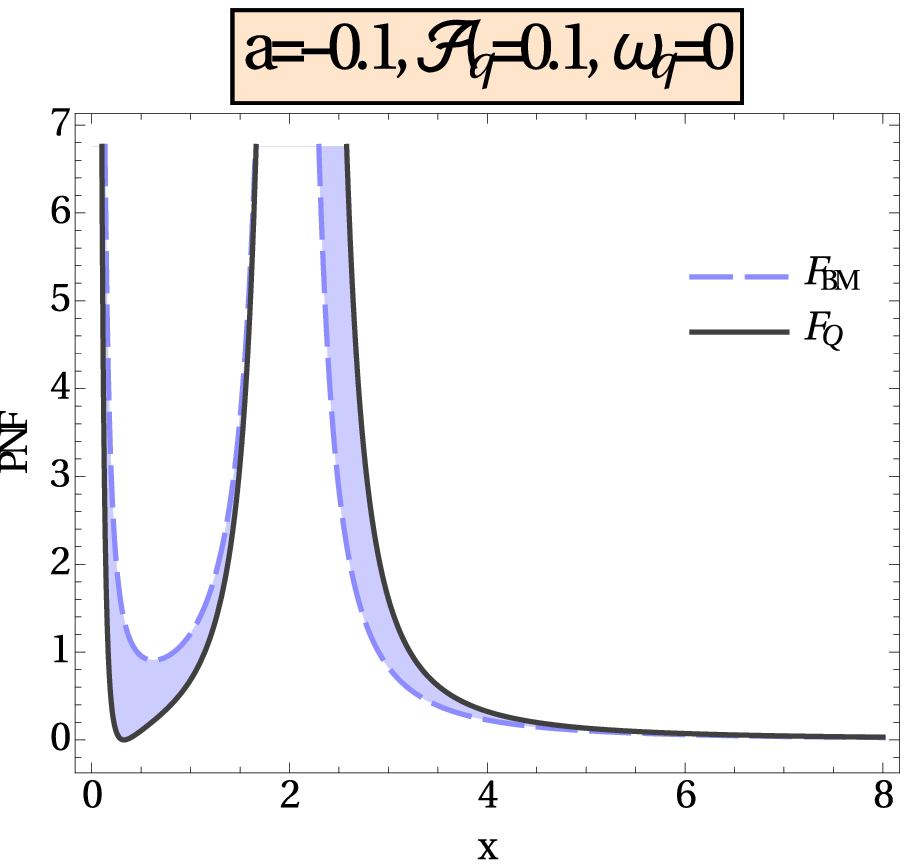}~~
\hspace*{0.5 cm}
\includegraphics*[scale=0.3]{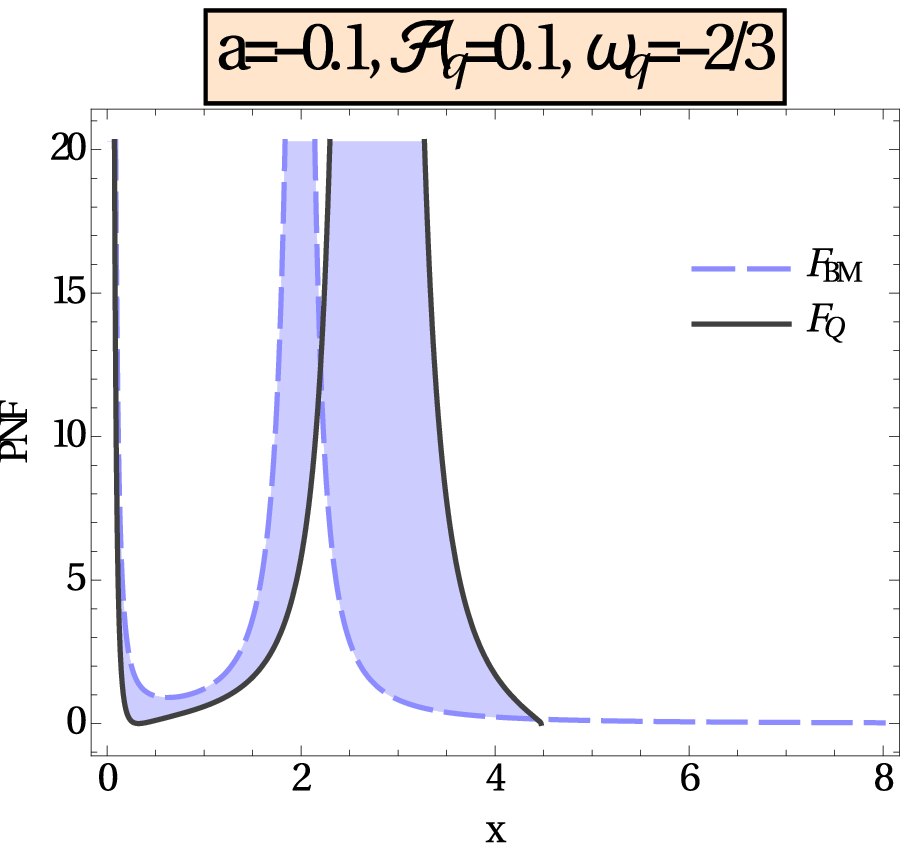}~~
\hspace*{0.5 cm}
\includegraphics*[scale=0.3]{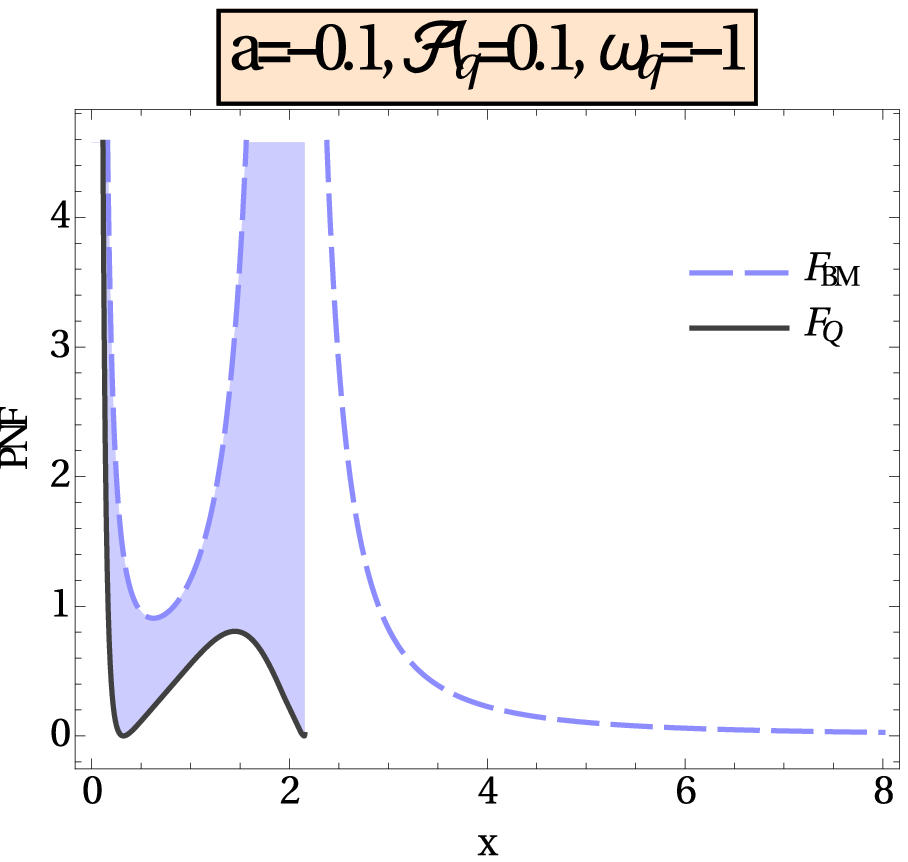}~~\\

\it{Here $a=-0.1$, for each values of ${\cal A}_q$, i.e., $10^{-3}$, $10^{-2}$ and $10^{-1}$ we have drawn a relative results for $\omega_q=\frac{1}{3}$ for radiation, $\omega_q=0$ for dust, $\omega_q=-\frac{2}{3}$ for quintessence and $\omega_q=-1$ for phantom barrier.}
\end{figure}

For the 3rd set of graphs, i.e., figures $1.8.a$ to $1.7.d$, we further decreased the angular momentum $a$ to $-0.3$ keeping the direction of rotation unaltered. Although for high $x$ and low $x$ both the above-mentioned forces are similar like previous results but for $1<x<4$ both the forces blow up once instead of two. Except for last two cases as we increased $x$ from beginning our force is dominating, suddenly Mukhopadhyay's PNF has replaced its nature and both the forces blow up near $x=2$, after passing through that singularity our force has become dominating again. For $a=-0.3,~{\cal A}_q=0.1,~\omega_q=-2/3$, our PNF stops giving values after $x\approx 4.5$, on the other hand, Mukhopadhyay's force just attends one local maximum apart from having a point of singularity. Lastly, for $a=-0.3,~{\cal A}_q=0.1~\&~ \omega_q=-1$, magnitude of our force is very low but with a local maxima for $x$ in $(1.5,2)$ but after being higher than 2 it stops giving physical value. The quintessential effect on PNF is very similar to that of a rotating Kerr BH. 

Here we observed that at the shatter region our force makes an envelope to the PNF represented by Mukhopadhyay. This leads to the conclusion that for a DE effect in BH metric, there is a wide range of radii where the attractive force is huge. This range of radius remains the same for any type of this exotic energy, even if we increase the scale of this effect it also remains invariant.

For next set of graphs, we have taken $a=-0.1$, a very low counter rotation. For all values of $a,~{\cal A}_q$ and $\omega_q$ both the forces are nearly equal, like previous results. But if we start a comparison between both the forces, Mukhopadhyay's force remains dominating in the first continuous region $x<2$ then both the forces blow up at a time and after that our force dominates for $x>2$. Finally, for two special cases, where $a=-0.1,~{\cal A}_q=0.1,~\omega_q=-2/3~\&~ \omega_q=-1$, both the forces are been nullified. For the first one, i.e., for BH embedded in quintessence it vanishes after $x\approx 4.2$ and for the second one, i.e., for the phantom barrier, it vanishes where $x>2$ but with a local maximum on $(1,2)$.

Since the rotation is very slow, we can expect that the effects of DE must lead to a different result. But here we have chosen the impact so small that it should not react too strange. For a relatively high impact we observed that the strongly accreting region has been increased.  

\begin{figure}
\centering

Fig $1.13.a$ \hspace{1.6 cm} Fig $1.13.b$ \hspace{1.6 cm} Fig $1.13.c$ \hspace{1.6 cm} Fig $1.13.d$
\includegraphics*[scale=0.3]{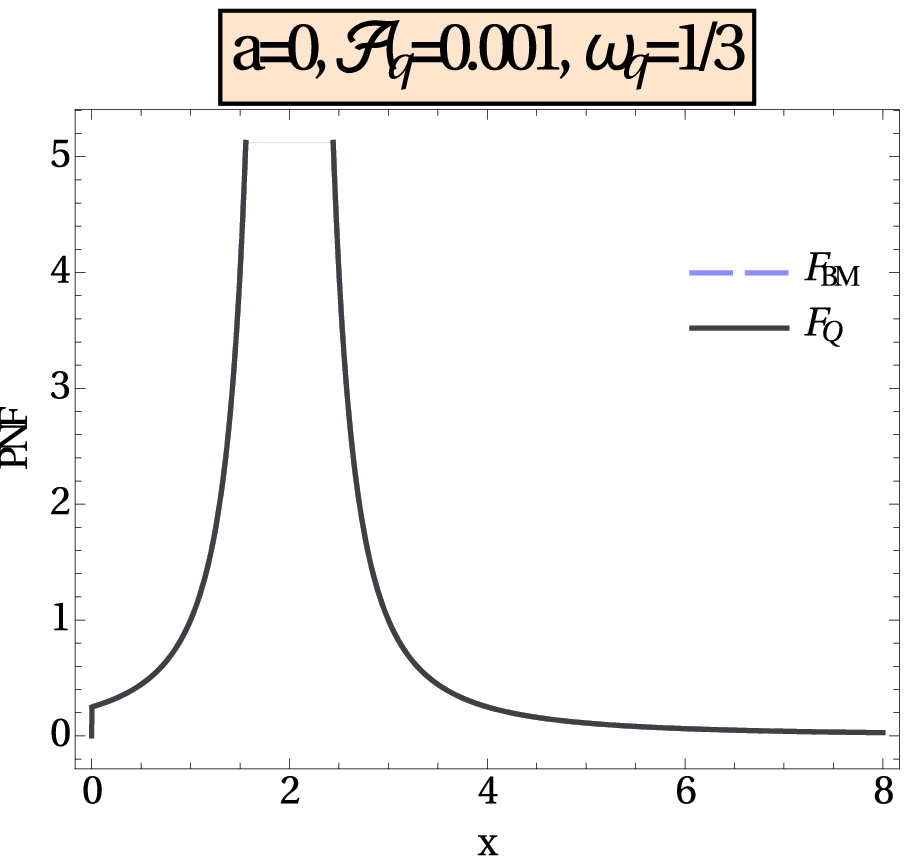}~~
\hspace*{0.5 cm}
\includegraphics*[scale=0.3]{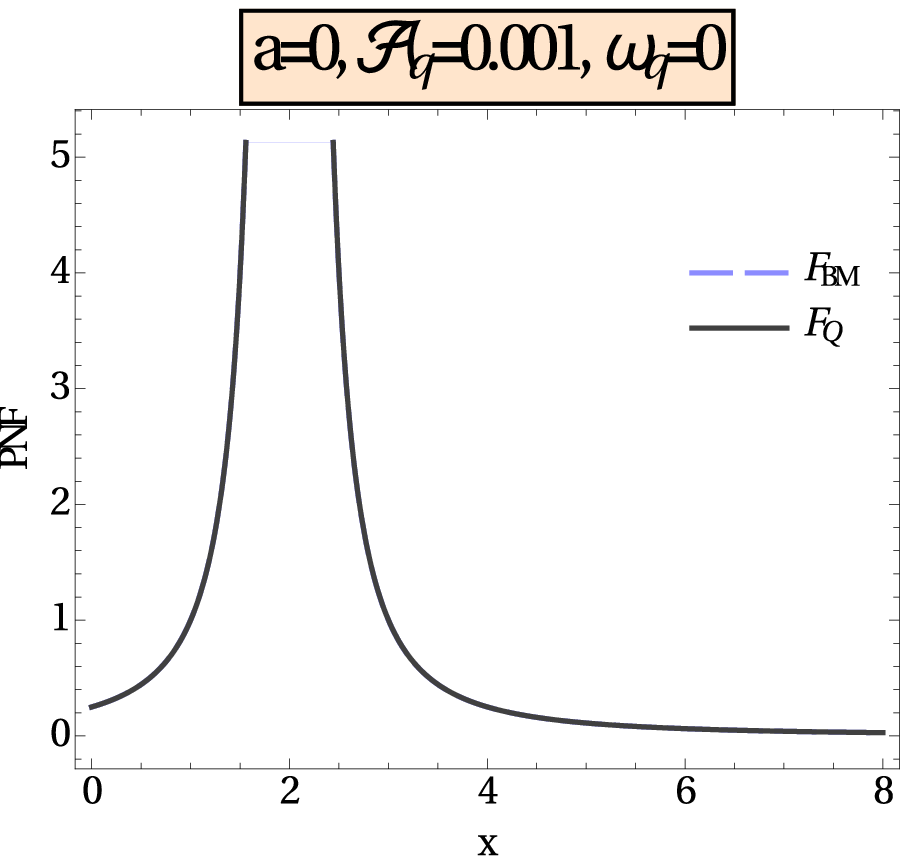}~~
\hspace*{0.5 cm}
\includegraphics*[scale=0.3]{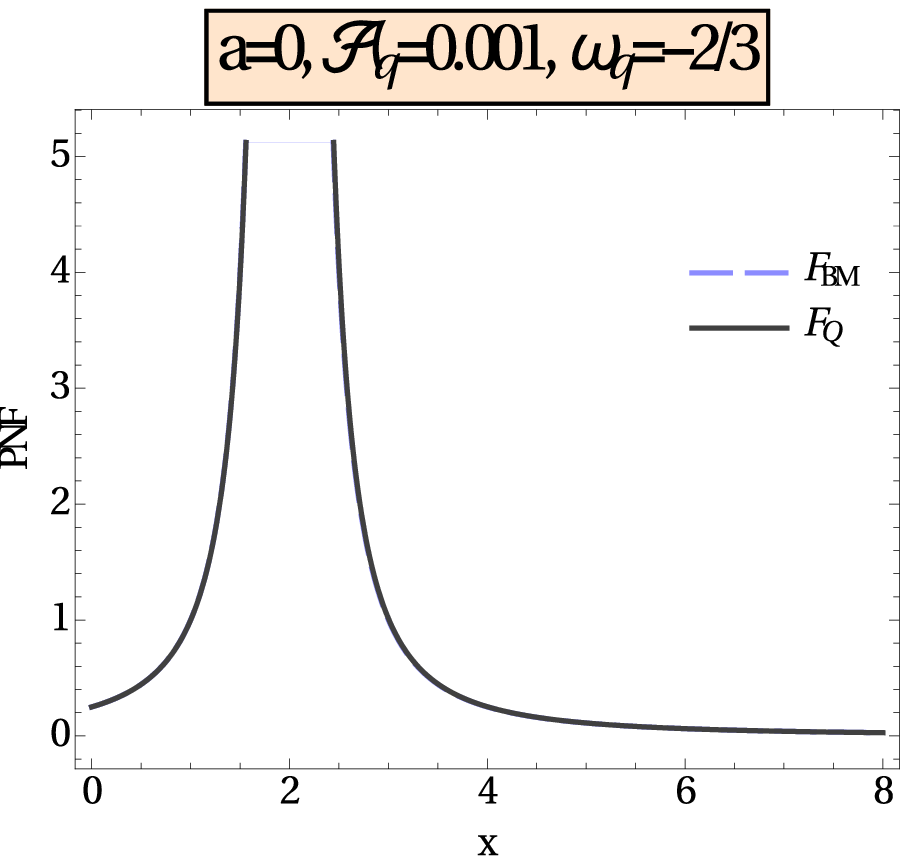}~~
\hspace*{0.5 cm}
\includegraphics*[scale=0.3]{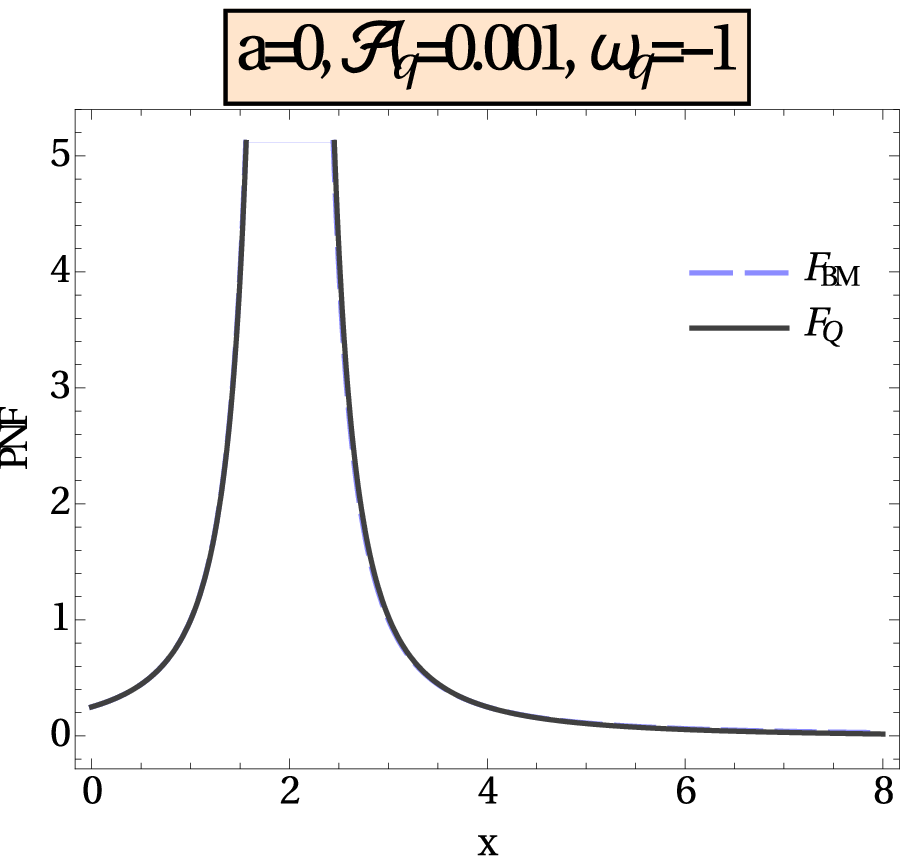}~~\\

Fig $1.14.a$ \hspace{1.6 cm} Fig $1.14.b$ \hspace{1.6 cm} Fig $1.14.c$ \hspace{1.6 cm} Fig $1.14.d$
\includegraphics*[scale=0.3]{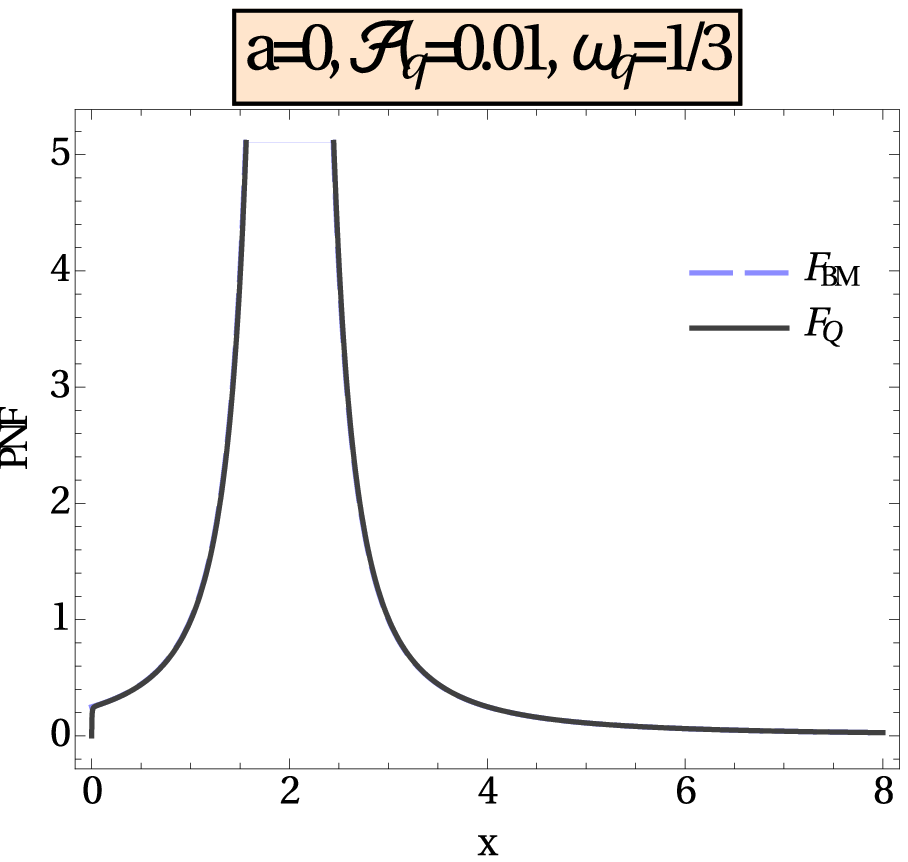}~~
\hspace*{0.5 cm}
\includegraphics*[scale=0.3]{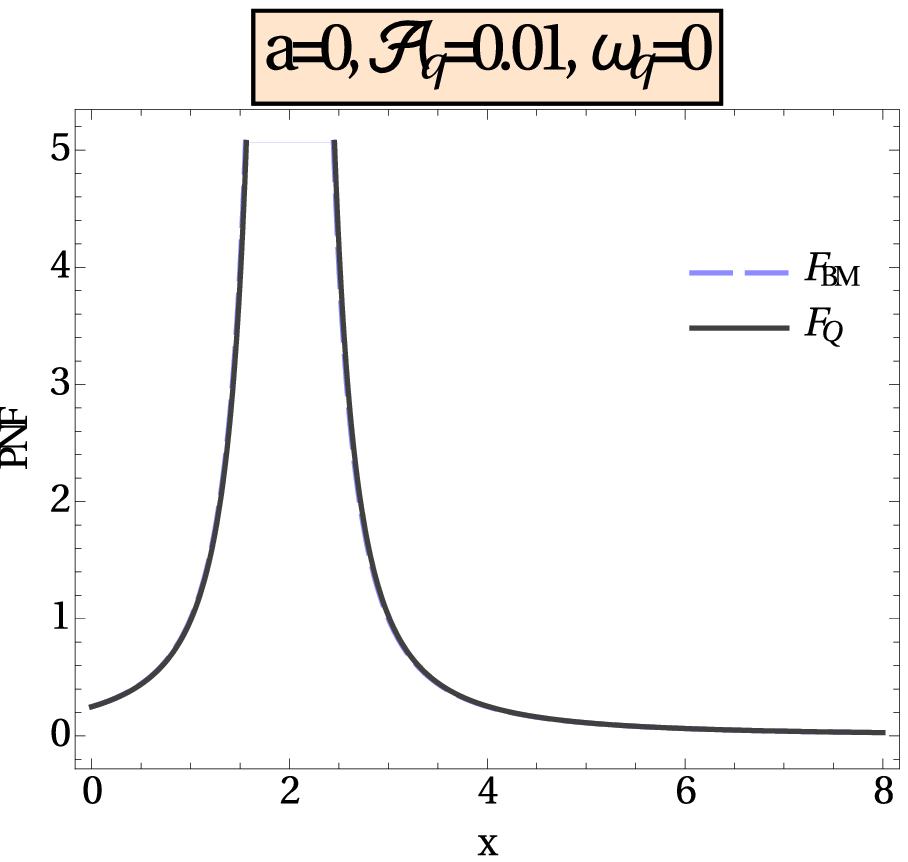}~~
\hspace*{0.5 cm}
\includegraphics*[scale=0.3]{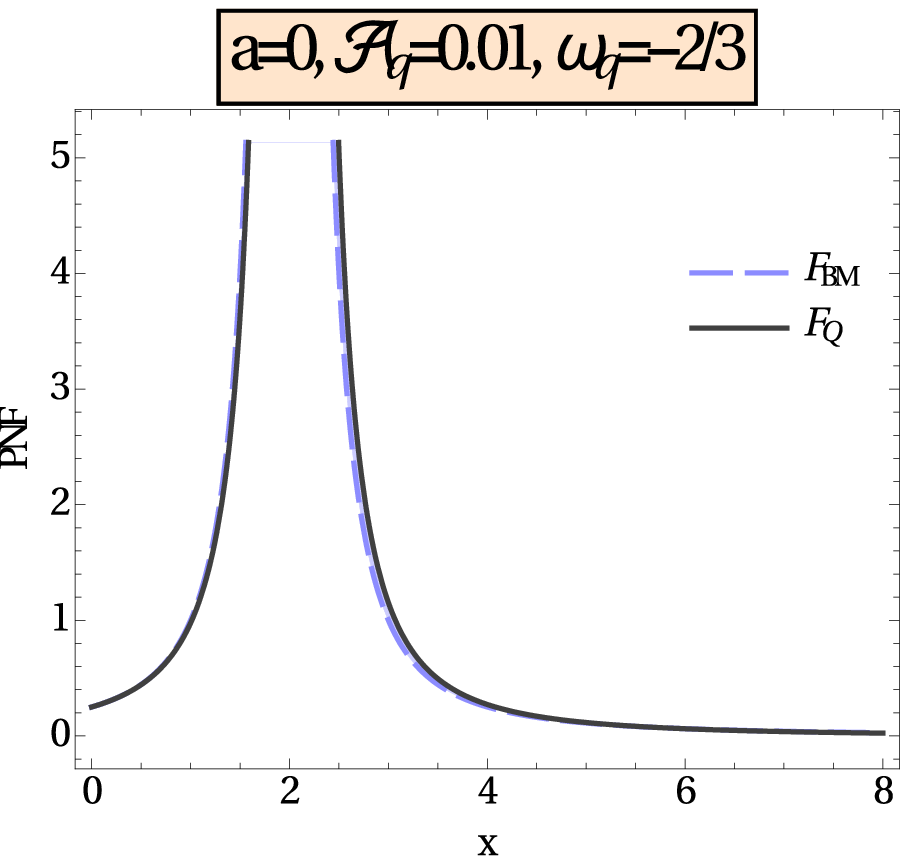}~~
\hspace*{0.5 cm}
\includegraphics*[scale=0.3]{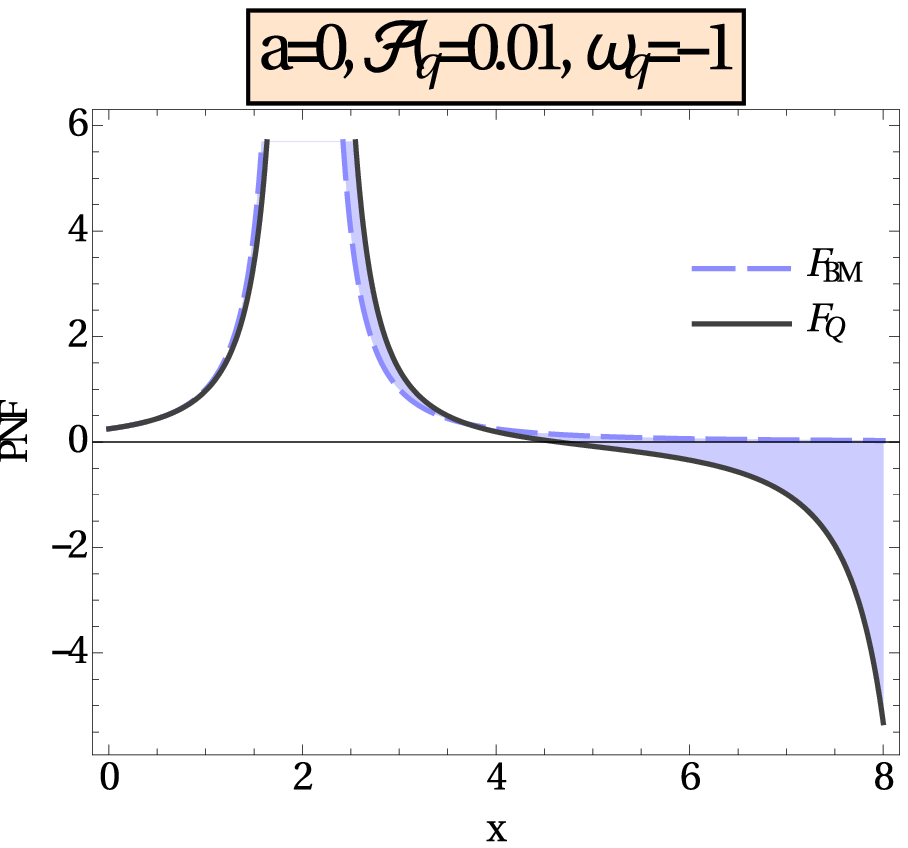}~~\\

Fig $1.15.a$ \hspace{1.6 cm} Fig $1.15.b$ \hspace{1.6 cm} Fig $1.15.c$ \hspace{1.6 cm} Fig $1.15.d$
\includegraphics*[scale=0.3]{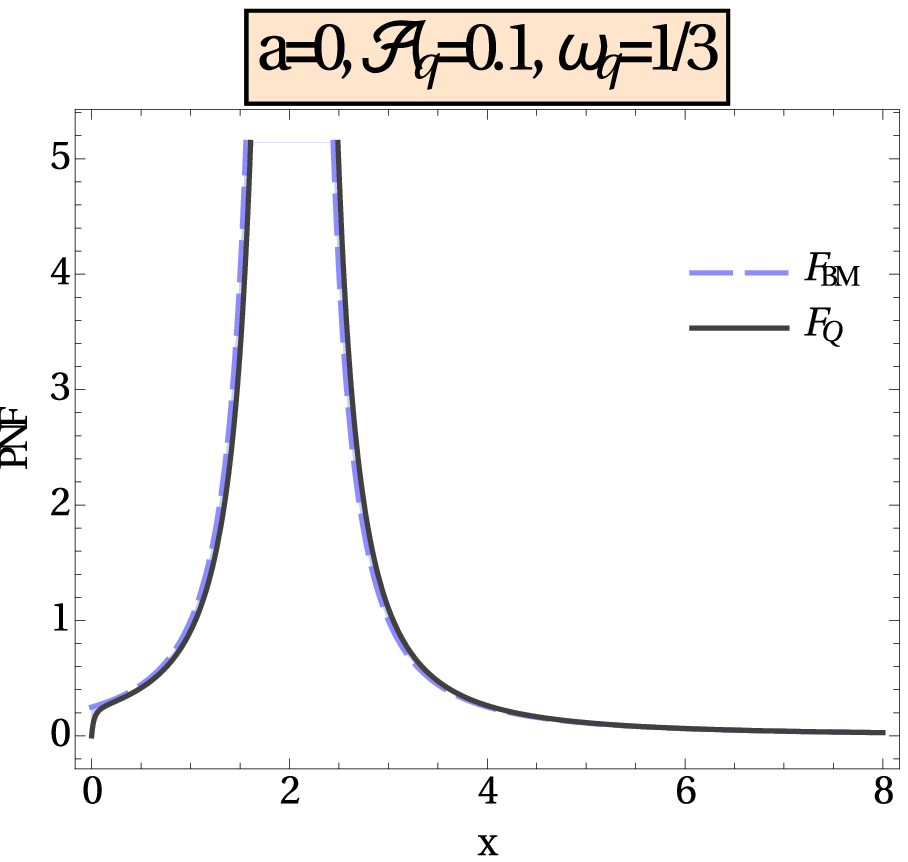}~~
\hspace*{0.5 cm}
\includegraphics*[scale=0.3]{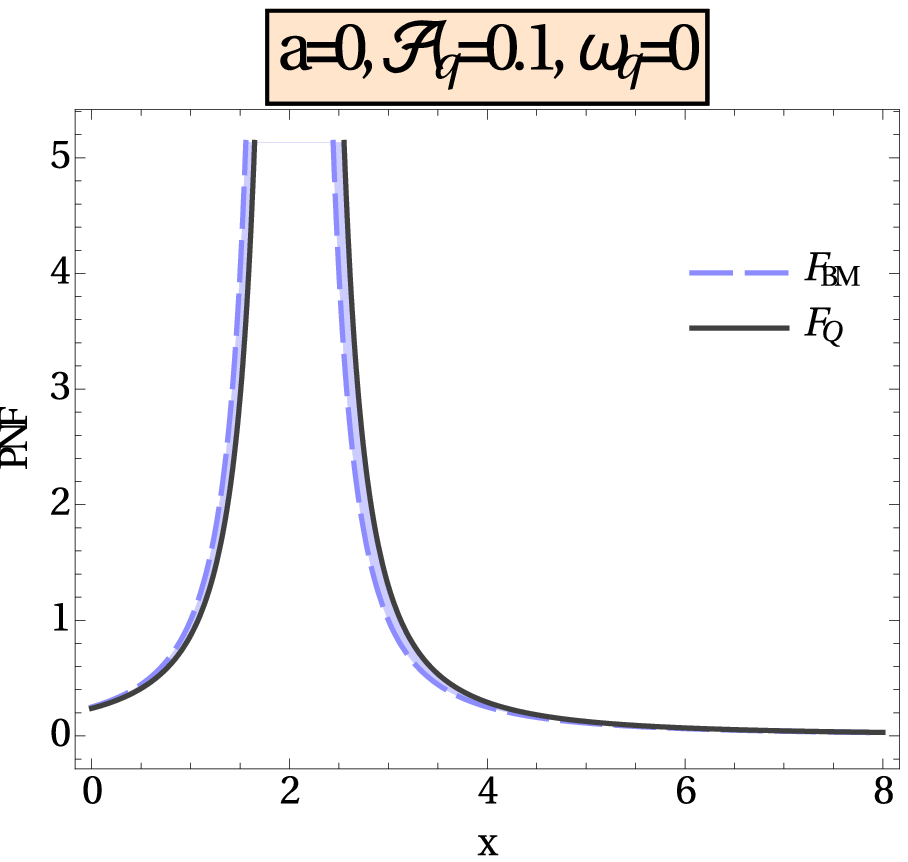}~~
\hspace*{0.5 cm}
\includegraphics*[scale=0.3]{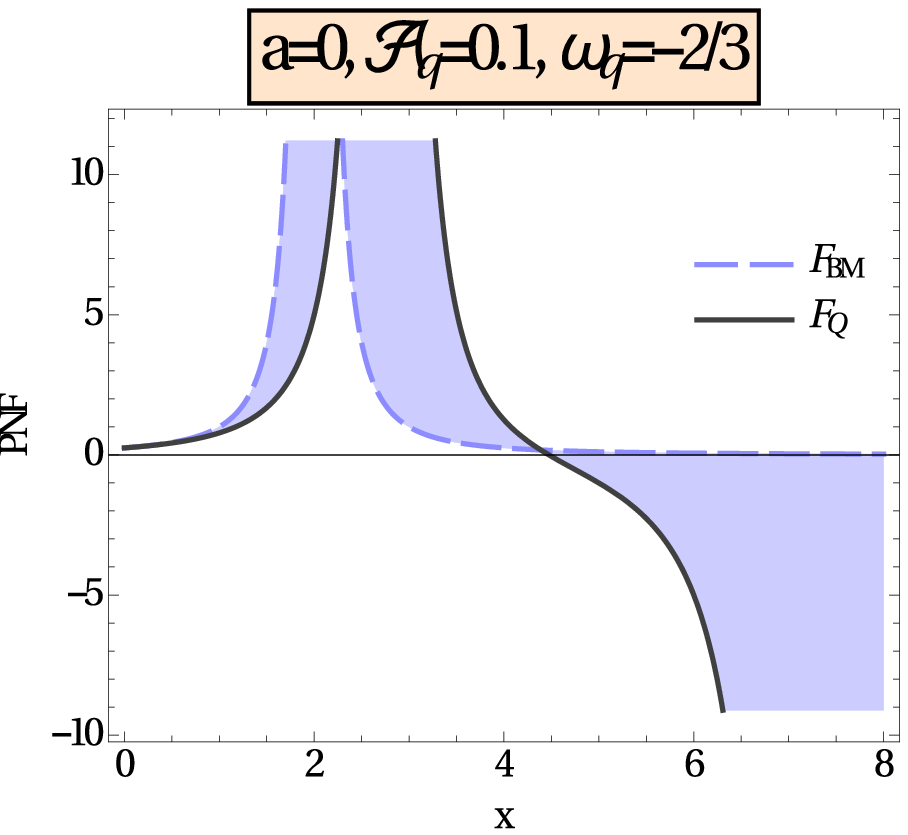}~~
\hspace*{0.5 cm}
\includegraphics*[scale=0.3]{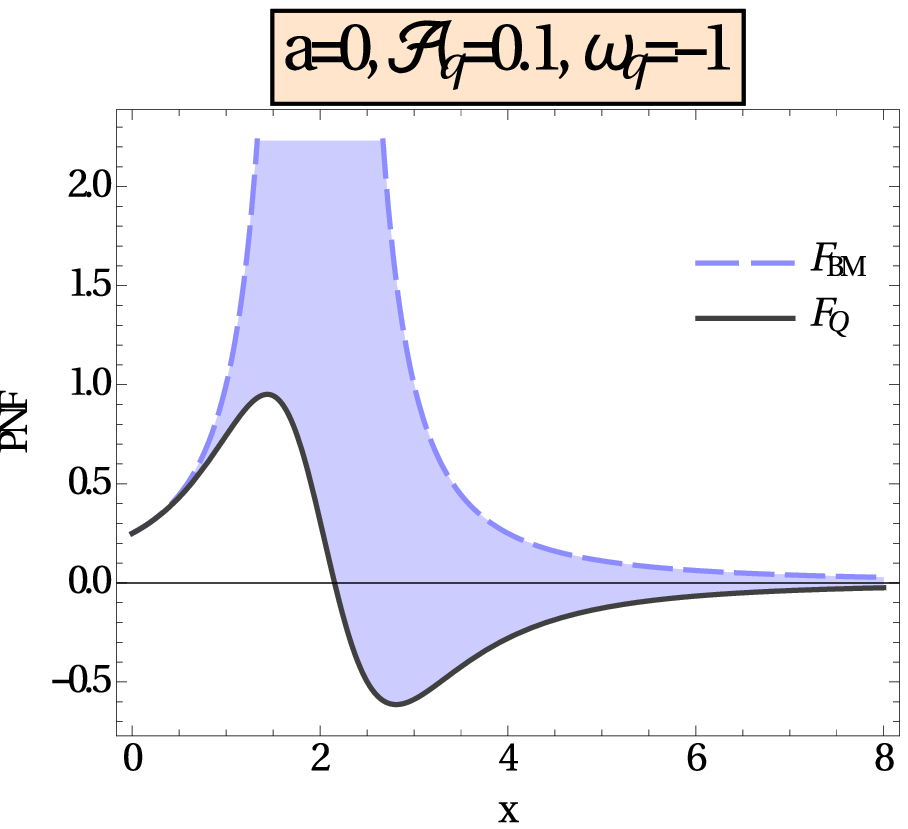}~~\\

\it{Here $a$ is taken as $0$ so that both PNFs reduced to that of a Schwarzschild like BH. For each values of ${\cal A}_q$, i.e., $10^{-3}$, $10^{-2}$ and $10^{-1}$ we have drawn the relative results for $\omega_q=\frac{1}{3}$ for radiation, $\omega_q=0$ for dust, $\omega_q=-\frac{2}{3}$ for quintessence and $\omega_q=-1$ for phantom barrier.}\\

Fig $1.16.a$ \hspace{1.6 cm} Fig $1.16.b$ \hspace{1.6 cm} Fig $1.16.c$ \hspace{1.6 cm} Fig $1.16.d$
\includegraphics*[scale=0.3]{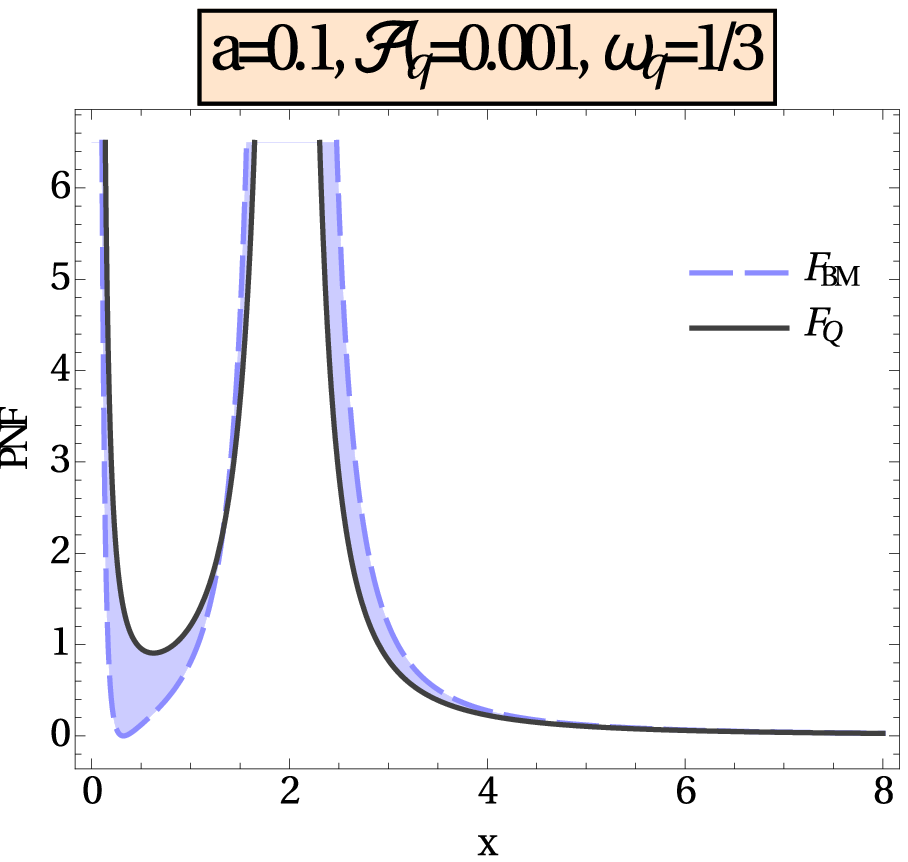}~~
\hspace*{0.5 cm}
\includegraphics*[scale=0.3]{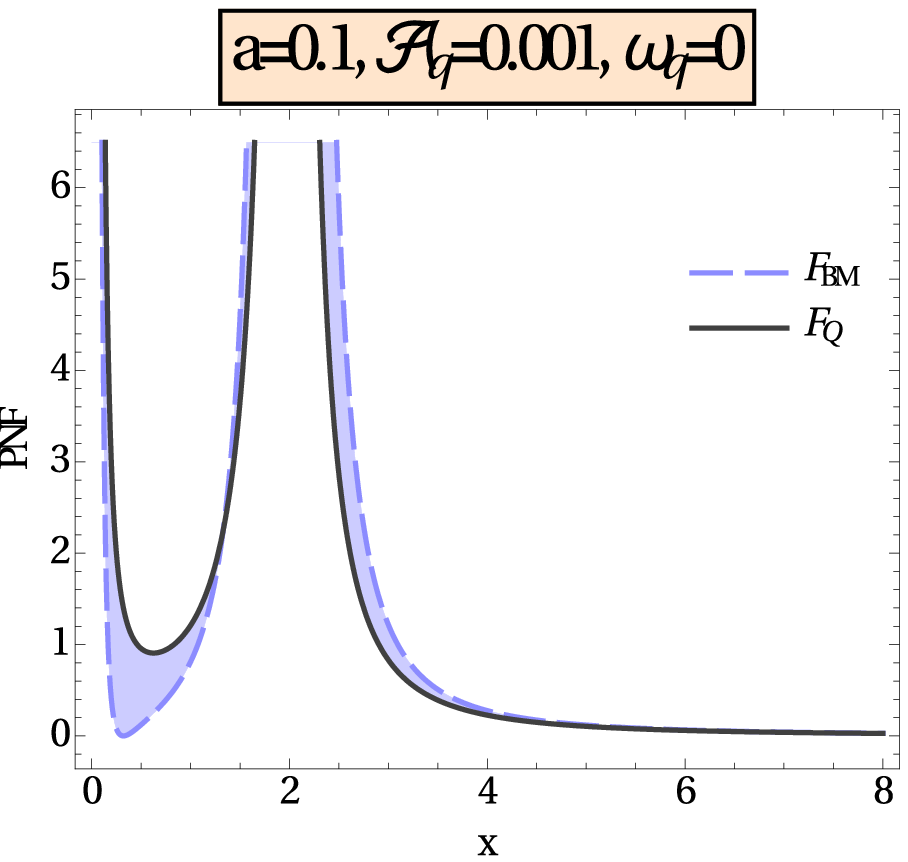}~~
\hspace*{0.5 cm}
\includegraphics*[scale=0.3]{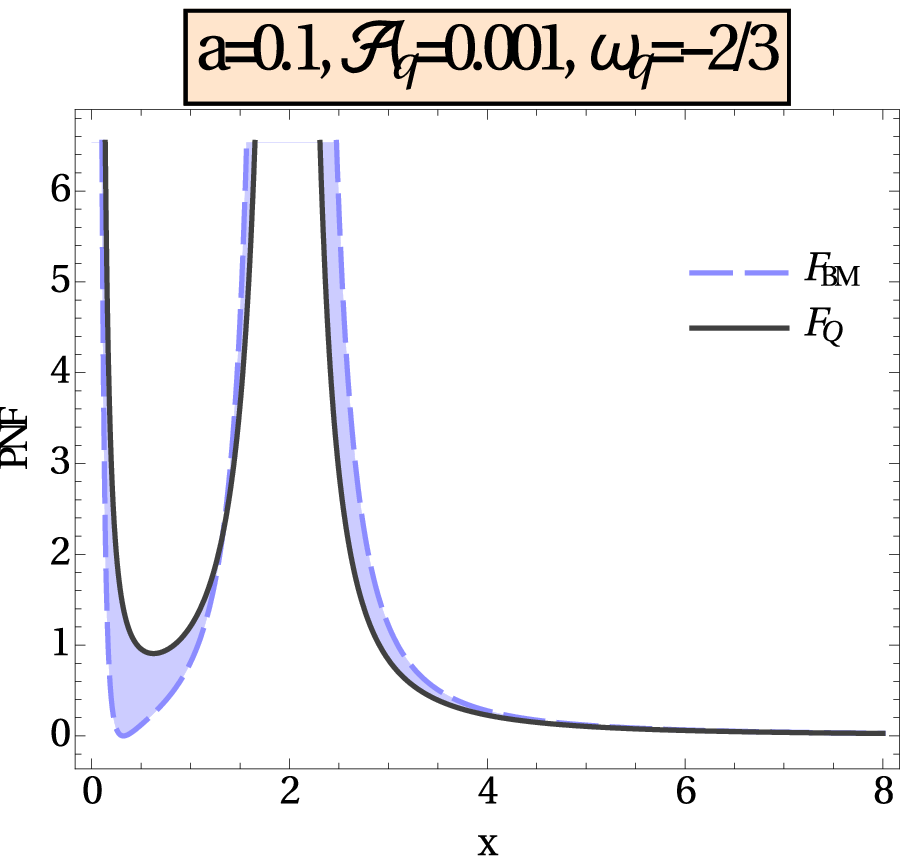}~~
\hspace*{0.5 cm}
\includegraphics*[scale=0.3]{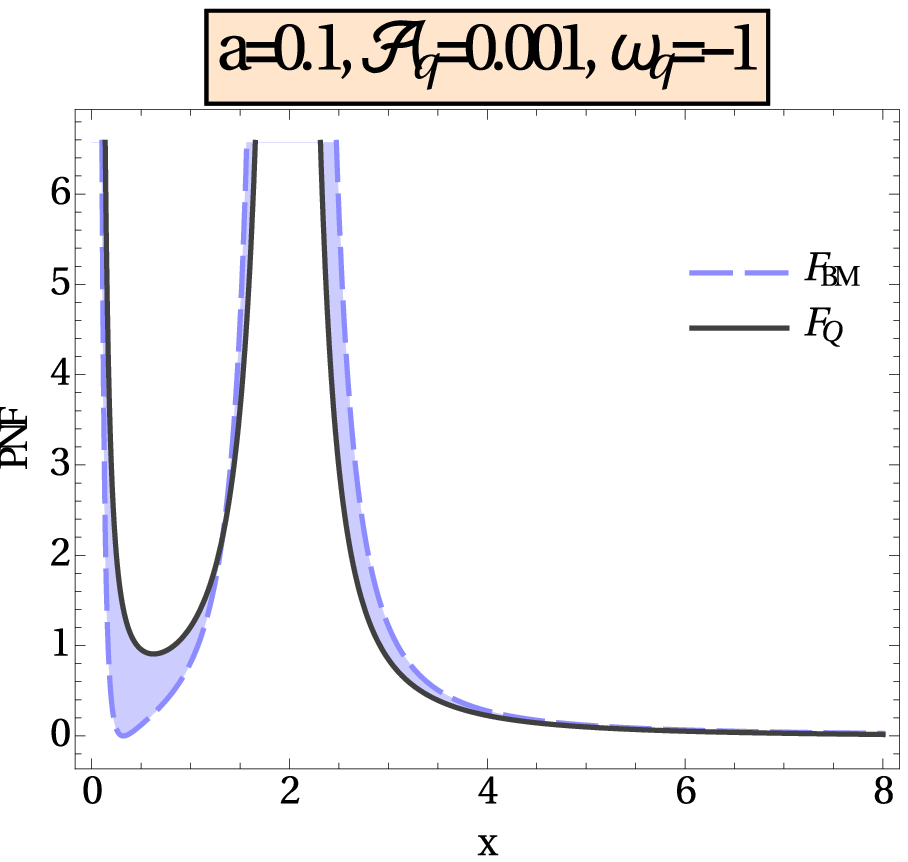}~~\\

Fig $1.17.a$ \hspace{1.6 cm} Fig $1.17.b$ \hspace{1.6 cm} Fig $1.17.c$ \hspace{1.6 cm} Fig $1.17.d$
\includegraphics*[scale=0.3]{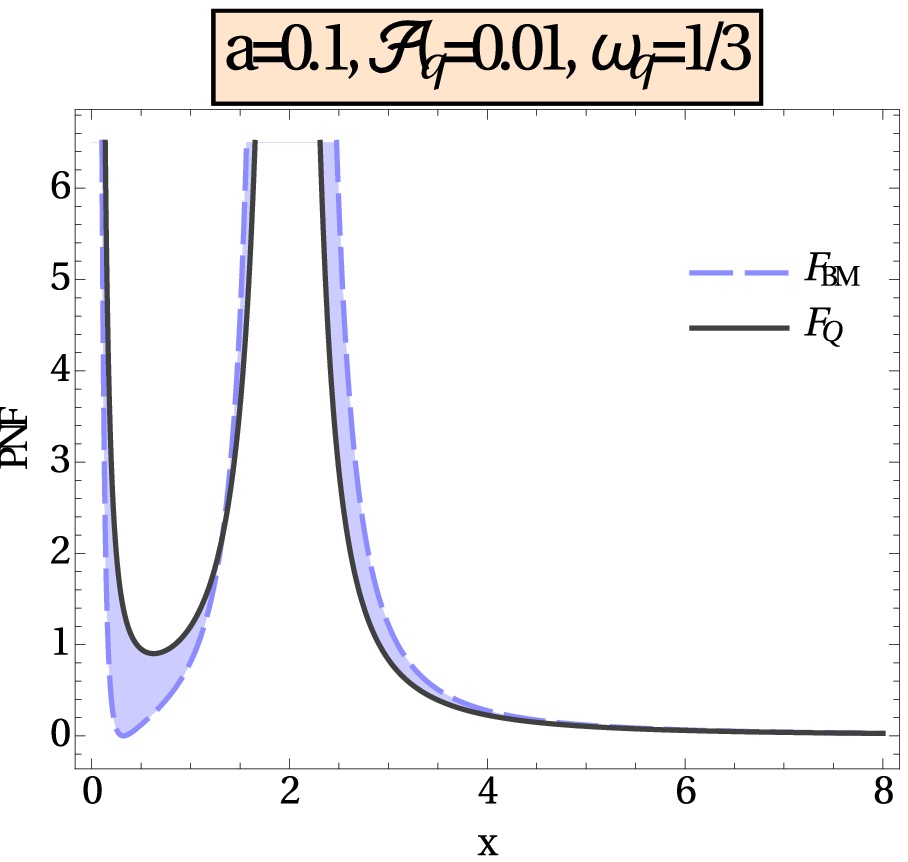}~~
\hspace*{0.5 cm}
\includegraphics*[scale=0.3]{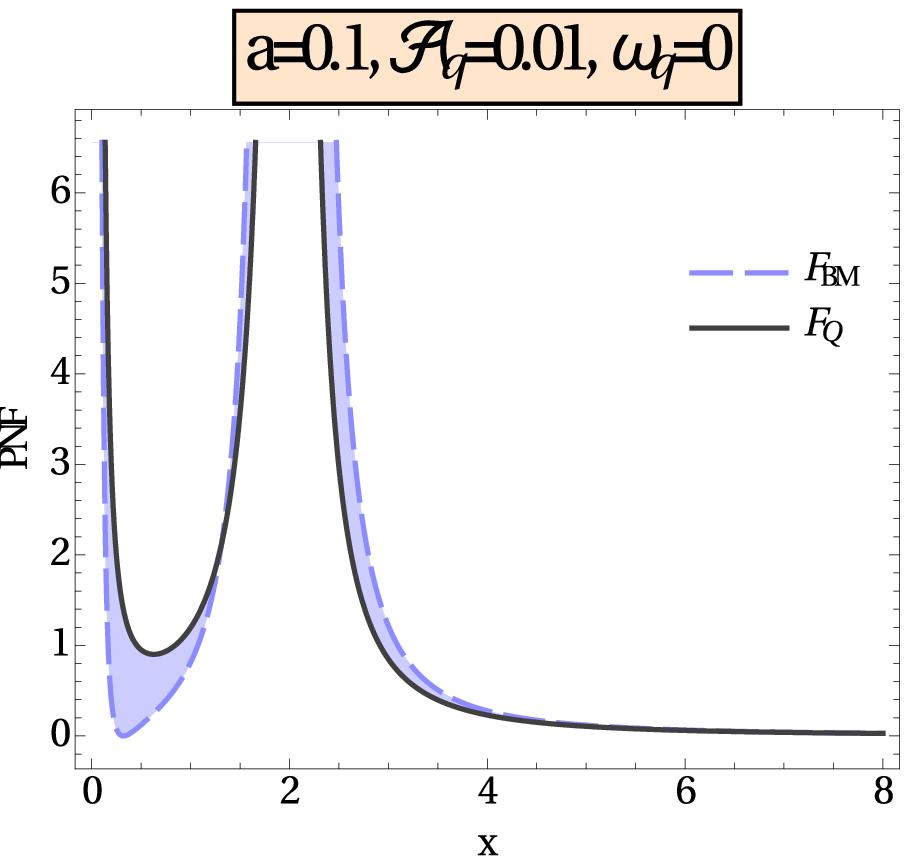}~~
\hspace*{0.5 cm}
\includegraphics*[scale=0.3]{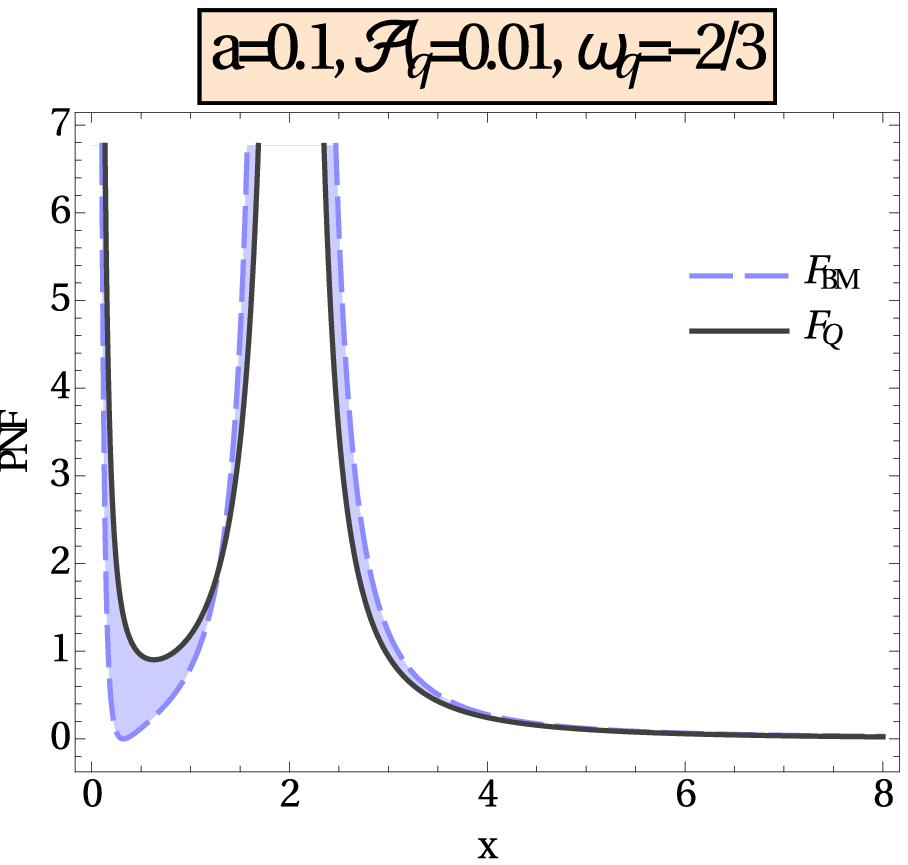}~~
\hspace*{0.5 cm}
\includegraphics*[scale=0.3]{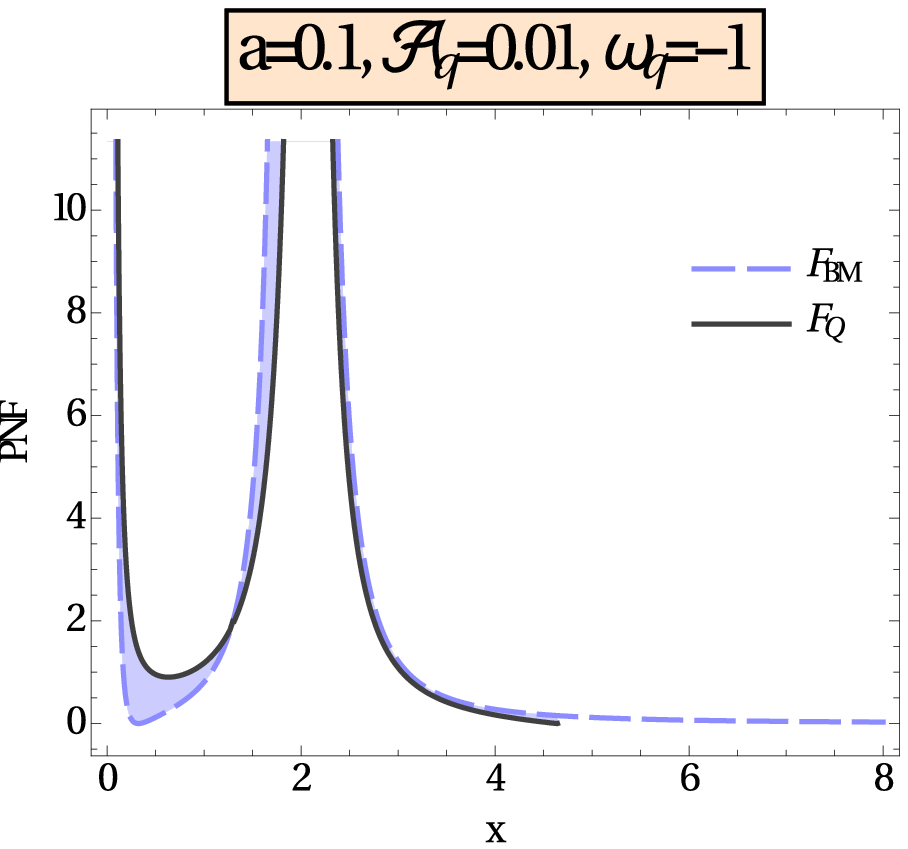}~~\\

Fig $1.18.a$ \hspace{1.6 cm} Fig $1.18.b$ \hspace{1.6 cm} Fig $1.18.c$ \hspace{1.6 cm} Fig $1.18.d$
\includegraphics*[scale=0.3]{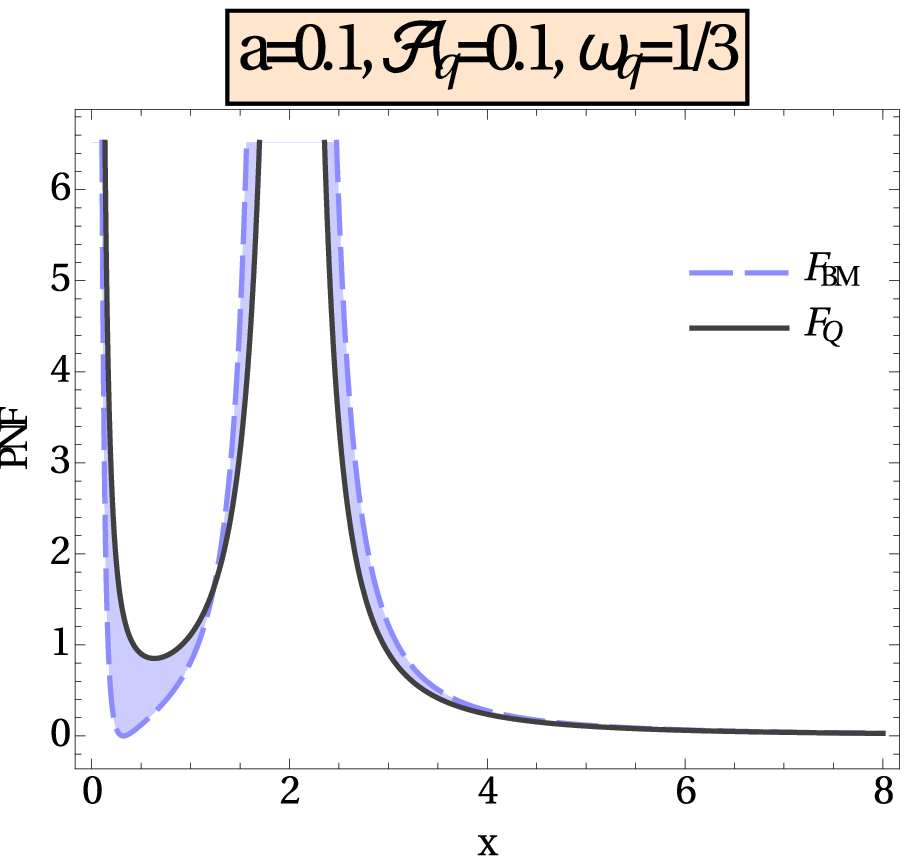}~~
\hspace*{0.5 cm}
\includegraphics*[scale=0.3]{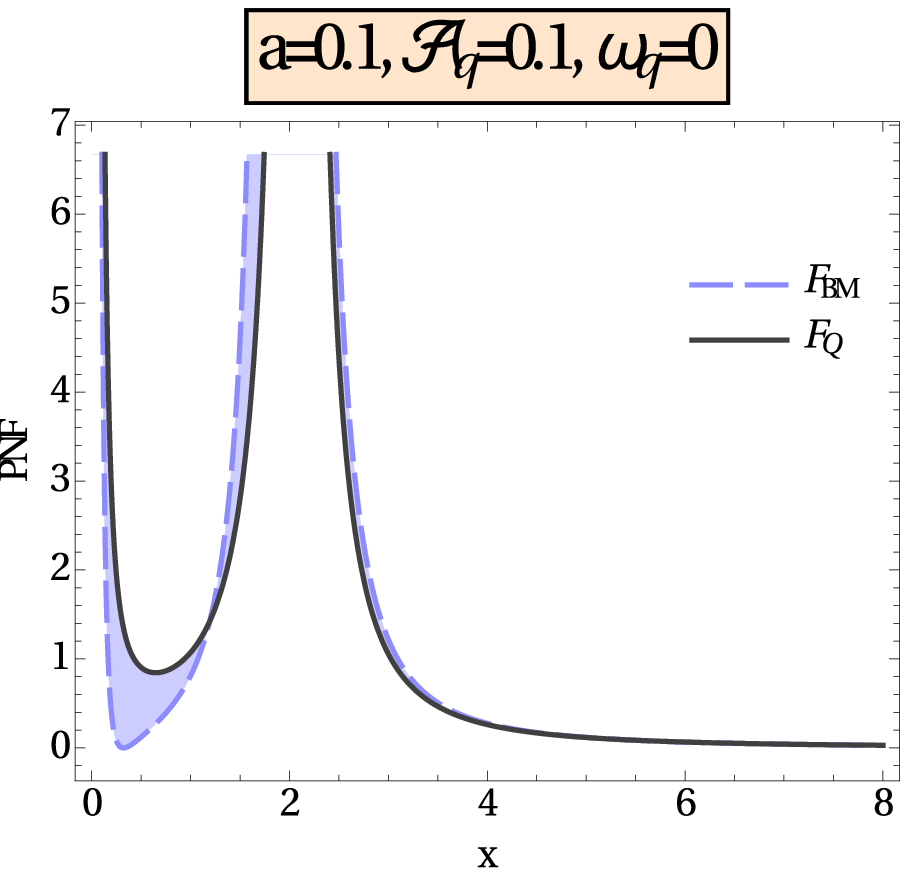}~~
\hspace*{0.5 cm}
\includegraphics*[scale=0.3]{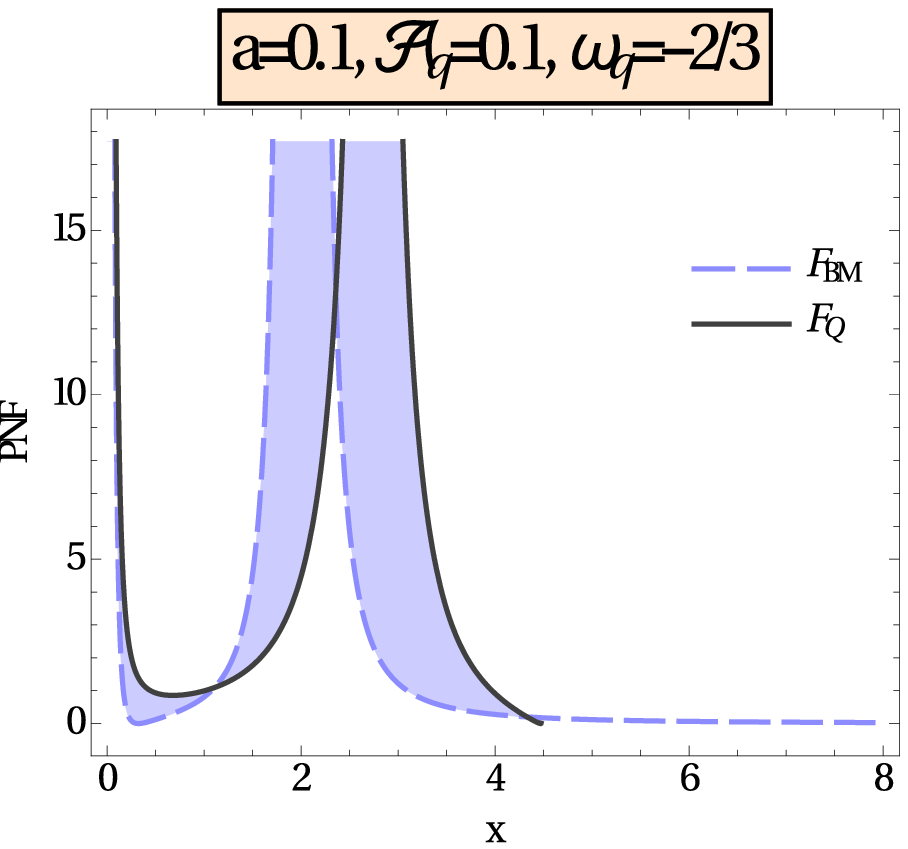}~~
\hspace*{0.5 cm}
\includegraphics*[scale=0.3]{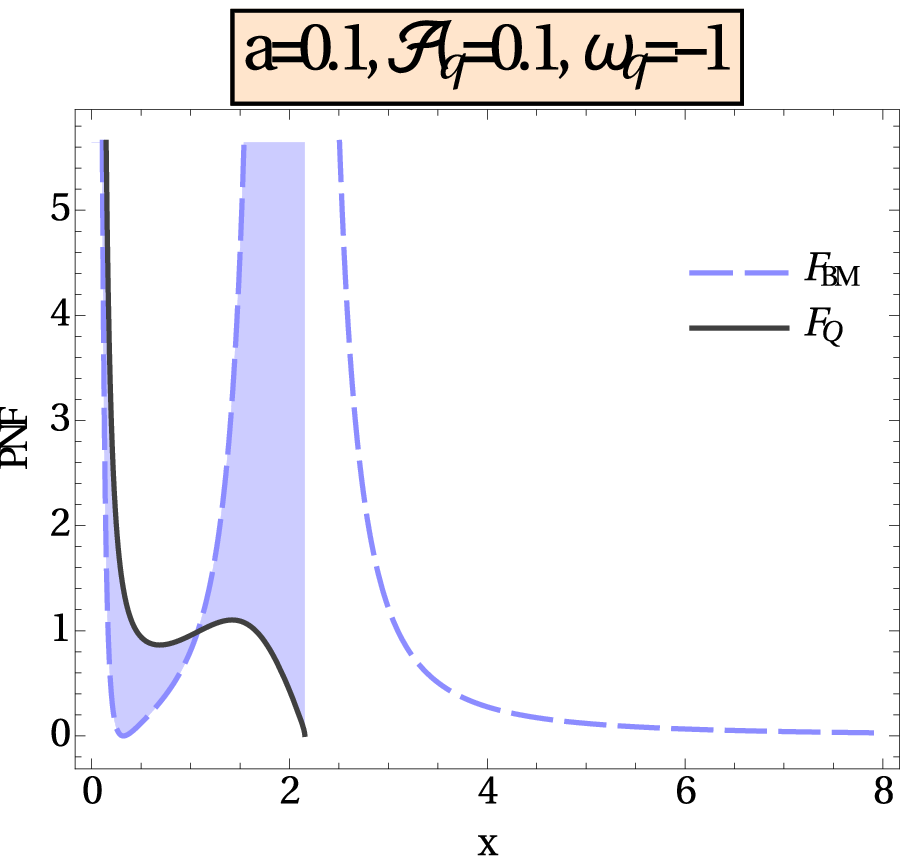}~~\\

\it{Here $a=0.1$, for each values of ${\cal A}_q$, i.e., $10^{-3}$, $10^{-2}$ and $10^{-1}$ we have drawn a relative results for $\omega_q=\frac{1}{3}$ for radiation, $\omega_q=0$ for dust, $\omega_q=-\frac{2}{3}$ for quintessence and $\omega_q=-1$ for phantom barrier.}

\end{figure}

For figures $1.13.a$ to $1.15.d$ we have considered $a$ to be zero, i.e., we are taking BH with angular momentum to be zero so the reduced BH is a non-rotating Schwarzschild BH. For radiation $(\omega_q=1/3)$ and dust $(\omega_q=0)$ both the forces are identically equal and relevant to previous results where both the PNF have a singularity at $x=2$. But for quintessence and phantom barrier, we have more interesting results. Especially, for $a=0,~{\cal A}_q=0.01,~\omega_q=-1$, our PNF changes its sign near $x=4.5$, i.e., instead of attraction the central object starts to repulse. For $a=0,~{\cal A}_q=0.1,~\omega_q=-2/3$, i.e., Schwarzschild BH embedded in quintessence there is a singularity at $x=2$, further if the radius of BH is greater than $4.5$, it starts to repulse with a significant amount of force. Even this force becomes stronger when we increase $x$. Finally, for $a=0,~{\cal A}_q=0.1,~\omega_q=-1$, our force does not blow up anywhere but obtains a maxima in $(1,2)$ and minima in $(2,4)$. But in this case, i.e., for phantom barrier a particle feels attractive force toward center when $x<2$ and when it crosses the horizon $x=2$ it feels a repulsive force, though this repulsive force diminishes as $x$ increases.  

Where DE effect is relatively high and we have taken the DE agent as quintessence and phantom barrier, we have observed only a repulsive force experienced by an accreting particle. This is very interesting fact that DE accretion starts to repulse which may be a possible explanation of accelerated expansion of the universe.

For the figures $1.16.a$ to $1.18.d$, we have considered $a=0.1$, i.e., we are taking a BH with a small amount of co-rotation. As we increase the radius of the BH, the force starts with a high amount, then starts to decrease and after obtaining minima they start to increase then blows up together and similarly for higher $x$ both of them converges to zero. One thing to notice is that Mukhopadhyay's PNF blows up early than that of us and initially our force remains dominating but in the second half Mukhopadhyay's force turns to be dominating. Finally, for $a=0.1,~{\cal A}_q=0.1,~\omega_k=-2/3$, i.e., for slowly co-rotating BH with a little amount of quintessence effect our PNF follows the other one but after reaching $4.5$ it stops giving values. Similarly for $a=0.1,~{\cal A}_q=0.1,~\omega_q=-1$, i.e., for phantom barrier our force does not even follow its predecessor and attains a maxima and then stops giving real and physical value after $x>2$.

If we dig too much through our figures we will get a point near $x=1$ where both the forces have coincided with each other, i.e., we have observed particular radius (other than large $x$) where both BHs attract with same forces irrespective of any DE to be considered. Also for quintessence, i.e., for $a=0.1,~{\cal A}_q=0.1,~\omega_q=-2/3$, our force blows up lately which indicates that the normal rotating BHs become saturated early than BHs with DE. 
\begin{figure}
\centering

Fig $1.19.a$ \hspace{1.6 cm} Fig $1.19.b$ \hspace{1.6 cm} Fig $1.19.c$ \hspace{1.6 cm} Fig $1.19.d$
\includegraphics*[scale=0.3]{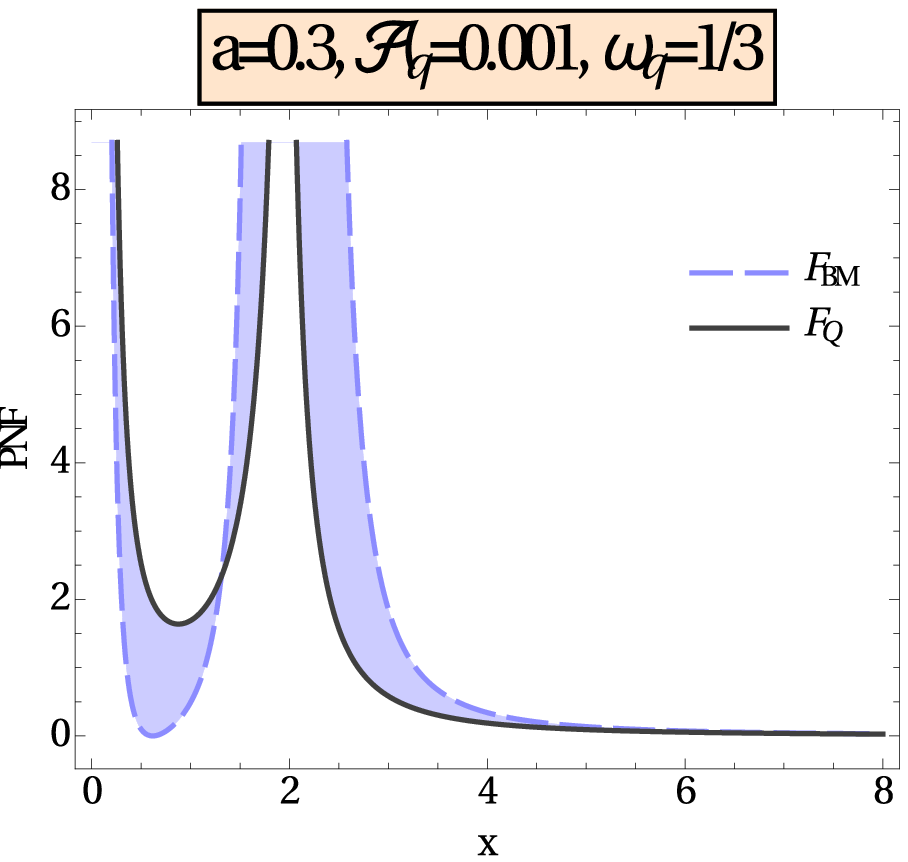}~~
\hspace*{0.5 cm}
\includegraphics*[scale=0.3]{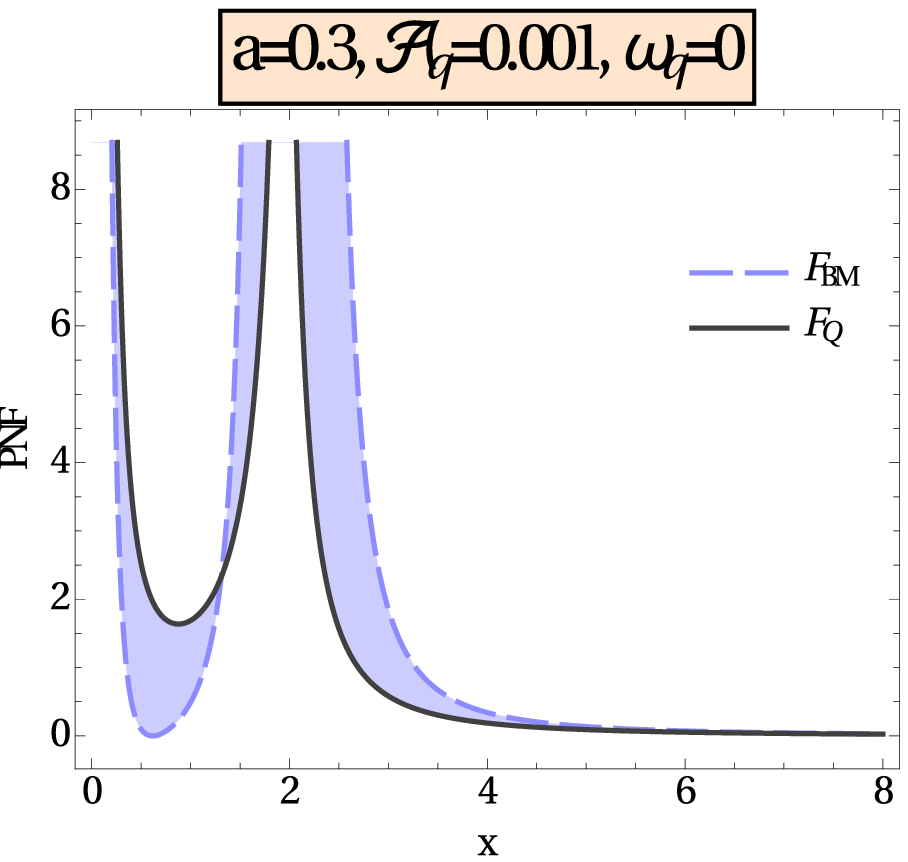}~~
\hspace*{0.5 cm}
\includegraphics*[scale=0.3]{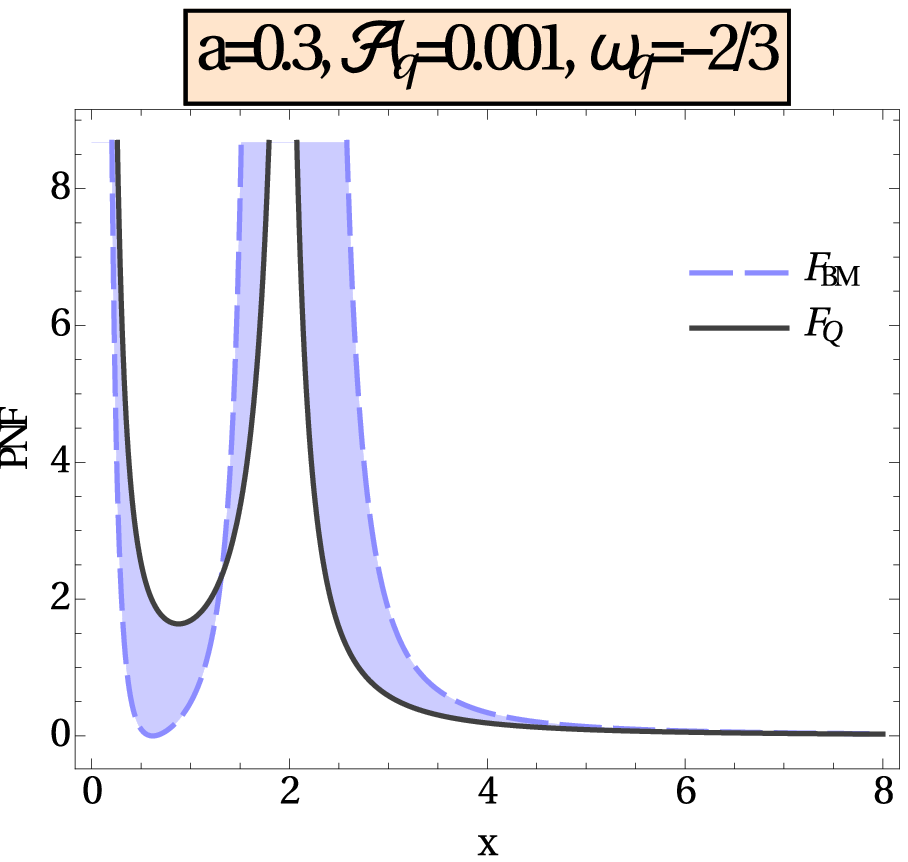}~~
\hspace*{0.5 cm}
\includegraphics*[scale=0.3]{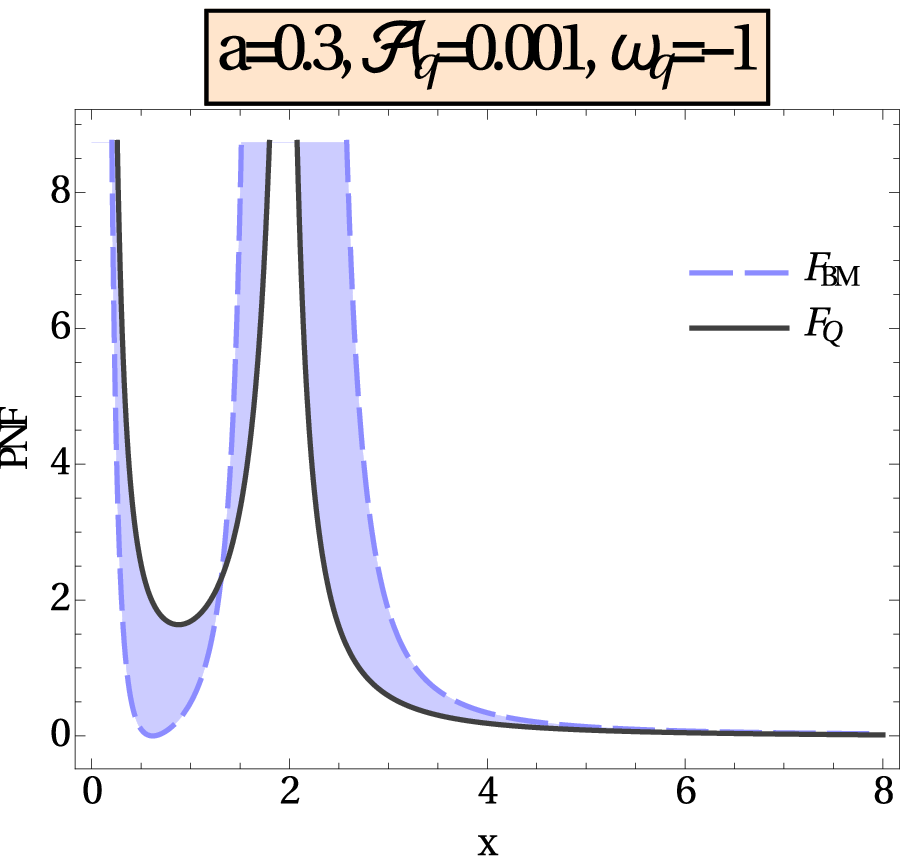}~~\\

Fig $1.20.a$ \hspace{1.6 cm} Fig $1.20.b$ \hspace{1.6 cm} Fig $1.20.c$ \hspace{1.6 cm} Fig $1.20.d$
\includegraphics*[scale=0.3]{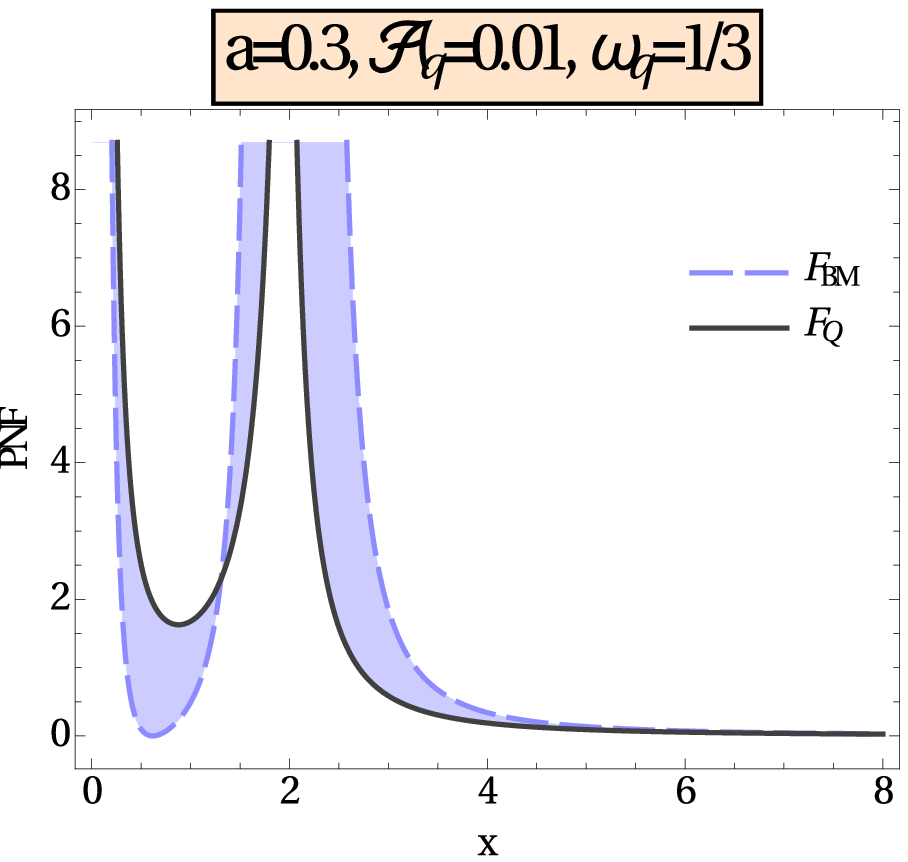}~~
\hspace*{0.5 cm}
\includegraphics*[scale=0.3]{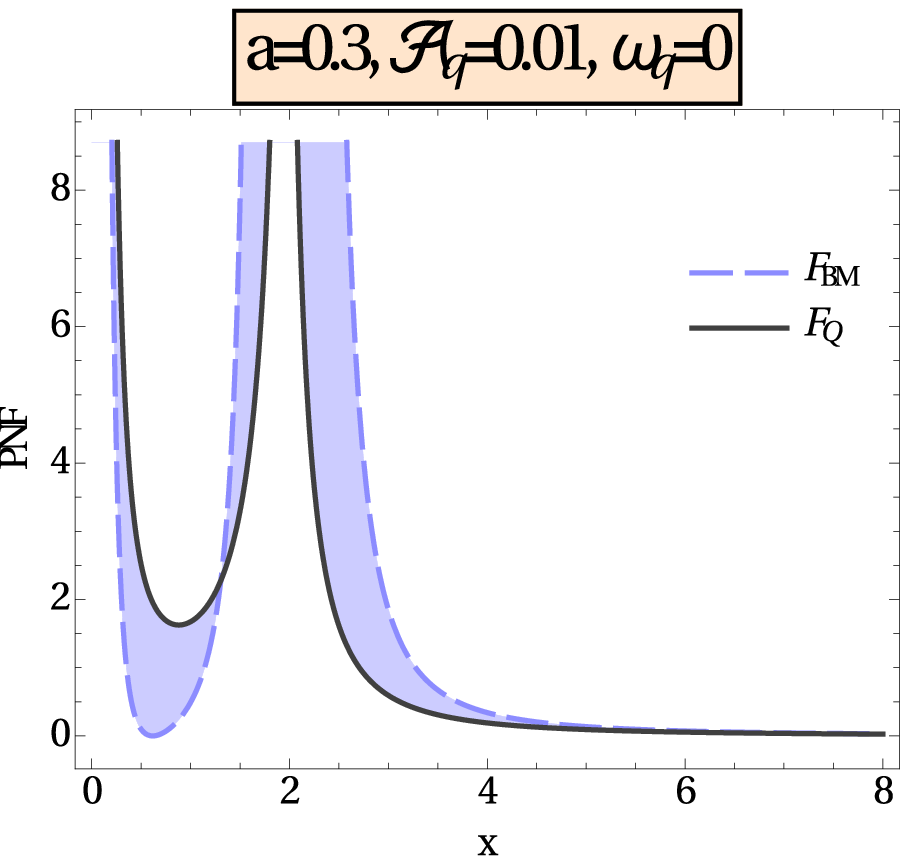}~~
\hspace*{0.5 cm}
\includegraphics*[scale=0.3]{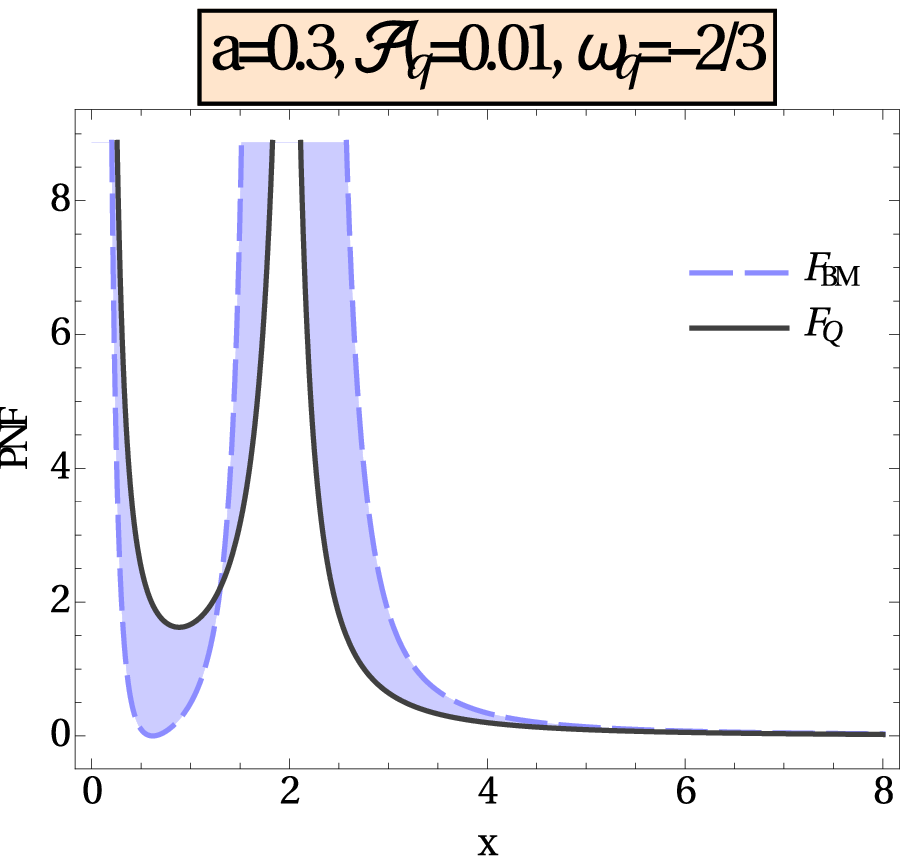}~~
\hspace*{0.5 cm}
\includegraphics*[scale=0.3]{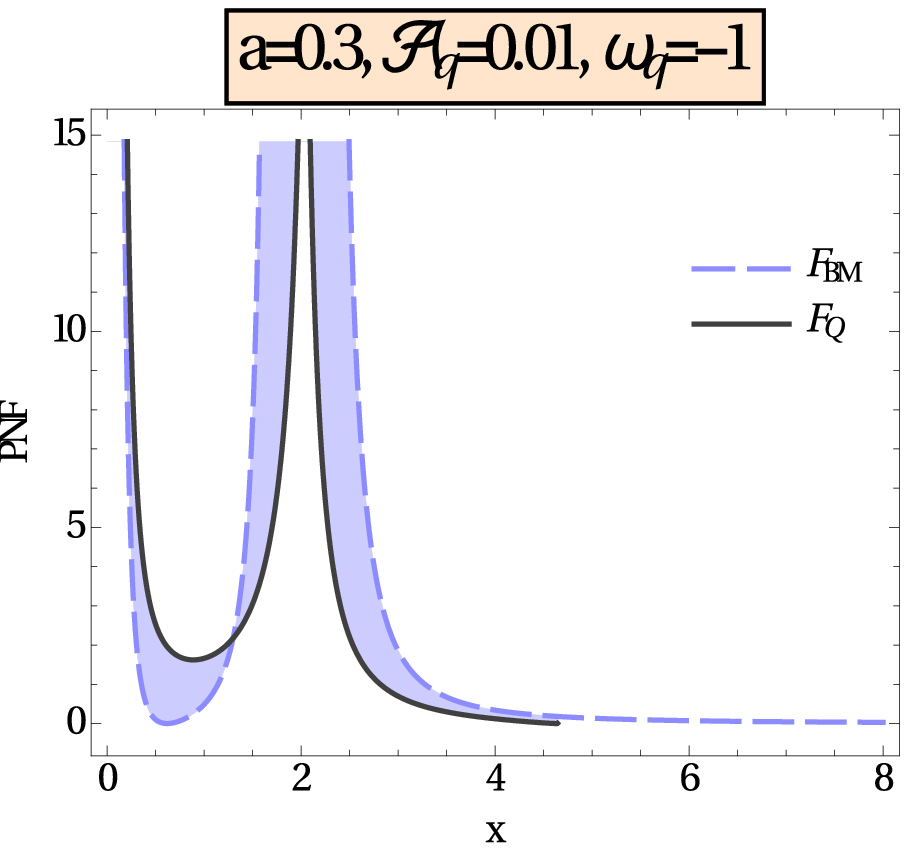}~~\\

Fig $1.21.a$ \hspace{1.6 cm} Fig $1.21.b$ \hspace{1.6 cm} Fig $1.21.c$ \hspace{1.6 cm} Fig $1.21.d$
\includegraphics*[scale=0.3]{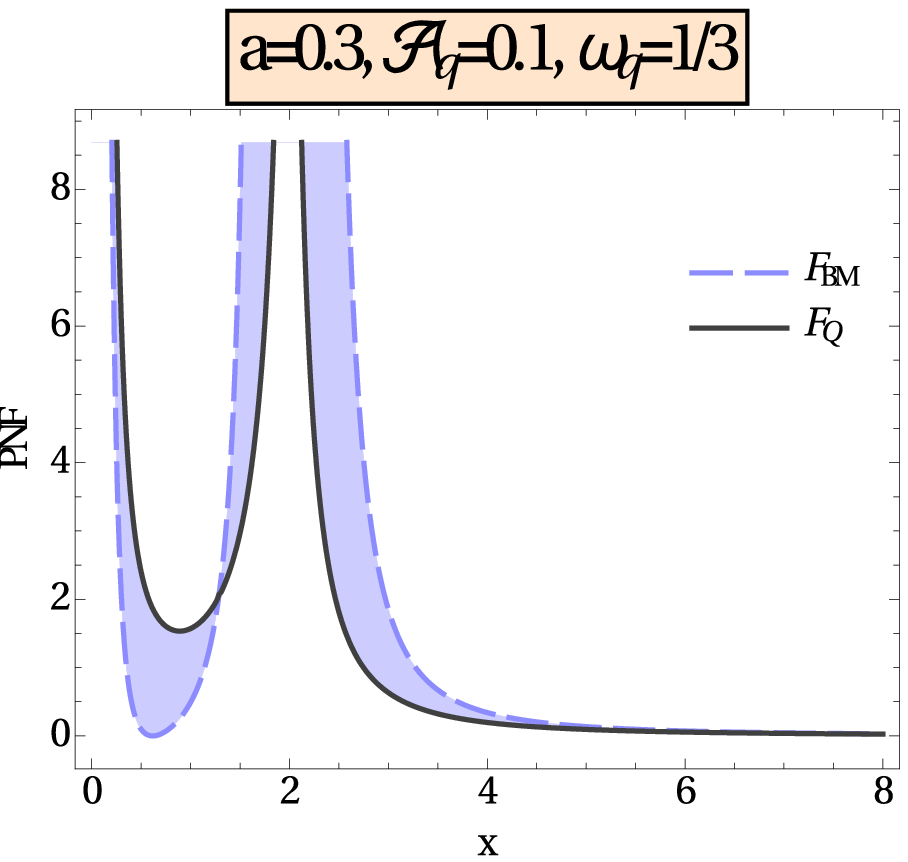}~~
\hspace*{0.5 cm}
\includegraphics*[scale=0.3]{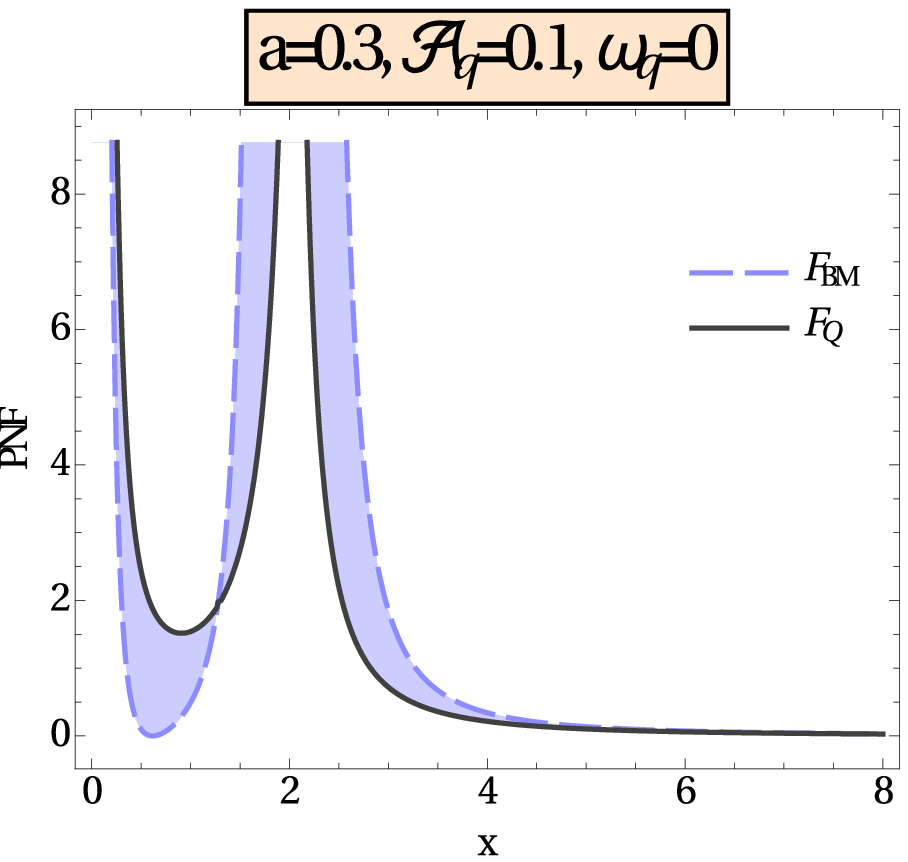}~~
\hspace*{0.5 cm}
\includegraphics*[scale=0.3]{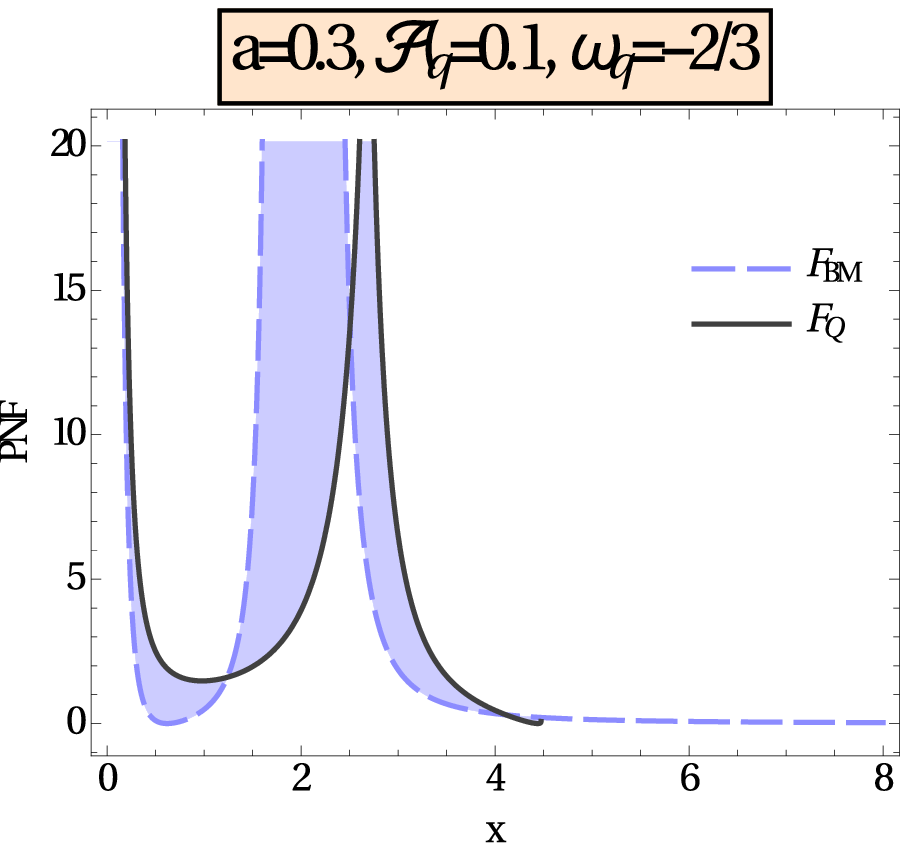}~~
\hspace*{0.5 cm}
\includegraphics*[scale=0.3]{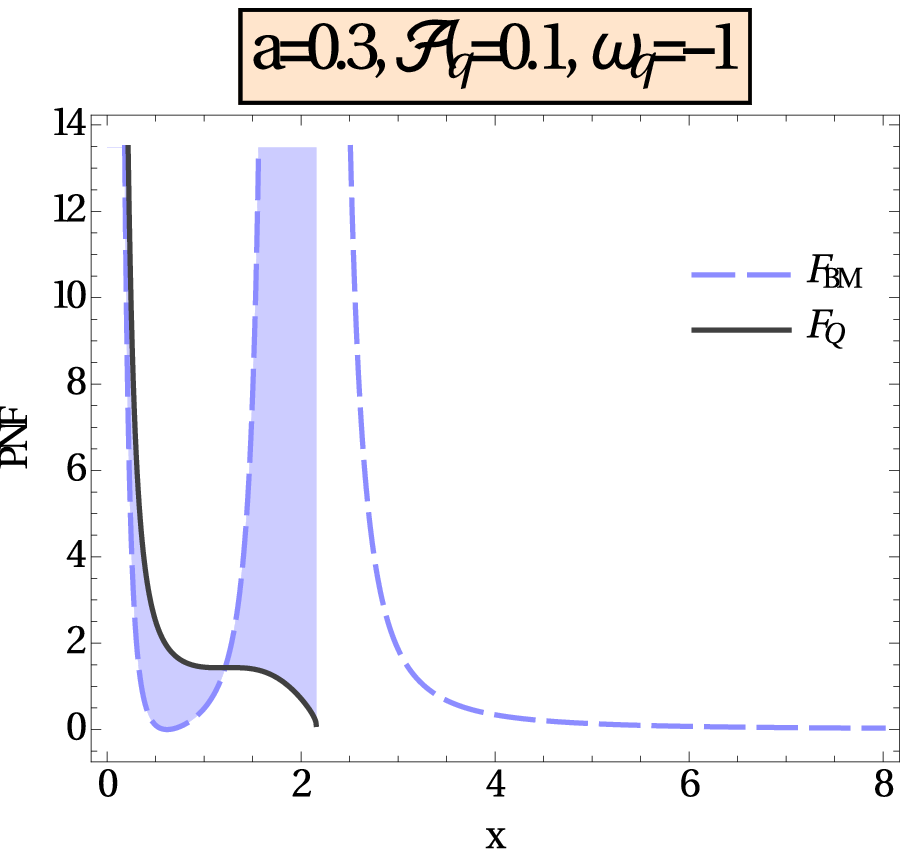}~~\\

\it{Here $a=0.3$, for each values of ${\cal A}_q$, i.e., $10^{-3}$, $10^{-2}$ and $10^{-1}$ we have drawn a relative results for $\omega_q=\frac{1}{3}$ for radiation, $\omega_q=0$ for dust, $\omega_q=-\frac{2}{3}$ for quintessence and $\omega_q=-1$ for phantom barrier.}\\

Fig $1.22.a$ \hspace{1.6 cm} Fig $1.22.b$ \hspace{1.6 cm} Fig $1.22.c$ \hspace{1.6 cm} Fig $1.22.d$
\includegraphics*[scale=0.3]{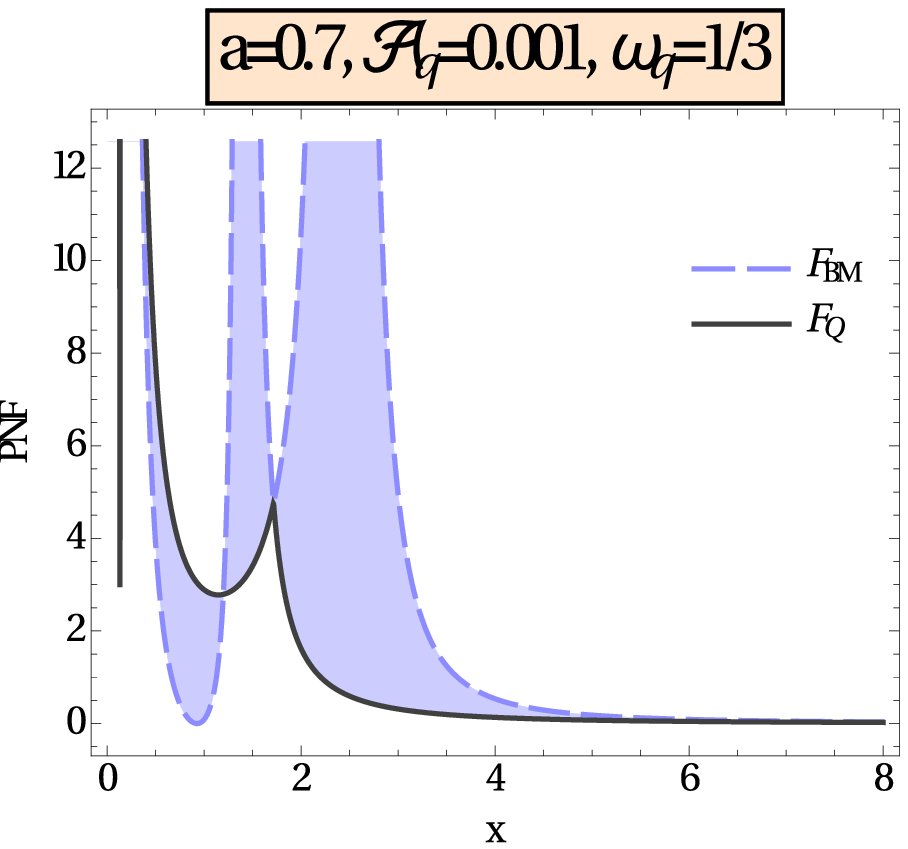}~~
\hspace*{0.5 cm}
\includegraphics*[scale=0.3]{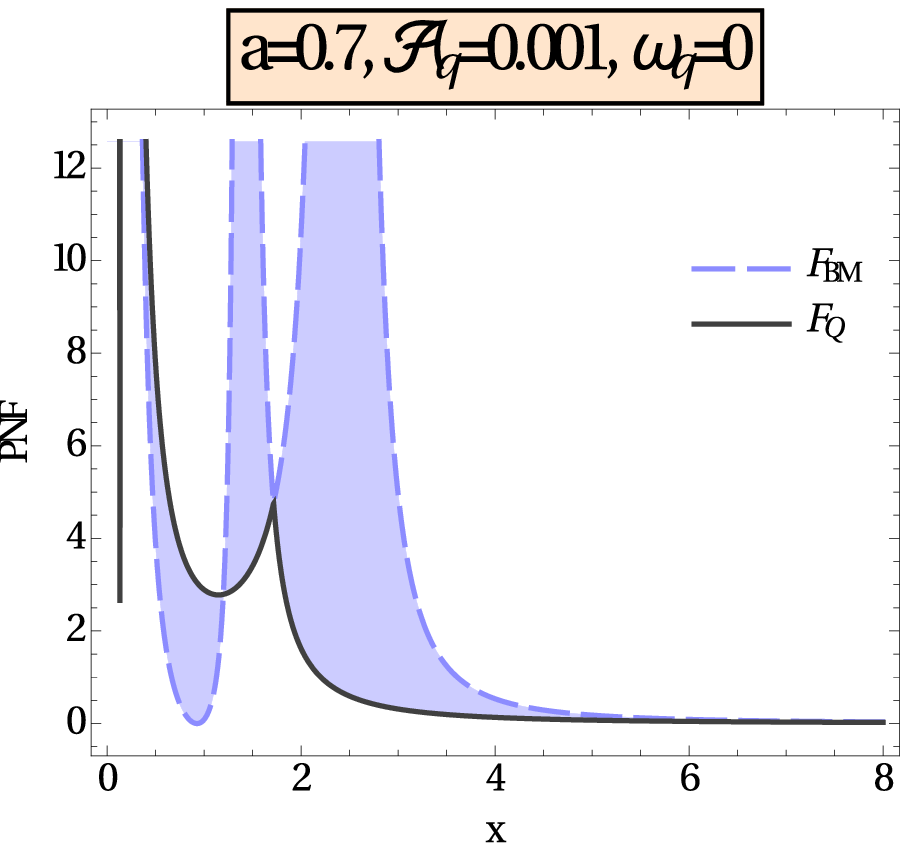}~~
\hspace*{0.5 cm}
\includegraphics*[scale=0.3]{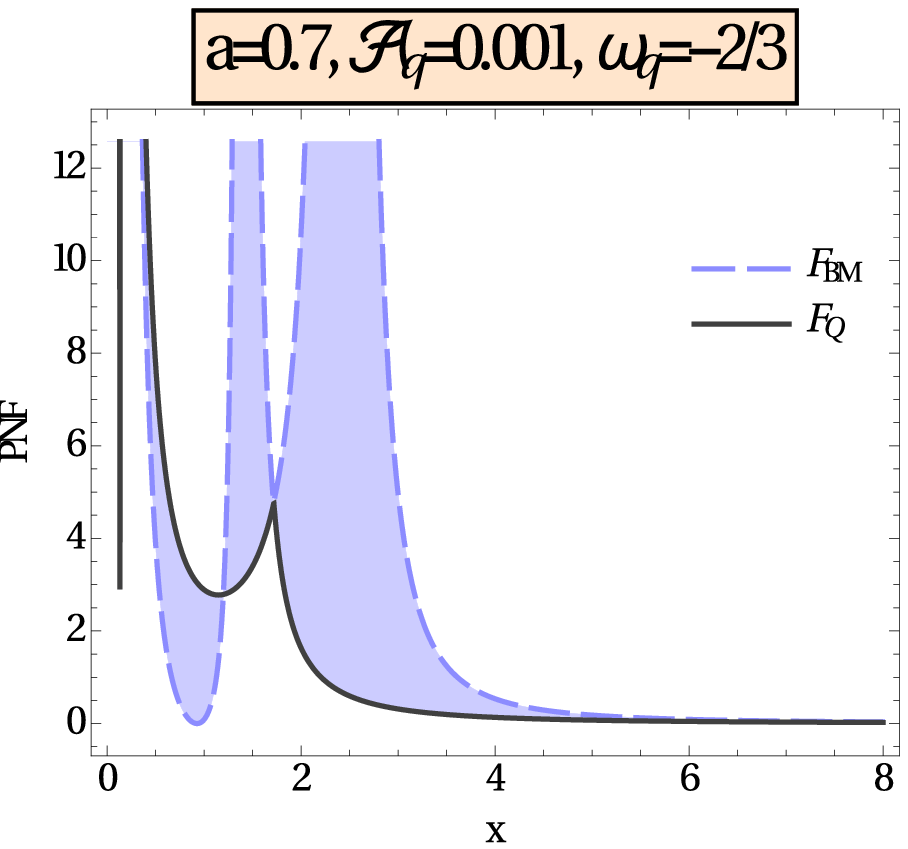}~~
\hspace*{0.5 cm}
\includegraphics*[scale=0.3]{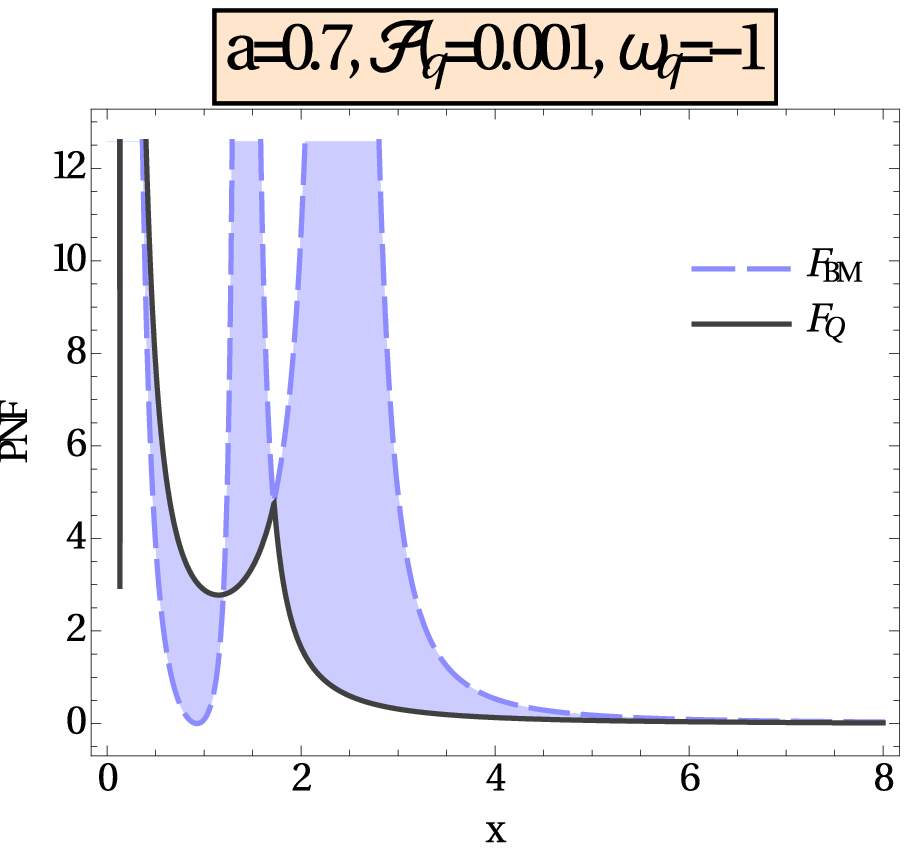}~~\\

Fig $1.23.a$ \hspace{1.6 cm} Fig $1.23.b$ \hspace{1.6 cm} Fig $1.23.c$ \hspace{1.6 cm} Fig $1.23.d$
\includegraphics*[scale=0.3]{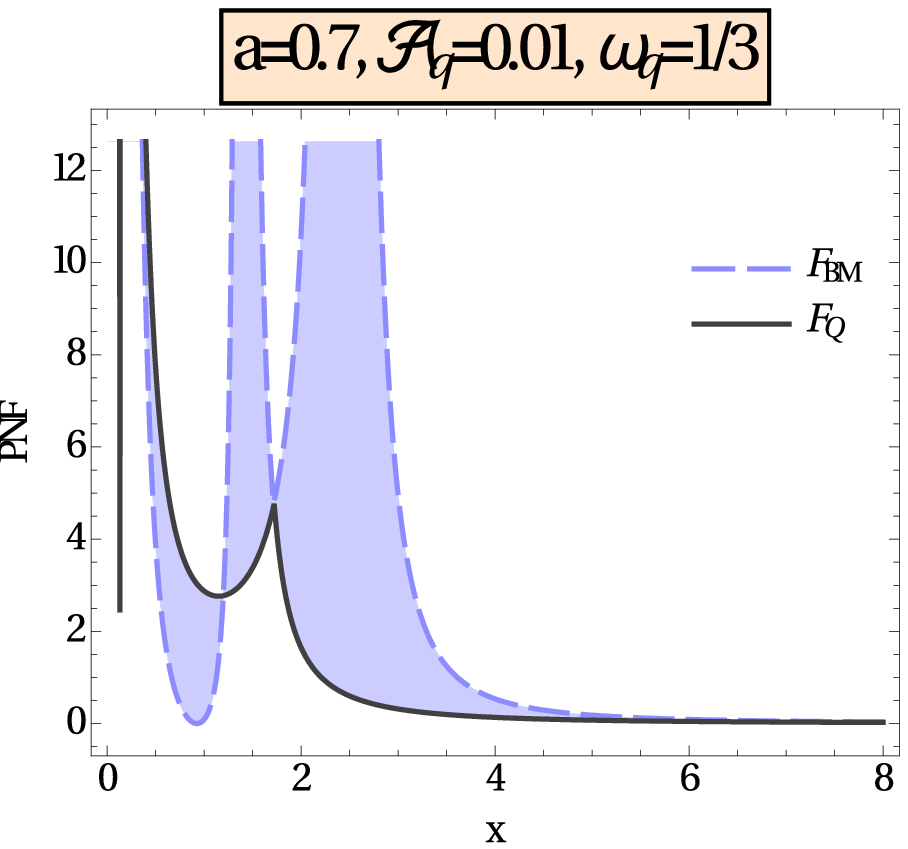}~~
\hspace*{0.5 cm}
\includegraphics*[scale=0.3]{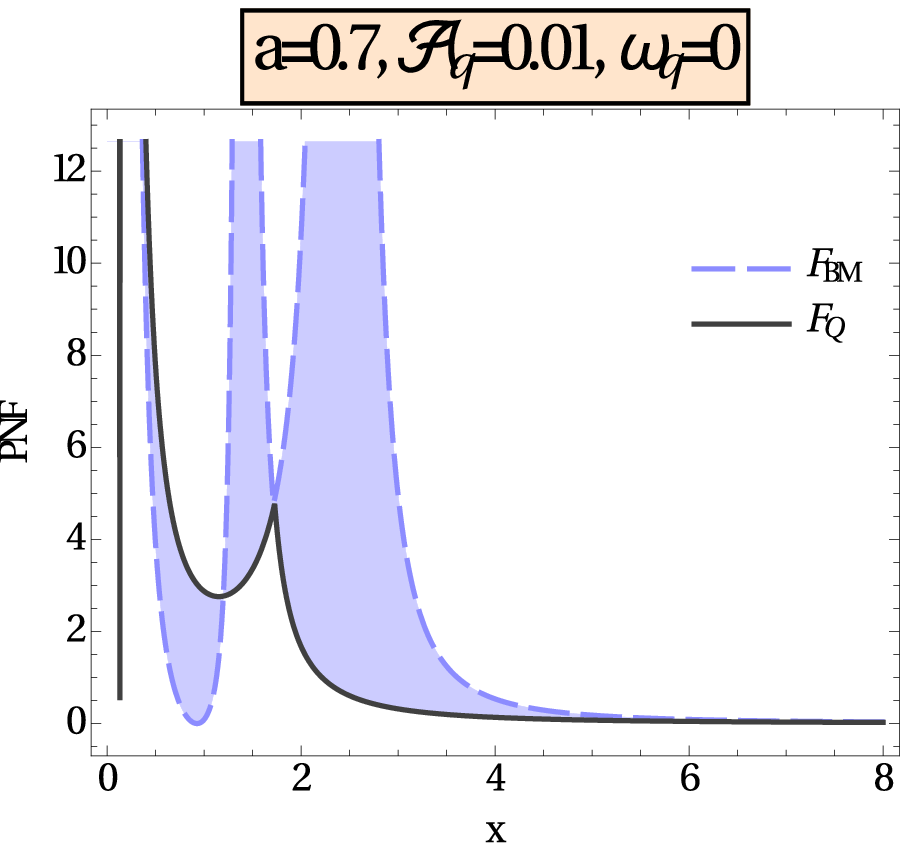}~~
\hspace*{0.5 cm}
\includegraphics*[scale=0.3]{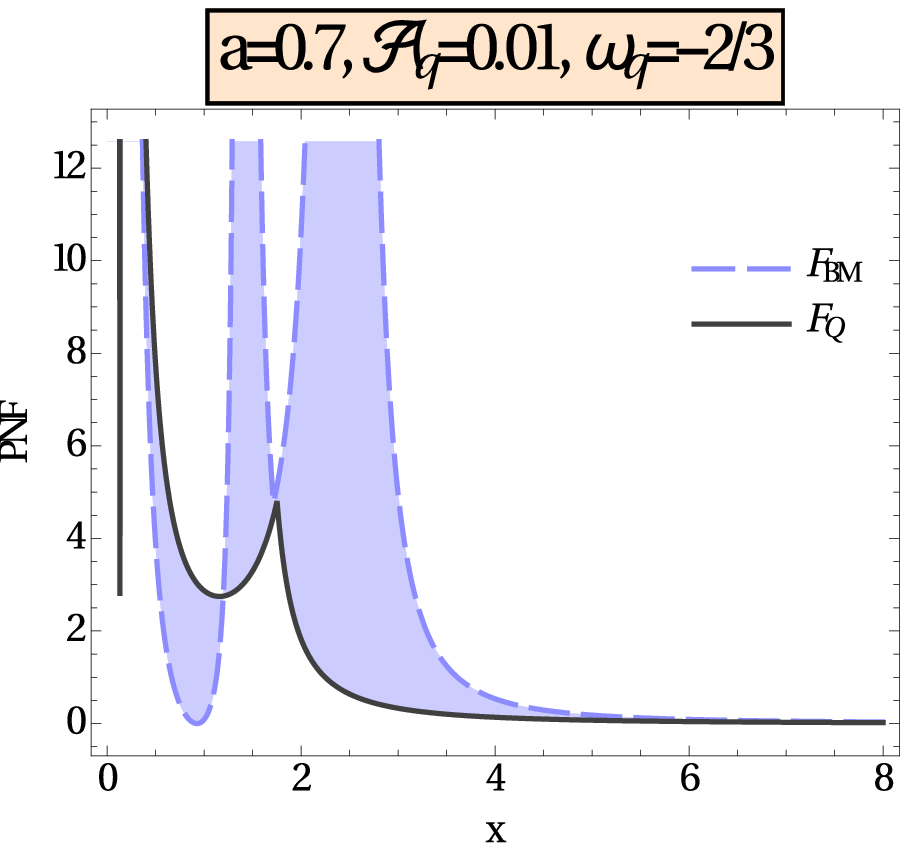}~~
\hspace*{0.5 cm}
\includegraphics*[scale=0.3]{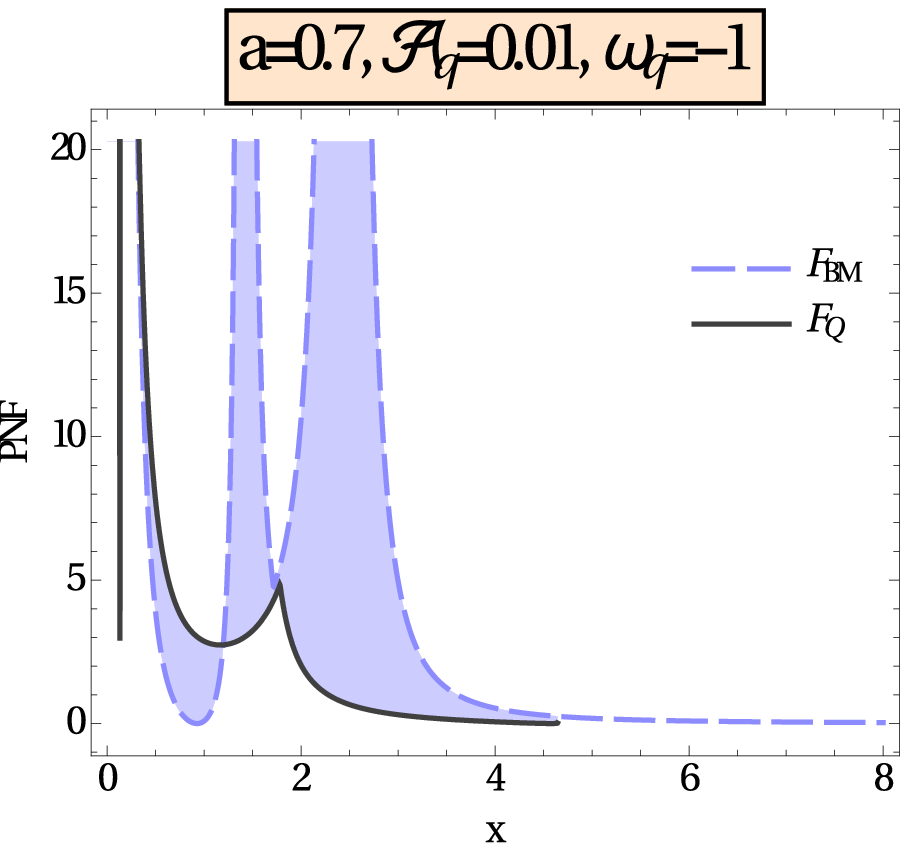}~~\\

Fig $1.24.a$ \hspace{1.6 cm} Fig $1.24.b$ \hspace{1.6 cm} Fig $1.24.c$ \hspace{1.6 cm} Fig $1.24.d$
\includegraphics*[scale=0.3]{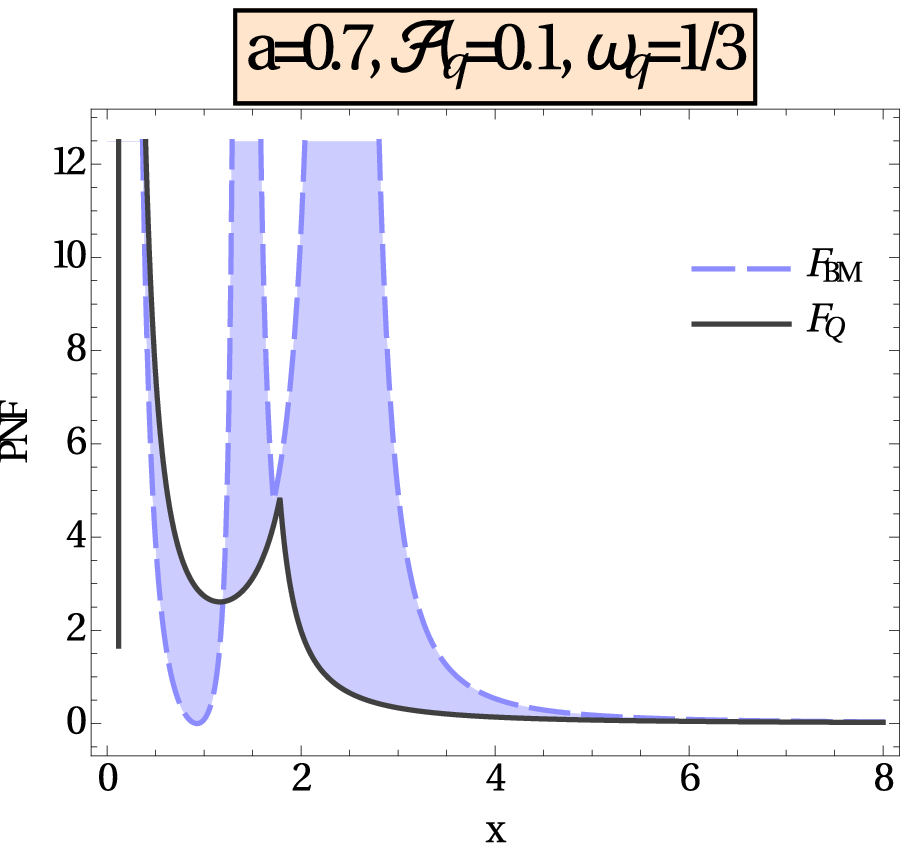}~~
\hspace*{0.5 cm}
\includegraphics*[scale=0.3]{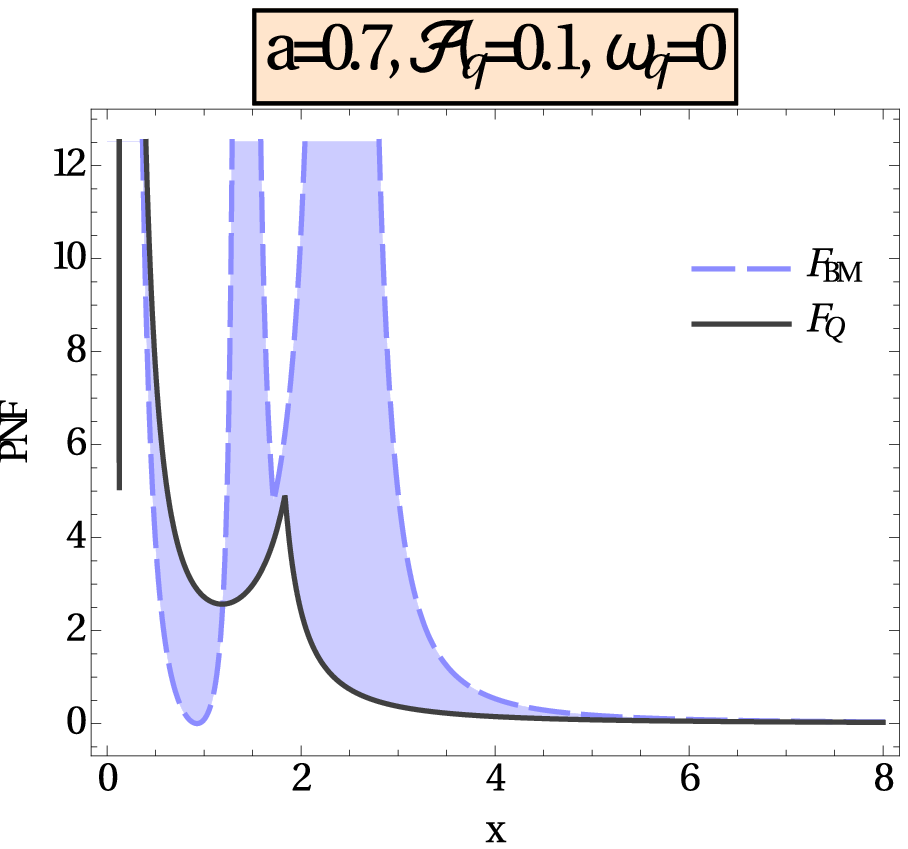}~~
\hspace*{0.5 cm}
\includegraphics*[scale=0.3]{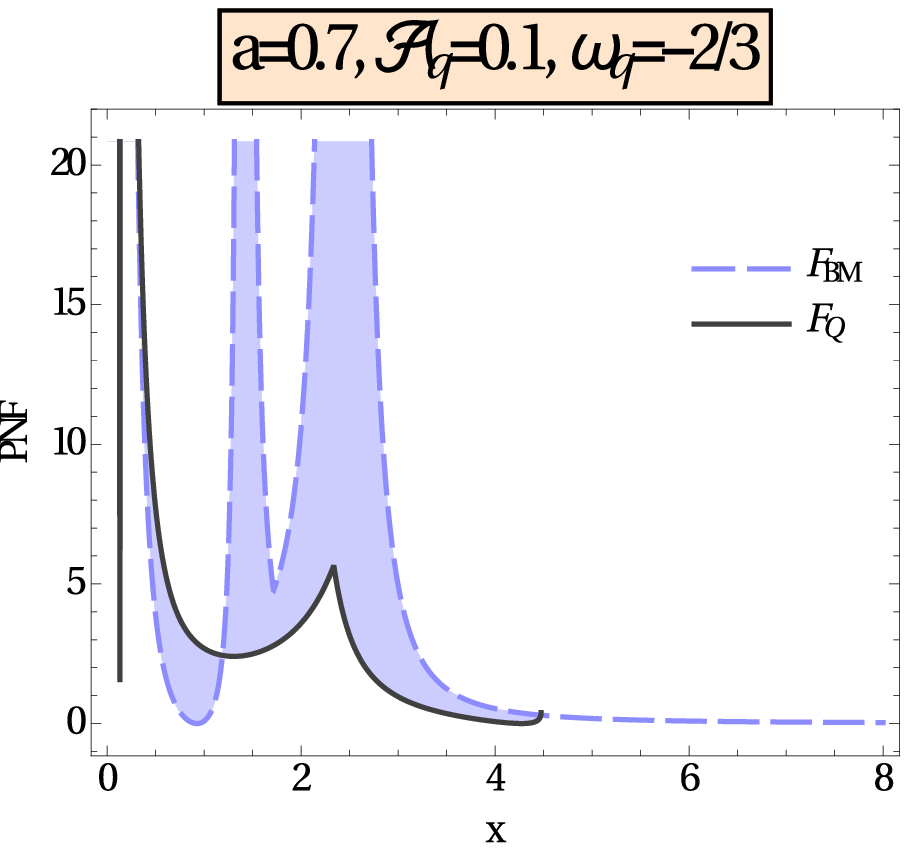}~~
\hspace*{0.5 cm}
\includegraphics*[scale=0.3]{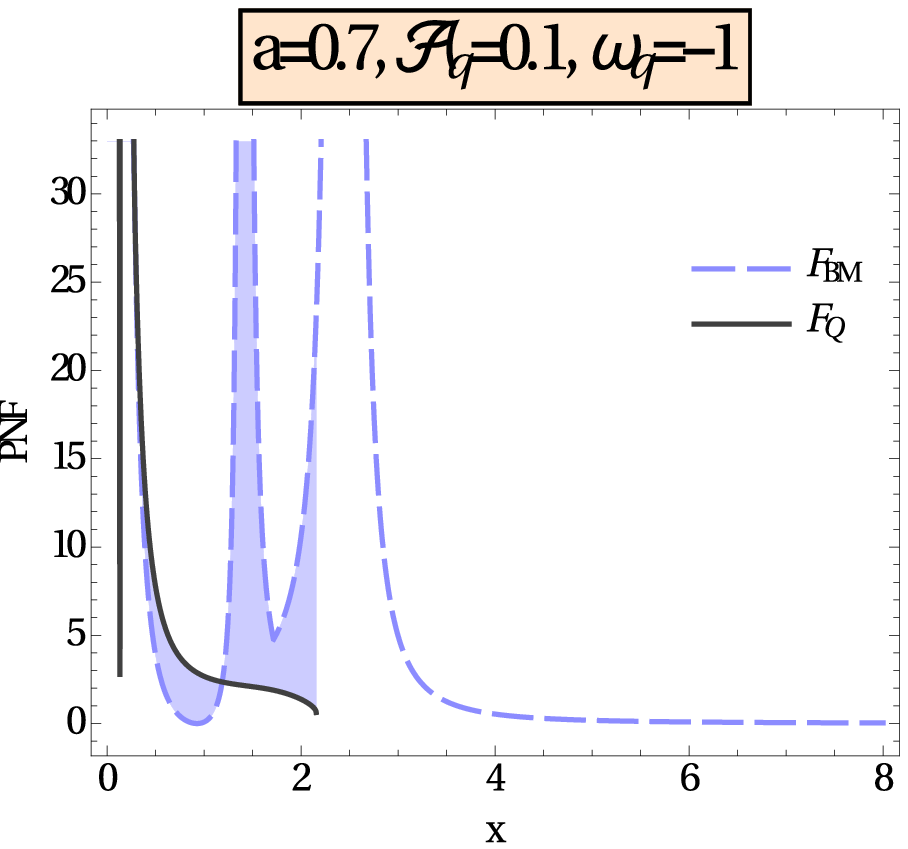}~~\\

\it{Here $a=0.7$, for each values of ${\cal A}_q$, i.e., $10^{-3}$, $10^{-2}$ and $10^{-1}$ we have drawn a relative results for $\omega_q=\frac{1}{3}$ for radiation, $\omega_q=0$ for dust, $\omega_q=-\frac{2}{3}$ for quintessence and $\omega_q=-1$ for phantom barrier.}

\end{figure}

Next, we have increased our co rotational value little bit higher than the previous one, i.e., taking $a=0.3$ and then following the same pattern of values of ${\cal A}_q$ and $\omega_q$, we obtained figures relevant to the previous set of graphs. Here our shatter region is low in breadth in comparison to that of the Mukhopadhyay's. Unlike previous one Mukhopadhyay's PNF creates an envelope in the shattering region. Before that, i.e., for low $x$, our force remains dominating and after that, i.e., for high $x$ other one remains over-top. For last two cases when $a=0.3,~{\cal A}_q=0.1,~\omega_q=-2/3$, our PNF ended up too early, i.e., for $x>4.5$ it stops generating any value. For $a=0.3,~{\cal A}_q=0.1,~\omega_q=-1$, our force ends up near $x=2$ maintaining the gradually decreasing nature.

DE in a lower-mediocre co-rotational BH plays nearly same behavior with that of a relatively slow co-rotating BH. Near the blowup region, the width of the region is relatively small comparing to that of the other one. This fact suggests that for a very small range of radius the force is very high when we assume the DE effect in a co-rotating BH. Like previous results near $x=1$ we have a special value of radius where the attracting force is independent of DE, i.e, at that vary radius attraction force is independent of DE effects.

For figures $1.22.a$ to $1.24.d$, we have plotted for a BH with angular momentum $a=0.7$, i.e., mediocre co-rotation. Here Mukhopadhyay's PNF blows up twice and between two blow up regions, there is a double point whereas our force attains a double point with the finite amount of attractive force everywhere. Except for the last two cases, the position of both the double point is the same but the direction of the double point obtained by our PNF is directed vertically upward. On the other hand, that of Mukhopadhyay's PNF points toward vertically downwards. For the first half, our force remains dominating and for last half other one takes charge, but like previous results, both the forces are same when we consider a small or massive BH. Finally for last two cases where $a=0.7,~{\cal A}_q=0.1,~\omega_q=-2/3$, our force vanishes, i.e., for $x>4.5$, it stops giving values but maintaining its previous nature. For $a=0.7,~{\cal A}_q=0.1,~\omega_q=-1$, it is completely different, i.e., it decreases monotonically and then stops near $x=2$. 

For a mediocre co-rotating BH the DE effect is very low in comparison to a simply co-rotating BH without any DE effect. Even for a DE, it does not even have a point of singularity but it also attains a double point. On the other hand, a simply rotational BH generates more attraction force even it blows up twice which indicates that DE actually turns a BH weak.

\begin{figure}
\centering
Fig $1.25.a$ \hspace{1.6 cm} Fig $1.25.b$ \hspace{1.6 cm} Fig $1.25.c$ \hspace{1.6 cm} Fig $1.25.d$
\includegraphics*[scale=0.3]{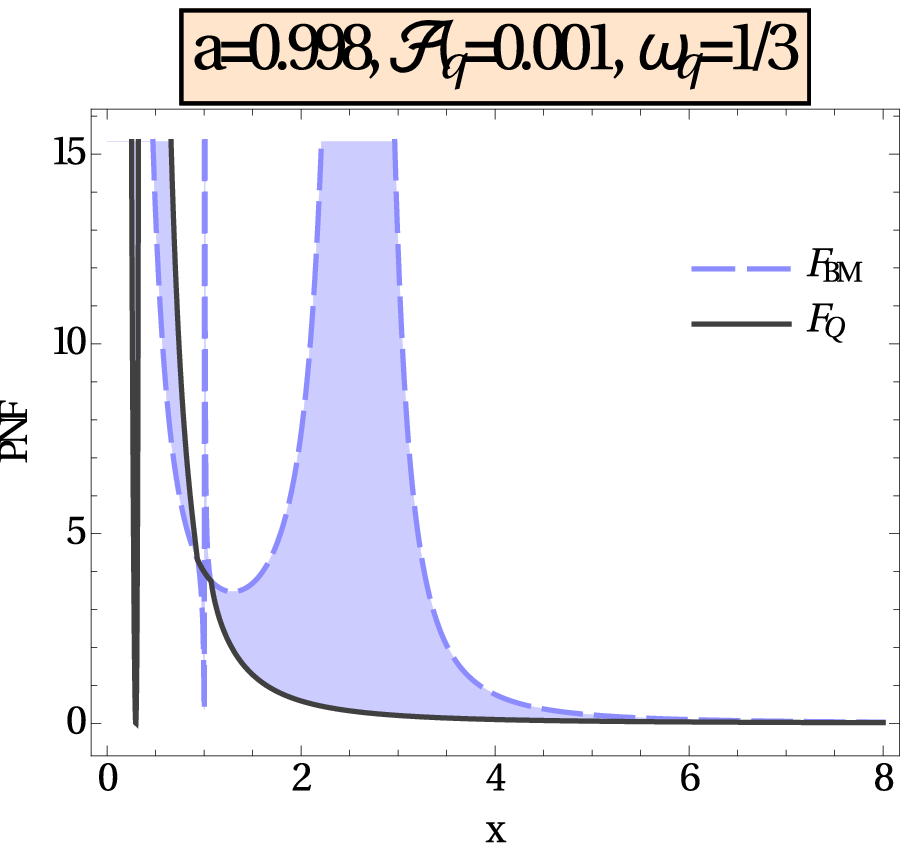}~~
\hspace*{0.5 cm}
\includegraphics*[scale=0.3]{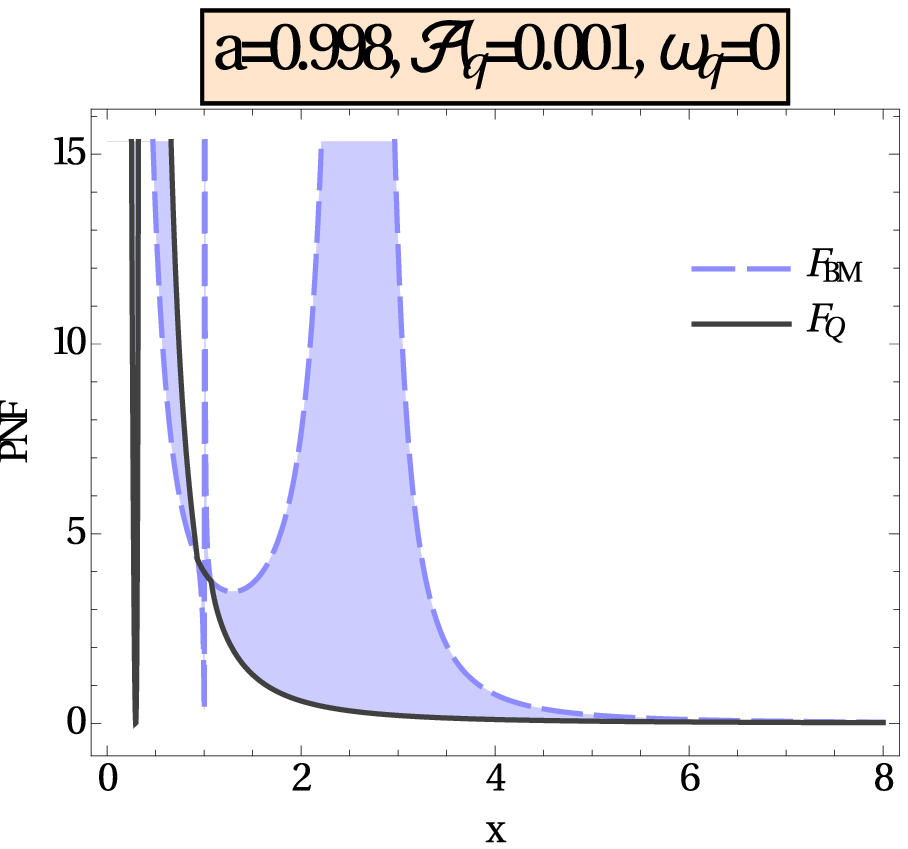}~~
\hspace*{0.5 cm}
\includegraphics*[scale=0.3]{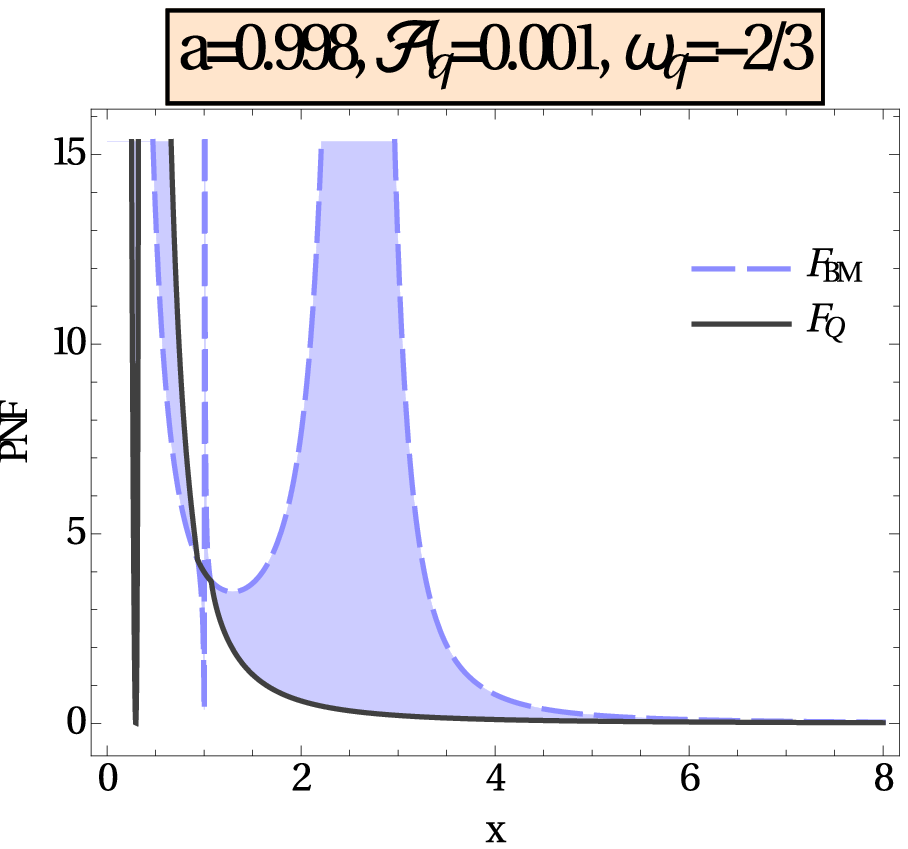}~~
\hspace*{0.5 cm}
\includegraphics*[scale=0.3]{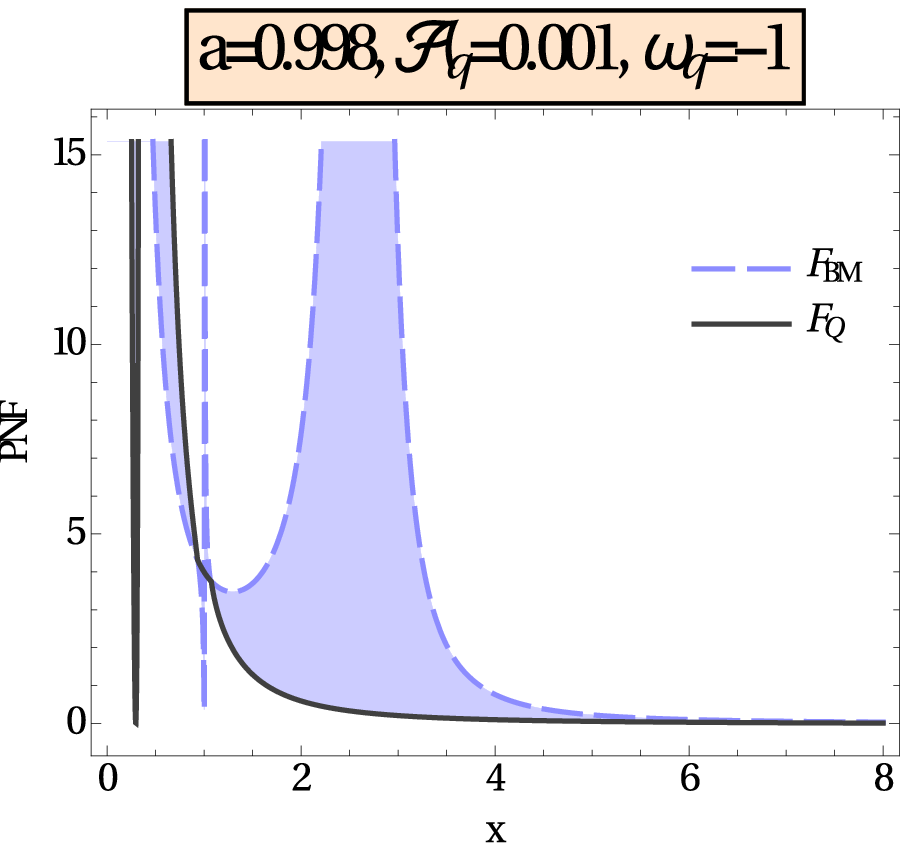}~~\\

Fig $1.26.a$ \hspace{1.6 cm} Fig $1.26.b$ \hspace{1.6 cm} Fig $1.26.c$ \hspace{1.6 cm} Fig $1.26.d$
\includegraphics*[scale=0.3]{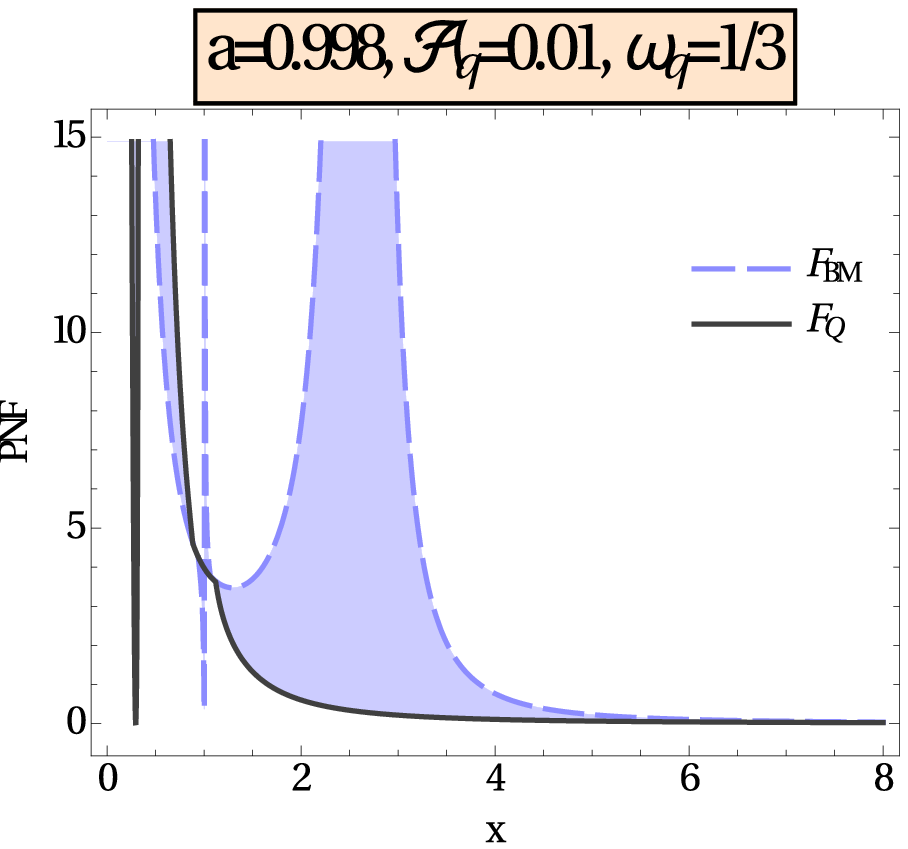}~~
\hspace*{0.5 cm}
\includegraphics*[scale=0.3]{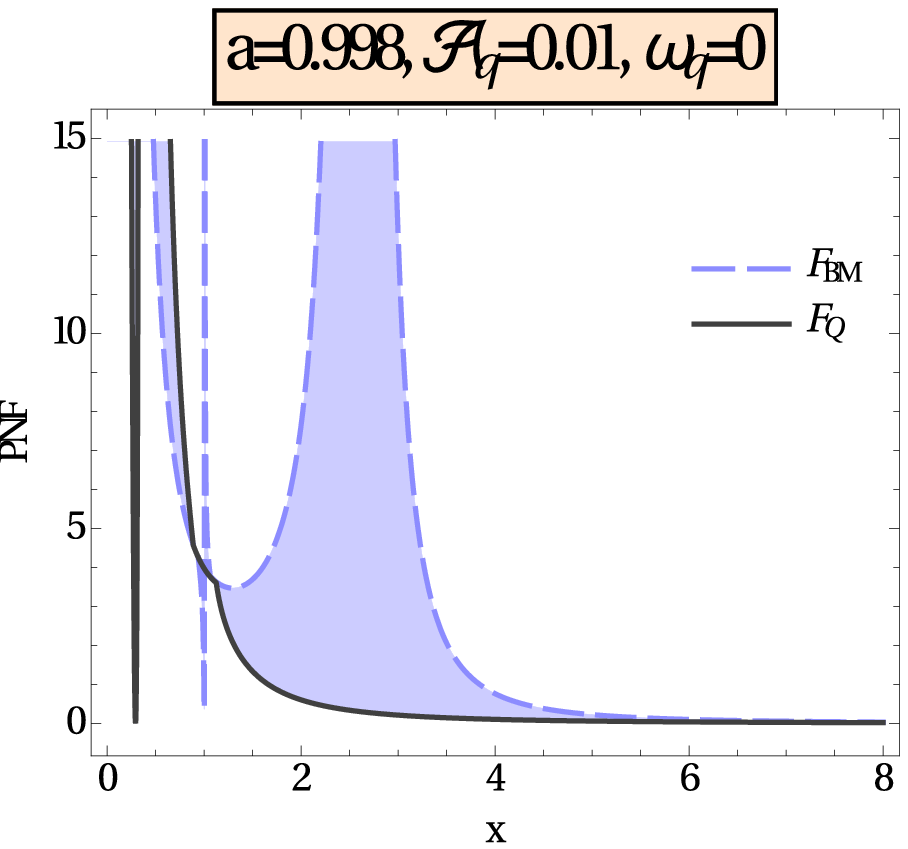}~~
\hspace*{0.5 cm}
\includegraphics*[scale=0.3]{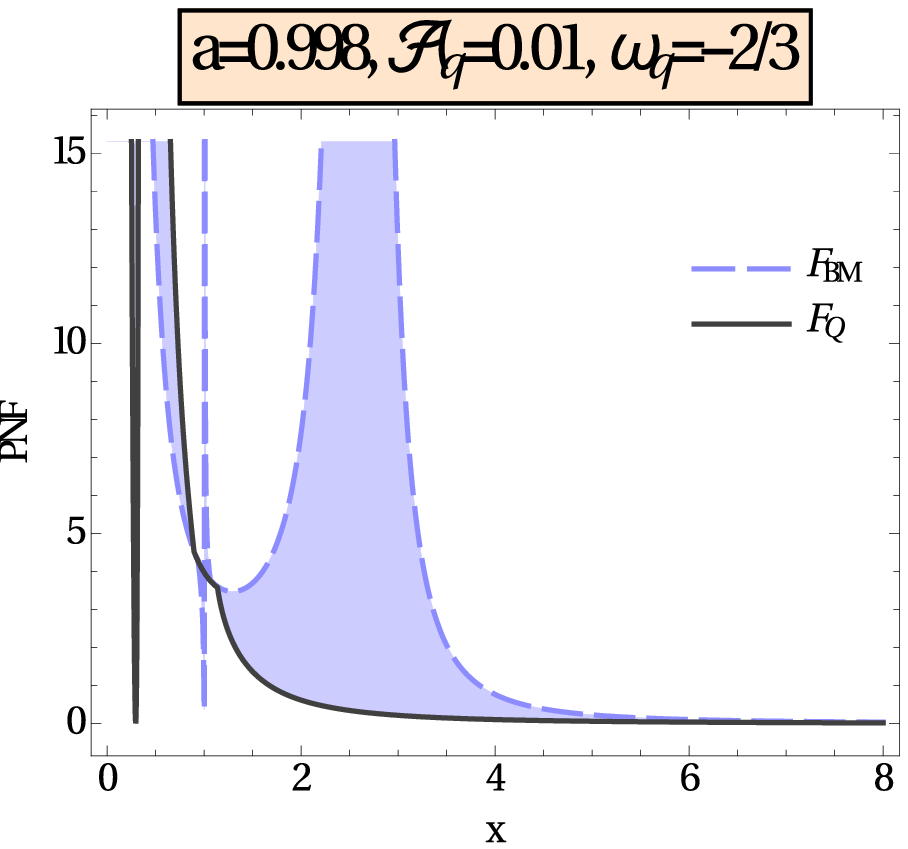}~~
\hspace*{0.5 cm}
\includegraphics*[scale=0.3]{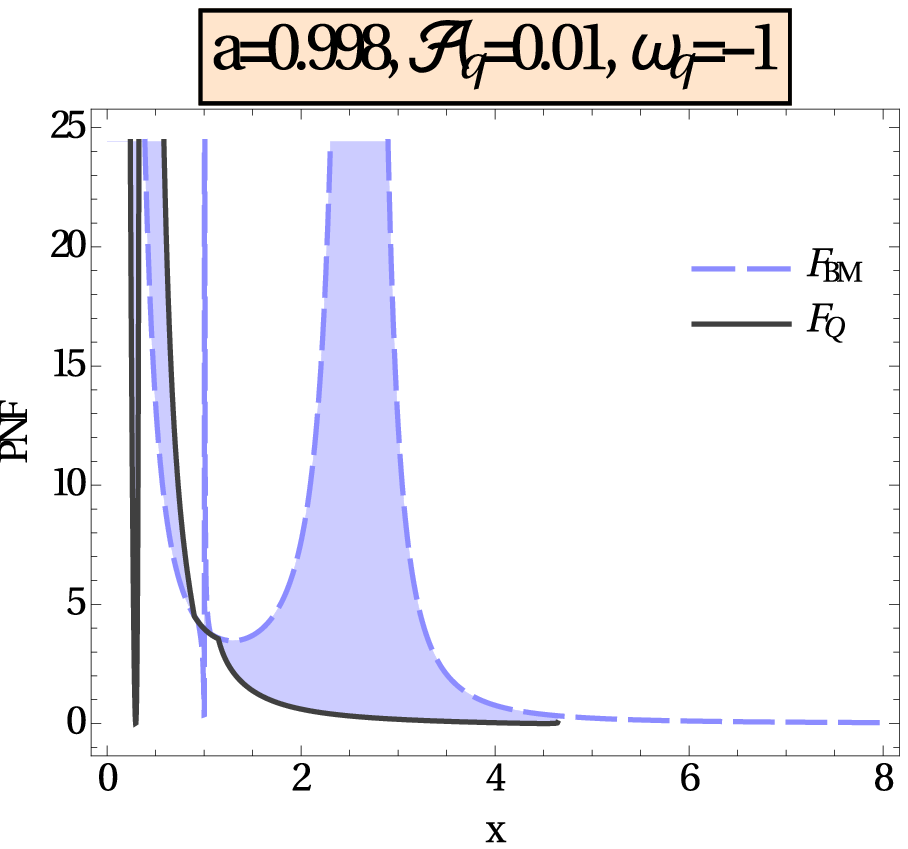}~~\\

Fig $1.27.a$ \hspace{1.6 cm} Fig $1.27.b$ \hspace{1.6 cm} Fig $1.27.c$ \hspace{1.6 cm} Fig $1.27.d$
\includegraphics*[scale=0.3]{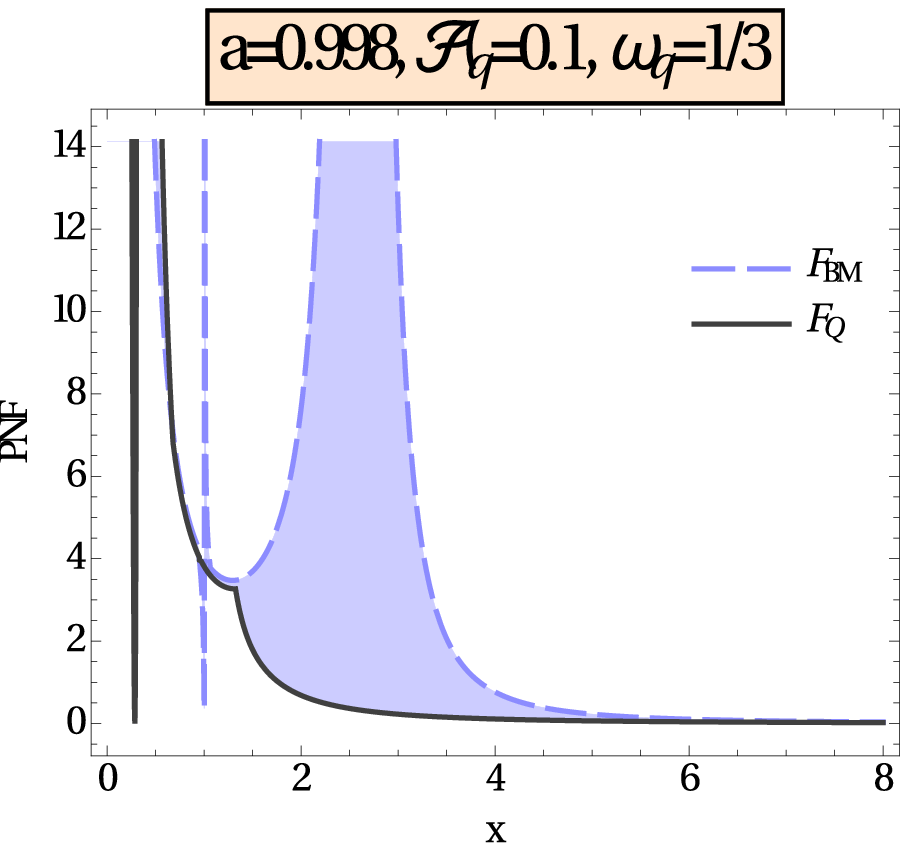}~~
\hspace*{0.5 cm}
\includegraphics*[scale=0.3]{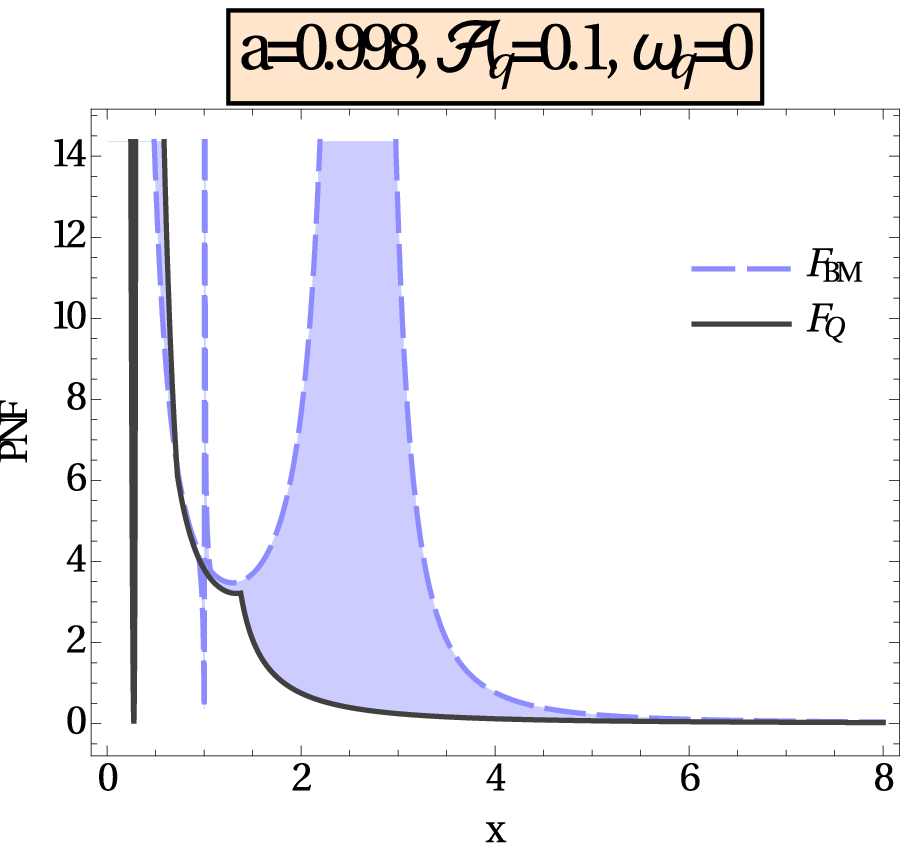}~~
\hspace*{0.5 cm}
\includegraphics*[scale=0.3]{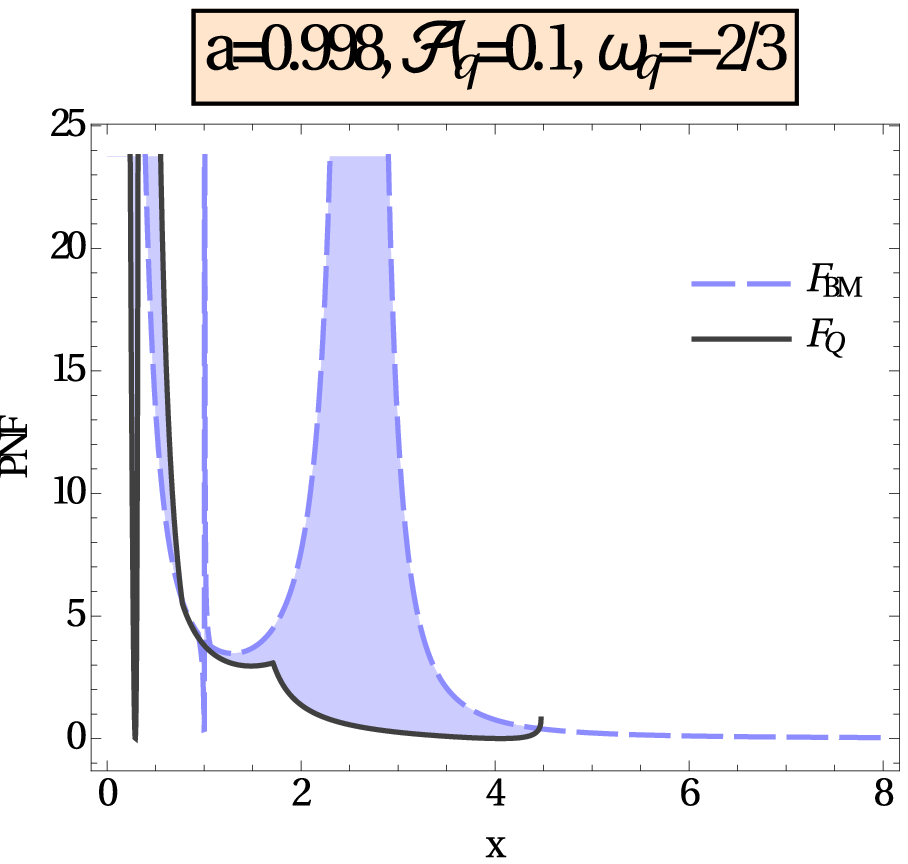}~~
\hspace*{0.5 cm}
\includegraphics*[scale=0.3]{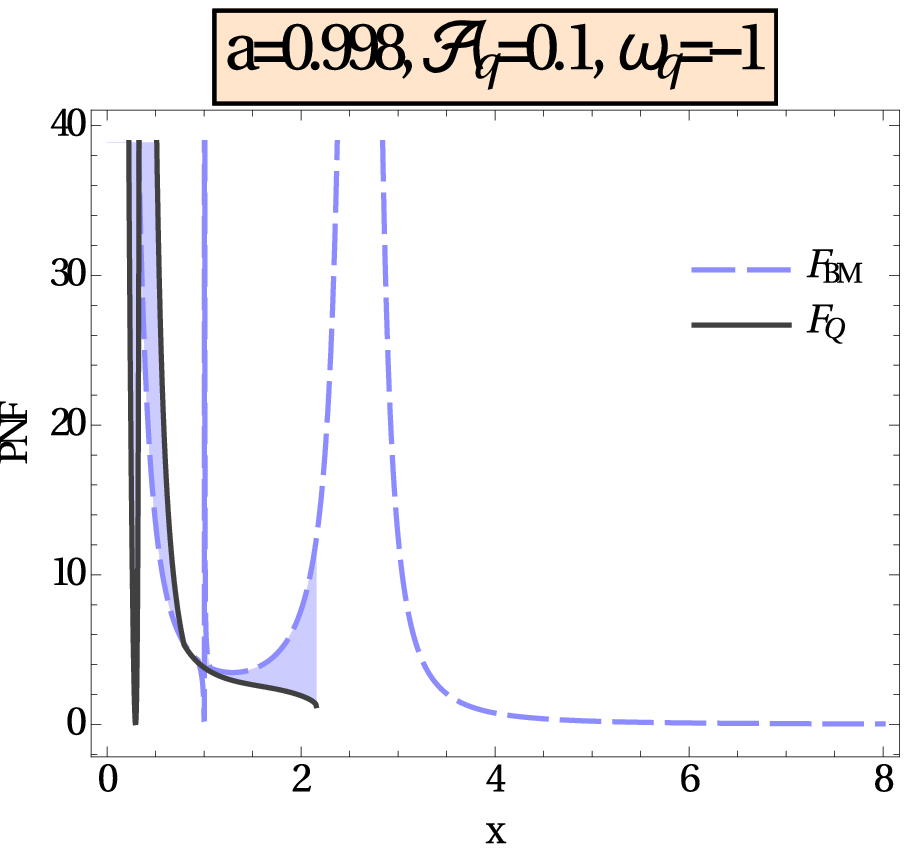}~~\\

\it{Here $a=0.998$, for each values of ${\cal A}_q$, i.e., $10^{-3}$, $10^{-2}$ and $10^{-1}$ we have drawn a relative results for $\omega_q=\frac{1}{3}$ for radiation, $\omega_q=0$ for dust, $\omega_q=-\frac{2}{3}$ for quintessence and $\omega_q=-1$ for phantom barrier.}

\end{figure}

In this final discussion we have considered our angular momentum of co rotation to be very high, i.e., we have taken $a=0.998$ and then following same pattern of values of ${\cal A}_q$ and $\omega_q$, we have obtained figures relevant to previous set of graphs. Here our PNF shatters once in between the region $(0,1)$ but comparing to of that of Mukhopadhyay's, his PNF blows up twice. Though Mukhopadhyay's PNF shatters twice, the first blown up region is very thin and the last shattering region is quite far from that of ours. After that, i.e., for high $x$ Mukhopadhyay's PNF remains over-top but both the forces converge when we increases $x$ further. For last two cases when we considered quintessence as DE agent, i.e., $a=0.998,~{\cal A}_q=0.1,~\omega_q=-2/3$, our PNF stops generating any value after $x>4.5$. For phantom barrier, i.e., $a=0.3,~{\cal A}_q=0.1,~\omega_q=-1$, our force ends up near $x=2$ maintaining the gradually decreasing nature.

Throughout this procedure graphs for co-rotating BH represent a reverse character than that for a counter-rotating one. DE accretion replaces its nature when we take account of the direction of rotation. For counter rotational BH effect of quintessence is clearly remarkable, it makes a BH more hungry, on the other hand for co-rotation it gradually decreases intake rate. When rotation is nullified DE shows up with a repulsive nature.

We have constructed a tabular representation  of our potential for BH embedded in quintessence and a rotating regular one. Here first table represents a comparison between the PNP $V_4(=\int F(x) dx)$ obtained by B Mukhopadhyay and the potential derived by us

\textbf{Table 1:} Comparison of two PNPs, i.e., one derived previously by Mukhopadhyay and another derived by us, without any DE effect at all
\begin{center}
\begin{tabular}{llll}  
\toprule
\multicolumn{2}{c}{For ${\cal A}_q=0$} \\
\cmidrule(r){1-2}
a    & Kerr & $V_{4~BM}$(error \%) & $V_{4~Q}$(difference \%) \\
\midrule
\multicolumn{2}{c}{For co-rotation} \\
\cmidrule(r){1-2}
0		&	4		& 4(0)							& 4(0)							\\
0.1 	&	3.797	& 3.788(-0.237029233605475)		& 2.92719(-22.907821964709)		\\
0.3		&	3.373	& 3.347(-0.770827156833686)		& 1.81651(-46.1455677438482)	\\
0.5		&	2.914	& 2.87(-1.50995195607413)		& 1.27254(-56.3301304049417)	\\
0.7		&	2.395	& 2.333(-2.58872651356993)		& 0.96533(-59.6939457202505)	\\
0.998	&	1.091	& 1.037(-4.94958753437214)		& 0.70917(-34.9981668194317)	\\
\midrule
\multicolumn{2}{c}{For counter-rotation} \\
\cmidrule(r){1-2}
-0.1	&	4.198	& 4.206(0.190566936636494)		& 5.95769(41.9173415912339)		\\
-0.3	&	4.58	& 4.606(0.567685589519646)		& 27.3216(496.541484716157)		\\
-0.5	&	4.949	& 4.993(0.889068498686613)		& 11828.1(238899.797938978)		\\
-0.7	&	5.308	& 5.368(1.1303692539563)		& 3130.35(58874.1899020347)		\\
-0.998	&	5.911	& 5.991(1.35340889866351)		& 5817.07(98310.9287768567)		\\

\bottomrule
\end{tabular}

\end{center}

In Table 1, we have (enlisted) the values of the forces calculated from the equation (\ref{PNP_quint}). We compared the same with the PNF value calculated by Mukhopadhyay\cite{Mukhopadhyay B PNP 2002}. The force decreases as we increase the co-rotation parameter and increases as we increase the magnitude of Counter-rotation parameter. In this table, we have considered the case where ${\cal A}_q=0$, i.e., we have not taken any dark energy into account.
 
Since our potential is too flexible to adjust the intensity of DE, we have taken ${\cal A}_q=0.001$, i.e., low amount of DE intensity and for different type of DE agents, we obtained the data set of PNP, which has been compared with that of Mukhopadhyay, in the following table 2.1-2.4

\textbf{Table 2.1:} Comparison of two PNPs for very low DE effect in radiation era
\begin{center}
\begin{tabular}{llll}  
\toprule
\multicolumn{2}{c}{For ${\cal A}_q=0.001,~\omega_q=\frac{1}{3}$} \\
\cmidrule(r){1-2}
a    & Kerr & $V_{4~BM}$(error \%) & $V_{4~Q}$(difference \%) \\
\midrule
\multicolumn{2}{c}{For co-rotation} \\
\cmidrule(r){1-2}
0		&	4		& 4(0)							& 4.00805(0.201249999999997)	\\
0.1 	&	3.797	& 3.788(-0.237029233605475)		&2.9325(-22.7679747168817)		\\
0.3		&	3.373	& 3.347(-0.770827156833686)		& 1.81845(-46.0880521790691)	\\
0.5		&	2.914	& 2.87(-1.50995195607413)		& 1.27355(-56.2954701441318)	\\
0.7		&	2.395	& 2.333(-2.58872651356993)		& 0.965932(-59.6688100208768)	\\
0.998	&	1.091	& 1.037(-4.94958753437214)		& 0.709505(-34.9674610449129)	\\
\midrule
\multicolumn{2}{c}{For counter-rotation} \\
\cmidrule(r){1-2}
-0.1	&	4.198	& 4.206(0.190566936636494)		& 5.95769(41.9173415912339)		\\
-0.3	&	4.58	& 4.606(0.567685589519646)		& 27.3216(496.541484716157)		\\
-0.5	&	4.949	& 4.993(0.889068498686613)		& 11828.1(238899.797938978)		\\
-0.7	&	5.308	& 5.368(1.1303692539563)		& 3130.35(58874.1899020347)		\\
-0.998	&	5.911	& 5.991(1.35340889866351)		& 5817.07(98310.9287768567)		\\

\bottomrule
\end{tabular}

\end{center}

In table 2.1 we have listed the values of the potential for ${\cal A}_q=0.001$ and $\omega_q=\frac{1}{3}$, i.e., for radiation. For co-rotation, as we increase rotation, we see the percentage of difference to decrease. But as we take counter rotation, the percentage of difference increases and modulus of this increase is very high. At $a=-0.5$ we have also noticed a abruptly huge positive value.

\textbf{Table 2.2:} Comparison of two PNPs for very low DE effect in pressure-less dust
\begin{center}
\begin{tabular}{llll}  
\toprule
\multicolumn{2}{c}{For ${\cal A}_q=0.001,~\omega_q=0$} \\
\cmidrule(r){1-2}
a    & Kerr & $V_{4~BM}$(error \%) & $V_{4~Q}$(difference \%) \\
\midrule
\multicolumn{2}{c}{For co-rotation} \\
\cmidrule(r){1-2}
0		&	4		& 4(0)							& 4.01815(0.453750000000008)	\\
0.1 	&	3.797	& 3.788(-0.237029233605475)		& 2.93836(-22.6136423492231)	\\
0.3		&	3.373	& 3.347(-0.770827156833686)		& 1.821(-46.0124518233027)		\\
0.5		&	2.914	& 2.87(-1.50995195607413)		& 1.27491(-56.2487989018531)	\\
0.7		&	2.395	& 2.333(-2.58872651356993)		& 0.966757(-59.634363256785)	\\
0.998	&	1.091	& 1.037(-4.94958753437214)		& 0.709963(-34.9254812098992)	\\
\midrule
\multicolumn{2}{c}{For counter-rotation} \\
\cmidrule(r){1-2}
-0.1	&	4.198	& 4.206(0.190566936636494)		& 5.97798(42.4006669842782)		\\
-0.3	&	4.58	& 4.606(0.567685589519646)		& 27.6538(503.794759825327)		\\
-0.5	&	4.949	& 4.993(0.889068498686613)		& 11076.3(223708.850272782)		\\
-0.7	&	5.308	& 5.368(1.1303692539563)		& 4563.43(85872.6827430294)		\\
-0.998	&	5.911	& 5.991(1.35340889866351)		& 11922.5(201600.219928946)		\\

\bottomrule
\end{tabular}

\end{center}
If the DE in which the BH is embedded is pressure-less dust, we have seen in table 2.2, the same pattern of the percentage of difference, i.e., for co-rotating BH attracting field lower its power, but for counter-rotating BH it increases its attracting nature when we consider this BH embedded in pressure-less dust. We have also received a highly jump in PNP at $a=-0.5$.

\textbf{Table 2.3:} Comparison of two PNPs for very low DE effect of a BH embedded in quintessence
\begin{center}
\begin{tabular}{llll}  
\toprule
\multicolumn{2}{c}{For ${\cal A}_q=0.001,~\omega_q=-\frac{2}{3}$} \\
\cmidrule(r){1-2}
a    & Kerr & $V_{4~BM}$(error \%) & $V_{4~Q}$(difference \%) \\
\midrule
\multicolumn{2}{c}{For co-rotation} \\
\cmidrule(r){1-2}
0		&	4		& 4(0)							& 4.03543(0.885749999999996)	\\
0.1 	&	3.797	& 3.788(-0.237029233605475)		& 2.92536(-22.9560179088754)	\\
0.3		&	3.373	& 3.347(-0.770827156833686)		& 1.78412(-47.1058404980729)	\\
0.5		&	2.914	& 2.87(-1.50995195607413)		& 1.22931(-57.8136582017845)	\\
0.7		&	2.395	& 2.333(-2.58872651356993)		& 0.917188(-61.7040501043841)	\\
0.998	&	1.091	& 1.037(-4.94958753437214)		& 0.657657(-39.7197983501375)	\\
\midrule
\multicolumn{2}{c}{For counter-rotation} \\
\cmidrule(r){1-2}
-0.1	&	4.198	& 4.206(0.190566936636494)		& 6.06751(44.5333492139114)		\\
-0.3	&	4.58	& 4.606(0.567685589519646)		& 30.036(555.807860262009)		\\
-0.5	&	4.949	& 4.993(0.889068498686613)		& 9982.34(201604.182663164)		\\
-0.7	&	5.308	& 5.368(1.1303692539563)		& 5071.26(95439.9397136398)		\\
-0.998	&	5.911	& 5.991(1.35340889866351)		& 3963.56(66953.9671798342)		\\

\bottomrule
\end{tabular}

\end{center}
This table 2.3 contains data for $\omega_q=-\frac{2}{3}$, i.e., BH embedded in quintessence. For co-rotation the percentage of difference is decreasing and our potential is nearly same with that of the standard Kerr metric. Although for counter-rotation this value increases with a high intensity for $a=-0.5$ and hereafter it increases normally but with a significant amount.

\textbf{Table 2.4:} Comparison of two PNPs for very low DE effect in phantom barrier
\begin{center}
\begin{tabular}{llll}  
\toprule
\multicolumn{2}{c}{For ${\cal A}_q=0.001,~\omega_q=-1$} \\
\cmidrule(r){1-2}
a    & Kerr & $V_{4~BM}$(error \%) & $V_{4~Q}$(difference \%) \\
\midrule
\multicolumn{2}{c}{For co-rotation} \\
\cmidrule(r){1-2}
0		&	4		& 4(0)							& -21375.5(-534487.5)		\\
0.1 	&	3.797	& 3.788(-0.237029233605475)		& 3.60605(-5.02897023966289)	\\
0.3		&	3.373	& 3.347(-0.770827156833686)		& 2.43791(-27.7227986955233)	\\
0.5		&	2.914	& 2.87(-1.50995195607413)		& 1.87887(-35.5226492793411)	\\
0.7		&	2.395	& 2.333(-2.58872651356993)		& 1.56151(-34.8012526096033)	\\
0.998	&	1.091	& 1.037(-4.94958753437214)		& 1.29607(18.7965169569203)	\\
\midrule
\multicolumn{2}{c}{For counter-rotation} \\
\cmidrule(r){1-2}
-0.1	&	4.198	& 4.206(0.190566936636494)		& 6.21526(48.0528823249166)		\\
-0.3	&	4.58	& 4.606(0.567685589519646)		& 34.9812(663.781659388646)		\\
-0.5	&	4.949	& 4.993(0.889068498686613)		& 6417.67(129576.096181047)		\\
-0.7	&	5.308	& 5.368(1.1303692539563)		& 13037.8(245525.470987189)		\\
-0.998	&	5.911	& 5.991(1.35340889866351)		& 13342.8(225628.303163593)		\\

\bottomrule
\end{tabular}
\end{center}

In this table 2.4 we have taken the DE agent as the phantom barrier, i.e., $\omega_q=-1$. Here we have noticed the conventional behavior, i.e., for co-rotation we got its decreasing nature and difference is also moderate. On the other hand, for counter-rotation this value increases without any abrupt changes, though the difference with a regular Kerr metric is significant.

For the next cluster of tables, i.e., form table 3.1 to table 3.4, we have taken ${\cal A}_q=0.001$, i.e., we have increased its value $10$ times more intense than previous observations.

\textbf{Table 3.1:} Comparison of two PNPs for moderate DE effect in radiation era
\begin{center}
\begin{tabular}{llll}  
\toprule
\multicolumn{2}{c}{For ${\cal A}_q=0.01,~\omega_q=\frac{1}{3}$} \\
\cmidrule(r){1-2}
a    & Kerr & $V_{4~BM}$(error \%) & $V_{4~Q}$(difference \%) \\
\midrule
\multicolumn{2}{c}{For co-rotation} \\
\cmidrule(r){1-2}
0		&	4		& 4(0)							& 4.08187(2.04675000000001)		\\
0.1 	&	3.797	& 3.788(-0.237029233605475)		& 2.97434(-21.6660521464314)	\\
0.3		&	3.373	& 3.347(-0.770827156833686)		& 1.83603(-45.5668544322562)	\\
0.5		&	2.914	& 2.87(-1.50995195607413)		& 1.2827(-55.9814687714482)		\\
0.7		&	2.395	& 2.333(-2.58872651356993)		& 0.971401(-59.4404592901879)	\\
0.998	&	1.091	& 1.037(-4.94958753437214)		& 0.712542(-34.6890925756187)	\\
\midrule
\multicolumn{2}{c}{For counter-rotation} \\
\cmidrule(r){1-2}
-0.1	&	4.198	& 4.206(0.190566936636494)		& 6.11024(45.551214864221)		\\
-0.3	&	4.58	& 4.606(0.567685589519646)		& 30.1366(558.004366812227)		\\
-0.5	&	4.949	& 4.993(0.889068498686613)		& 10010.4(202171.165892099)		\\
-0.7	&	5.308	& 5.368(1.1303692539563)		& 6986.94(131530.369253956)		\\
-0.998	&	5.911	& 5.991(1.35340889866351)		& 3972.07(67097.9360514295)		\\

\bottomrule
\end{tabular}
\end{center}

In this first table, i.e, in table 3.1, we have considered radiation era as usual, i.e., $\omega_q=\frac{1}{3}$. we have observed the monotonic nature of our potential along with that of Mukhopadhyay for co-rotational values. For counter-rotation these values are the exactly opposite, i.e., it strictly increases monotonically.

\textbf{Table 3.2:} Comparison of two PNPs for moderate DE effect in pressure-less dust
\begin{center}
\begin{tabular}{llll}  
\toprule
\multicolumn{2}{c}{For ${\cal A}_q=0.01,~\omega_q=0$} \\
\cmidrule(r){1-2}
a    & Kerr & $V_{4~BM}$(error \%) & $V_{4~Q}$(difference \%) \\
\midrule
\multicolumn{2}{c}{For co-rotation} \\
\cmidrule(r){1-2}
0		&	4		& 4(0)							& 4.18837(4.70925)				\\
0.1 	&	3.797	& 3.788(-0.237029233605475)		& 3.03537(-20.0587305767711)	\\
0.3		&	3.373	& 3.347(-0.770827156833686)		& 1.86223(-44.7900978357545)	\\
0.5		&	2.914	& 2.87(-1.50995195607413)		& 1.29657(-55.5054907343857)	\\
0.7		&	2.395	& 2.333(-2.58872651356993)		& 0.979763(-59.0913152400835)	\\
0.998	&	1.091	& 1.037(-4.94958753437214)		& 0.717171(-34.2648029330889)	\\
\midrule
\multicolumn{2}{c}{For counter-rotation} \\
\cmidrule(r){1-2}
-0.1	&	4.198	& 4.206(0.190566936636494)		& 6.32882(50.7579799904716)		\\
-0.3	&	4.58	& 4.606(0.567685589519646)		& 34.6793(657.189956331878)		\\
-0.5	&	4.949	& 4.993(0.889068498686613)		& 7226.93(145928.086482118)		\\
-0.7	&	5.308	& 5.368(1.1303692539563)		& 7523.04(141630.218538056)		\\
-0.998	&	5.911	& 5.991(1.35340889866351)		& 4965.37(83902.1992894603)		\\

\bottomrule
\end{tabular}
\end{center}

In table 3.2, things are pretty much same as the previous one for co-rotation, although we have considered the pressureless dust. But for counter rotation it increases slowly retaining its previous behavior unless for last case where this rotation is very high, i.e., $a=-0.998$, our potential drops its rate of increase.

\textbf{Table 3.3:} Comparison of two PNPs for moderate DE effect of a BH embedded in quintessence
\begin{center}
\begin{tabular}{llll}  
\toprule
\multicolumn{2}{c}{For ${\cal A}_q=0.01,~\omega_q=-\frac{2}{3}$} \\
\cmidrule(r){1-2}
a    & Kerr & $V_{4~BM}$(error \%) & $V_{4~Q}$(difference \%) \\
\midrule
\multicolumn{2}{c}{For co-rotation} \\
\cmidrule(r){1-2}
0		&	4		& 4(0)							& -146715(-3667975)			\\
0.1 	&	3.797	& 3.788(-0.237029233605475)		& 29.0542(665.188306557809)	\\
0.3		&	3.373	& 3.347(-0.770827156833686)		& 27.3158(709.836940409131)	\\
0.5		&	2.914	& 2.87(-1.50995195607413)		& 26.1397(797.0384351407)	\\
0.7		&	2.395	& 2.333(-2.58872651356993)		& 25.0532(946.062630480167)	\\
0.998	&	1.091	& 1.037(-4.94958753437214)		& 23.3593(2041.09074243813)	\\
\midrule
\multicolumn{2}{c}{For counter-rotation} \\
\cmidrule(r){1-2}
-0.1	&	4.198	& 4.206(0.190566936636494)		& 33.884(707.146260123868)		\\
-0.3	&	4.58	& 4.606(0.567685589519646)		& 1118.43(24319.8689956332)		\\
-0.5	&	4.949	& 4.993(0.889068498686613)		& 12873.8(260029.319054354)		\\
-0.7	&	5.308	& 5.368(1.1303692539563)		& 12710.5(239359.306706858)		\\
-0.998	&	5.911	& 5.991(1.35340889866351)		& 6515.27(110122.804939942)		\\

\bottomrule
\end{tabular}

\end{center}
Here, in table 3.3 for quintessence, $\omega_q=-\frac{2}{3}$ we have succeeded to derive a different result, i.e., without rotation the value of our PNP is enormously high and negative which represents the repulsive nature. This is again the perfect evidence to indicate the accelerated universe. On the other hand, for rest of the cases we have experienced the similar behavior of this potential with a slightly change, i.e., even for co-rotation the difference in magnitude is relatively higher than other cases.

\begin{center}
\textbf{Table 3.4:} Comparison of two PNPs for moderate DE effect in phantom barrier
\begin{tabular}{llll}  
\toprule
\multicolumn{2}{c}{For ${\cal A}_q=0.01,~\omega_q=-1$} \\
\cmidrule(r){1-2}
a    & Kerr & $V_{4~BM}$(error \%) & $V_{4~Q}$(difference \%) \\
\midrule
\multicolumn{2}{c}{For co-rotation} \\
\cmidrule(r){1-2}
0		&	4		& 4(0)							& -4890.03(-122350.75)			\\
0.1 	&	3.797	& 3.788(-0.237029233605475)		& 5.43543(43.150645246247)		\\
0.3		&	3.373	& 3.347(-0.770827156833686)		& 3.16053(-6.29914023124815)	\\
0.5		&	2.914	& 2.87(-1.50995195607413)		& 2.23484(-23.3067947838023)	\\
0.7		&	2.395	& 2.333(-2.58872651356993)		& 1.76629(-26.2509394572025)	\\
0.998	&	1.091	& 1.037(-4.94958753437214)		& 1.38036(26.5224564619615)		\\
\midrule
\multicolumn{2}{c}{For counter-rotation} \\
\cmidrule(r){1-2}
-0.1	&	4.198	& 4.206(0.190566936636494)		& 16.3139(288.611243449261)		\\
-0.3	&	4.58	& 4.606(0.567685589519646)		& 9845.98(214877.729257642)		\\
-0.5	&	4.949	& 4.993(0.889068498686613)		& 4921.32(99340.6950899172)		\\
-0.7	&	5.308	& 5.368(1.1303692539563)		& 2537.34(47702.1853805577)		\\
-0.998	&	5.911	& 5.991(1.35340889866351)		& 4338.42(73295.7029267468)		\\

\bottomrule
\end{tabular}
\end{center}

In table 3.4, for phantom barrier also when our black hole of consideration is non-rotating the repulsive behavior is present but not so intense like the previous one. Although for counter-rotation especially for $a=-0.3$ this value of our potential jumped up and after that it starts to decrees along with the increment of the magnitude of rotation until this value reaches $0.998$.

Finally, for table 4.1 to 4.4 we have further increase the intensity of dark energy effect parameter ${\cal A}$ $10$ times then made a tabular representation as usual to know the present scenario of our universe better.

\begin{center}
\textbf{Table 4.1:} Comparison of two PNPs for relatively high DE effect in radiation era
\begin{tabular}{llll}  
\toprule
\multicolumn{2}{c}{For ${\cal A}_q=0.1,~\omega_q=\frac{1}{3}$} \\
\cmidrule(r){1-2}
a    & Kerr & $V_{4~BM}$(error \%) & $V_{4~Q}$(difference \%) \\
\midrule
\multicolumn{2}{c}{For co-rotation} \\
\cmidrule(r){1-2}
0		&	4		& 4(0)							& 4.97853(24.46325)
		\\
0.1 	&	3.797	& 3.788(-0.237029233605475)		& 3.45767(-8.93679220437187)	\\
0.3		&	3.373	& 3.347(-0.770827156833686)		& 2.02851(-39.8603616958198)	\\
0.5		&	2.914	& 2.87(-1.50995195607413)		& 1.38015(-52.6372683596431)	\\
0.7		&	2.395	& 2.333(-2.58872651356993)		& 1.02863(-57.0509394572025)	\\
0.998	&	1.091	& 1.037(-4.94958753437214)		& 0.743825(-31.8217231897342)	\\
\midrule
\multicolumn{2}{c}{For counter-rotation} \\
\cmidrule(r){1-2}
-0.1	&	4.198	& 4.206(0.190566936636494)		& 8.14443(94.0073844687946)		\\
-0.3	&	4.58	& 4.606(0.567685589519646)		& 2908.82(63411.3537117904)		\\
-0.5	&	4.949	& 4.993(0.889068498686613)		& 22242.8(449340.290967872)		\\
-0.7	&	5.308	& 5.368(1.1303692539563)		& 3921.96(73787.716654107)		\\
-0.998	&	5.911	& 5.991(1.35340889866351)		& 10670.2(180414.295381492)		\\

\bottomrule
\end{tabular}
\end{center}

In table 4.1, we have listed the values of both the potentials for ${\cal A}_q=0.1$ which is relatively high and $\omega_q=\frac{1}{3}$. For co-rotation, as we increase rotation, we see the percentage of difference to decrease. But as we take counter rotation, for $a=-0.5$ our PNP is very large and after that it behaves normally.

\textbf{Table 4.2:} Comparison of two PNPs for relatively high DE effect in pressure-less dust
\begin{center}
\begin{tabular}{llll}  
\toprule
\multicolumn{2}{c}{For ${\cal A}_q=0.1,~\omega_q=0$} \\
\cmidrule(r){1-2}
a    & Kerr & $V_{4~BM}$(error \%) & $V_{4~Q}$(difference \%) \\
\midrule
\multicolumn{2}{c}{For co-rotation} \\
\cmidrule(r){1-2}
0		&	4		& 4(0)							& 7.01908(75.477)				\\
0.1 	&	3.797	& 3.788(-0.237029233605475)		& 4.43581(16.8240716355017)		\\
0.3		&	3.373	& 3.347(-0.770827156833686)		& 2.38029(-29.4310702638601)	\\
0.5		&	2.914	& 2.87(-1.50995195607413)		& 1.55044(-46.7934111187371)	\\
0.7		&	2.395	& 2.333(-2.58872651356993)		& 1.12587(-52.9908141962422)	\\
0.998	&	1.091	& 1.037(-4.94958753437214)		& 0.795028(-27.1285059578368)	\\
\midrule
\multicolumn{2}{c}{For counter-rotation} \\
\cmidrule(r){1-2}
-0.1	&	4.198	& 4.206(0.190566936636494)		& 14.2199(238.730347784659)		\\
-0.3	&	4.58	& 4.606(0.567685589519646)		& 7775.9(169679.475982533)		\\
-0.5	&	4.949	& 4.993(0.889068498686613)		& 6436.14(129949.302889473)		\\
-0.7	&	5.308	& 5.368(1.1303692539563)		& 48203.7(908033.006782216)		\\
-0.998	&	5.911	& 5.991(1.35340889866351)		& 6003.38(101462.848925732)		\\

\bottomrule
\end{tabular}
\end{center}

Here, in table 4.2 for pressure-less dust, we have observed the same behavior as it was in table 3.2, i.e., for co-rotation it is strictly monotone decreasing but in counter rotation it shows more of a highly increasing nature and for $a=-0.7$ it reaches the maxima.

\textbf{Table 4.3:} Comparison of two PNPs for relatively high DE effect of a BH embedded in quintessence
\begin{center}
\begin{tabular}{llll}  
\toprule
\multicolumn{2}{c}{For ${\cal A}_q=0.1,~\omega_q=-\frac{2}{3}$} \\
\cmidrule(r){1-2}
a    & Kerr & $V_{4~BM}$(error \%) & $V_{4~Q}$(difference \%) \\
\midrule
\multicolumn{2}{c}{For co-rotation} \\
\cmidrule(r){1-2}
0		&	4		& 4(0)							& -19343.3(-483682.5)			\\
0.1 	&	3.797	& 3.788(-0.237029233605475)		& 54.7781(342.66789570714)		\\
0.3		&	3.373	& 3.347(-0.770827156833686)		& 16.4267(387.005632967684)		\\
0.5		&	2.914	& 2.87(-1.50995195607413)		& 8.5581(193.689087165408)		\\
0.7		&	2.395	& 2.333(-2.58872651356993)		& 4.86478(103.122338204593)		\\
0.998	&	1.091	& 1.037(-4.94958753437214)		& 2.32699(113.289642529789)		\\
\midrule
\multicolumn{2}{c}{For counter-rotation} \\
\cmidrule(r){1-2}
-0.1	&	4.198	& 4.206(0.190566936636494)		& 146676(3493849.49976179)		\\
-0.3	&	4.58	& 4.606(0.567685589519646)		& 4743480(103569332.31441)		\\
-0.5	&	4.949	& 4.993(0.889068498686613)		& 6262.16(126433.845221257)		\\
-0.7	&	5.308	& 5.368(1.1303692539563)		& 66641.6(1255393.59457423)		\\
-0.998	&	5.911	& 5.991(1.35340889866351)		& 11818.7(199844.17188293)		\\

\bottomrule
\end{tabular}

\end{center}
Like previous results in table 3.3, in table 4.3 for quintessence, we also have a observation that without rotation the value of our PNP is enormously high and negative. This amount is even larger than that was given in table 3.3. On the other hand, for rest of the cases we have experienced the similar behavior of this potential but like table 4.2 it reaches the maximum point when $a=-0.7$.

\textbf{Table 4.4:} Comparison of two PNPs for relatively high DE effect in phantom barrier
\begin{center}
\begin{tabular}{llll}  
\toprule
\multicolumn{2}{c}{For ${\cal A}_q=0.01,~\omega_q=-1$} \\
\cmidrule(r){1-2}
a    & Kerr & $V_{4~BM}$(error \%) & $V_{4~Q}$(difference \%) \\
\midrule
\multicolumn{2}{c}{For co-rotation} \\
\cmidrule(r){1-2}
0		&	4		& 4(0)							& -1.24479(-131.11975)			\\
0.1 	&	3.797	& 3.788(-0.237029233605475)		& -1.24479(-132.783513299974)	\\
0.3		&	3.373	& 3.347(-0.770827156833686)		& -1.0873(-132.235398754818)	\\
0.5		&	2.914	& 2.87(-1.50995195607413)		& -0.792974(-127.212560054907)	\\
0.7		&	2.395	& 2.333(-2.58872651356993)		& -0.397126(-116.581461377871)	\\
0.998	&	1.091	& 1.037(-4.94958753437214)		& 0.293332(-73.1134738771769)	\\
\midrule
\multicolumn{2}{c}{For counter-rotation} \\
\cmidrule(r){1-2}
-0.1	&	4.198	& 4.206(0.190566936636494)		& -1.24479(-129.651977131968)		\\
-0.3	&	4.58	& 4.606(0.567685589519646)		& -1.0873(-123.740174672489)		\\
-0.5	&	4.949	& 4.993(0.889068498686613)		& -0.792974(-116.022913719943)		\\
-0.7	&	5.308	& 5.368(1.1303692539563)		& -0.397126(-107.481650339111)		\\
-0.998	&	5.911	& 5.991(1.35340889866351)		& 0.293332(-95.0375232617154)		\\

\bottomrule
\end{tabular}
\end{center}

In this table 4.4, most of the values of our potential is negative though the magnitude is very low. For phantom barrier this potential shows only repulsive nature except for two highly rotational cases, i.e., for $a=0.998$ and $-0.998$.

\section{Brief Discussions and Conclusions}
We have shown a brand new and simple calculation scheme to evaluate PNF. We have started to obtain the PNF for a rotating Kerr BH. Our derived PNF matches exactly with that of the previously derived one by Mukhopadhyay\cite{Mukhopadhyay B PNP 2002}, shown in graphs and tables, which justifies our calculation scheme. Here we succeeded to obtain a PNF for a rotating Kerr BH embedded in quintessence, which may be the more general case of study, as our universe is undergoing with late time cosmic acceleration and this DE model may explain its justification. We have shown that our potential to be efficient enough to match exactly with Mukhopadhyay's PNF for terminal values. 

Here we have taken a Kerr black hole embedded in quintessence, which was produced by S. Ghosh in 2016, so he included current aspects of our accelerating universe. If we observe this metric carefully, we see $$
ds^2=-\frac{\Delta-a^2sin^2\theta}{\Sigma}dt^2+\frac{\Sigma}{\Delta}dr^2-2asin^2\theta\left(1-\frac{\Delta-a^2sin^2\theta}{\Sigma}\right)dt d\phi$$ $$+\Sigma d\theta^2+sin^2\theta\left[\Sigma+a^2sin^2\theta\left(2-\frac{\Delta-a^2sin^2\theta}{\Sigma}\right)\right]d\phi^2,$$ where $\Sigma=r^2+a^2cos^2\theta$ and $\Delta=r^2+a^2+2Mr-\frac{{\cal A}_q}{r^{3\omega_q-1}}$, if we put ${\cal A}_q=0$ this metric turns uot to be a regular Kerr metric. Again if we put $\alpha=0$ also, it reduces to a Kiselev metric\cite{Kiselev 2003}. 

For different values of $\omega_q$, we have a different type of dark energy agents. If we start at $\omega_q=\frac{1}{3}$ then we are in the radiation era, then $\omega_q=0$ handles the particular case of pressureless dust and $\omega_q=-\frac{2}{3}$ is our main concern of study, i.e., the quintessence and finally $\omega_q=-1$ signifies the phantom barrier. If we look more carefully our obtained PNF has the same nature with B Mukhopadhyay’s one if we choose ${\cal A}_q=0$(which was also shown in the graphs and tables). Again if we put both $a=0,~{\cal A}_q=0$ then it reduces to that of Paczynski-Wiita PNF.

We have also discussed the nature of this new PNF by comparing the plots of two PNFs, i.e., for a rotating Kerr BH and a rotating Kerr BH embedded in quintessence, with respect to the radius of the central massive object. We compared both the graphs for various values of angular momentum, i.e., highly rotation, mediocre rotation and slow rotation, a various option of the intensity of DE agent. Since we have obtained a general form of PNF, we are still able to calculate the same for different aspects of DE agents like radiation era, dust, quintessence and even for phantom barriers. Comparing two sets of graphs we have shown that central BH with high angular momentum (either co-rotating or counter-rotating) possesses two specific value of radius where attraction value is massive when it's embedded in quintessence. Whereas without any DE effect we got a single region where this PNF blows up. For a mediocre value of angular momentum we also showed nearly same character but with a slight difference rather than this region being thinner in breadth, it also obtained a double point though with opposite orientation when embedded in quintessence. For high and medium rotation we have shown that BH which is embedded in quintessence attracts with a bigger value of attraction force rather than that it also possesses two regions where this force too large to determine. Effects of DE seems to be weak for a slow rotation as the comparison came up with nearly the same results although the highly accreting region is relatively fat. The most interesting result was for very slow rotation. We have shown that for quintessence and phantom barrier PNF is negative, i.e., it came up with a repulsive nature. For some particular values of ${\cal A}_q$ and $\omega_q$, this repulsive nature increases as radius increases. This type of behavior may explain the accelerating universe. Since this nature shows only for slow or no rotation it is more relevant to assume that universe consists a maximum number of very slow rotating BH than a higher or moderated one.

We have also derived the values of PNP for both the cases with DE and without DE and then tabular comparison with the observational value of that for standard Kerr BH to show a maximum error of our computational scheme in the second and third column in Table-1. For co-rotation, this error is at most $4.95 \%$ whereas for counter-rotation it's just $1.35 \%$. In the fourth column of each table we obtained the values of PNP for different aspects of DE effects, i.e., different intensity of DE effects ${\cal A}_q$ and different values of $\omega_q$. It shows in most of the cases that rotation actually increases the value of the PNP. For co-rotation, the value of PNP is less than that of the observed value of Kerr whereas for counter-rotation we observed a different scenario. Again for counter-rotating BH, this difference is enormously high in values in comparison of co-rotating one. For medium and higher values of ${\cal A}_q$ along with quintessence and phantom barrier as DE agents this PNP is repulsive for non-rotating or slowly rotating cases.

There are various aspects of this study. If we see the viewpoint of the further opportunities we have three points to consider.  One, this is a general case since other cases turn out to be the particular and terminal cases. Two, in our PNF we have too much flexibility to choose the values of $\omega_q$. So anyone can parameterize this DE agent differently to study the nature of PNF for any kind of background in which the central black hole is embedded. Three, we also have the flexibility to choose the value of ${\cal A}_q$, i.e., we can also regulate the impact of DE in a user-defined way. 

On the other hand, if we study it in a a broad sense then previous studies have no theoretical explanation to the accelerating universe since we are able to provide theoretical support in our graphical and tabular studies so our PNP is much more reliable.

Finally, for a future scope, we can use this PNF to study the accretion around a black hole surrounded by DE(and the effect of central gravitating object effect on it) or the nature of the orbits around such black hole.

\section{Acknowledgments}
This research is supported by the project grant of state Government of West Bengal, Department of higher education, Science and Technology and biotechnology (File No.: $ST/P/S\&T/16G-19/2017$). 
SSS thanks WBDST and West Bengal State Government for Senior Research Fellowship.
RB thanks IUCAA, Pune for Visiting Associateship.

RB dedicates this article to his Ph.D. supervisor Prof. Subenoy Chakraborty as a tribute on his $60^{th}$ birth year.

\end{document}